\documentclass[12pt,preprint]{aastex}

\newcommand{\WMAP}{\textsl{WMAP}}

\newcommand{\uK}{\ensuremath{\mu{\rm K}}}

\newcommand{\TauA}{Tau~A}

\newcommand{\twobytwo}[4]
{\left(\begin{array}{cc}#1&#2\\#3&#4\end{array}\right)}

\shortauthors{Page et al.}
\shorttitle{WMAP 3-Year Polarization Maps}
\slugcomment{ApJ, in press, January 5, 2007}

\begin{document}
\title{Three Year \textsl{Wilkinson Microwave 
Anisotropy Probe} (\WMAP) 
Observations:\\ Polarization Analysis} 

\author{
L. Page\altaffilmark{1},
G. Hinshaw\altaffilmark{2},
E. Komatsu \altaffilmark{12},
M. R. Nolta \altaffilmark{9},
D. N. Spergel \altaffilmark{5},
C. L. Bennett\altaffilmark{10},
C. Barnes\altaffilmark{1},
R. Bean\altaffilmark{5,8},
O. Dor\'{e}\altaffilmark{5,9},
J. Dunkley\altaffilmark{1,5},
M. Halpern \altaffilmark{3},
R. S. Hill\altaffilmark{2},
N. Jarosik \altaffilmark{1},
A. Kogut \altaffilmark{2},
M. Limon \altaffilmark{2},
S. S. Meyer \altaffilmark{4},
N. Odegard \altaffilmark{2},
H. V. Peiris \altaffilmark{4,14},
G. S. Tucker \altaffilmark{6},
L. Verde \altaffilmark{13},
J. L. Weiland\altaffilmark{2},
E. Wollack \altaffilmark{2},
E. L. Wright \altaffilmark{7}
}
 
\altaffiltext{1}{Dept. of Physics, Jadwin Hall, 
Princeton University, Princeton, NJ 08544-0708}
\altaffiltext{2}{Code 665, NASA/Goddard Space Flight Center, 
Greenbelt, MD 20771}
\altaffiltext{3}{Dept. of Physics and Astronomy, University of 
British Columbia, Vancouver, BC  Canada V6T 1Z1}
\altaffiltext{4}{Depts. of Astrophysics and Physics, KICP and EFI, 
University of Chicago, Chicago, IL 60637}
\altaffiltext{5}{Dept. of Astrophysical Sciences, 
Peyton Hall, Princeton University, Princeton, NJ 08544-1001}
\altaffiltext{6}{Dept. of Physics, Brown University, 
182 Hope St., Providence, RI 02912-1843}
\altaffiltext{7}{UCLA Astronomy, PO Box 951562, Los Angeles, CA 90095-1562}
\altaffiltext{8}{612 Space Sciences Building, 
Cornell University, Ithaca, NY  14853}
\altaffiltext{9}{Canadian Institute for Theoretical Astrophysics, 
60 St. George St, University of Toronto, 
Toronto, ON  Canada M5S 3H8}
\altaffiltext{10}{Dept. of Physics \& Astronomy, 
The Johns Hopkins University, 3400 N. Charles St., 
Baltimore, MD  21218-2686}
\altaffiltext{12}{Univ. of Texas, Austin, Dept. of Astronomy, 
2511 Speedway, RLM 15.306, Austin, TX 78712}
\altaffiltext{13}{Univ. of Pennsylvania, Dept. of Physics and Astronomy, 
Philadelphia, PA  19104}
\altaffiltext{14}{Hubble Fellow}
\email{page@princeton.edu}


\begin{abstract}

The Wilkinson Microwave Anisotropy Probe (\WMAP) has mapped the entire 
sky in five frequency bands between 23 and 94 GHz with polarization 
sensitive radiometers. We present three-year full-sky maps of the polarization 
and analyze them for foreground emission and cosmological implications. 
These observations open up a new window for understanding how the universe
began and help set a foundation for future observations. 

{\WMAP} observes significant levels of polarized foreground emission due
to both Galactic synchrotron radiation and thermal dust emission.
Synchrotron radiation is the dominant signal at $\ell<50$ 
and $\nu\la 40$~GHz, while thermal dust emission is evident at 94~GHz.
The least contaminated channel is at $61$~GHz. We present a model
of polarized foreground emission that captures the large angular
scale characteristics of the microwave sky.

After applying a Galactic mask that cuts 25.7\% of the sky, we show that 
the high Galactic latitude rms polarized foreground emission,
averaged over $\ell=4-6$, ranges from  $\approx 5~\mu$K at 22~GHz to
$\la 0.6~\mu$K at 61~GHz. By comparison, the levels of intrinsic 
CMB polarization for a $\Lambda$CDM model with an optical depth
of $\tau=0.09$ and assumed tensor 
to scalar ratio $r=0.3$ are $\approx 0.3~\mu$K for E-mode polarization
and $\approx 0.1~\mu$K for B-mode polarization. To analyze the maps for 
CMB polarization at $\ell<16$, we subtract a model of the foreground emission
that is based primarily on a scaling \WMAP's 23~GHz map.

In the foreground corrected maps, 
we detect $\ell(\ell+1)C^{EE}_{\ell=<2-6>}/2\pi = 0.086\pm0.029~(\mu K)^2$.
This is interpreted as the result of
rescattering of the CMB by free electrons released during 
reionization at $z_r =\ensuremath{11.0^{+ 2.6}_{- 2.5}}$ for 
a model with instantaneous reionization.
By computing the likelihood of just the EE data as a function of
$\tau$ we find $\tau=0.10\pm0.03$. When the same EE data
are used in the full six parameter fit to all \WMAP~data (TT, TE, EE), 
we find $\tau=0.09\pm0.03$. Marginalization over the foreground subtraction
affects this value by $\delta\tau<0.01$. 

We see no evidence for B-modes, limiting them to 
$\ell(\ell+1)C^{BB}_{\ell=<2-6>}/2\pi = -0.04\pm0.03~(\mu {\rm K})^2$.
We perform a template fit to the E-mode and B-mode data with an 
approximate model for the tensor scalar ratio. We find  
that the limit from the polarization signals
alone is $r<2.2$ (95\%~CL) where $r$ is evaluated at $k=0.002~{\rm Mpc}^{-1}$.
This corresponds to a limit on the cosmic density of 
gravitational waves of $\Omega_{GW}h^2<5\times10^{-12}$. From the full 
\WMAP~analysis, we find 
$r<\ensuremath{0.55}$ (95\% CL) 
corresponding to a limit of $\Omega_{GW}h^2<1\times 10^{-12}$
(95\% CL). The limit on $r$ is approaching 
the upper bound of predictions for some of the simplest models of inflation,
$r\sim 0.3$.
\end{abstract}

\section{Introduction}
\label{sec:intro}
 
The temperature anisotropy in the cosmic microwave background is
well established as a powerful constraint on theories of the
early universe. A related observable, 
the polarization anisotropy of the CMB, 
gives us a new window into the physical conditions of that era.
At large angular scales the polarization has the potential to  
be a direct probe of the universe at an age of $10^{-35}$~s
as well as to inform us about the ionization history of the universe.
This paper reports on the direct detection of CMB polarization 
at large angular scales and helps set a foundation for future observations. 
It is one of four related papers on the three-year \WMAP\ analysis:
\citet{jarosik/etal:prep} report on systematic errors and mapmaking,
\citet{hinshaw/etal:prep} on the temperature anisotropy and basic results,
and \citet{spergel/etal:prep} on the parameter estimation and cosmological
significance. 
 
The polarization of the CMB was predicted
soon after the discovery of the CMB \citep{rees:1968}.
Since then, considerable advances have been
made on both theoretical and observational fronts. The theoretical
development \citep{
basko/polnarev:1980,
kaiser:1983,
bond/efstathiou:1984,
polnarev:1985,
bond/efstathiou:1987,
crittenden/davis/steinhardt:1993,
harari/zaldarriaga:1993,
frewin/polnarev/coles:1994,
coulson/crittenden/turok:1994,
crittenden/coulson/turok:1995,
ng/ng:1995,
zaldarriaga/harari:1995,
kosowsky:1996,
seljak:1997,
zaldarriaga/seljak:1997,
kamionkowski/kosowsky/stebbins:1997}
has evolved to where there are precise predictions
and a common language to describe the
polarization signal. \citet{hu/white:1997} give a pedagogical 
overview.

The first limits on the polarization were 
placed by \citet{penzias/wilson:1965}, 
followed by \citet{caderni/etal:1978,nanos:1979,lubin/smoot:1979,
lubin/smoot:1981,lubin/melese/smoot:1983, wollack/etal:1993,
netterfield/etal:1997, sironi/etal:1997, torbet/etal:1999, keating/etal:2001}
and \citet{hedman/etal:2002}.
In 2002, the DASI team announced a detection of CMB
polarization at sub-degree angular scales
based on 9 months of data from a 13 
element 30~GHz interferometer \citep{kovac/etal:2002,leitch/etal:2002}.
The signal level was consistent with that
expected from measurements of the temperature spectrum. 
The DASI results were confirmed and extended \citep{leitch/etal:2005} 
almost contemporaneously with the release of the 
CBI \citep{readhead/etal:2004b}
and CAPMAP \citep{barkats/etal:2005} results. More recently, the Boomerang
team has released its measurement 
of CMB polarization \citep{montroy/etal:2005}. 
All of these measurements were made at small angular scales ($\ell>100$). 
Of the experiments that measure the polarization, 
the DASI, CBI, and Boomerang \citep{piacentini/etal:2005} teams
also report detections of the temperature-polarization cross
correlation. 

The CMB polarization probes the evolution of the decoupling 
and reionization epochs. The polarization signal is generated
by Thompson scattering of a local quadrupolar radiation 
pattern by free electrons. The scattering 
of the same quadrupolar pattern in a direction perpendicular to the 
line of sight to the observer has the effect of isotropizing the
quadrupolar radiation field. The net polarization results
from a competition between these two effects. 
We estimate the magnitude of the signal 
following Basko and Polnarev (1980).
By integrating the Boltzmann equation for the photon distribution 
they show that the ratio of the polarization anisotropy ($E_{rms}$) to 
the temperature ($T_{rms}$) signal in a flat cosmology is given by
\begin{equation}
{E_{rms}\over T_{rms}} 
= { \int_0^\infty [e^{-0.3\tau(z')} - e^{-\tau(z')}]\sqrt{1+z'}dz'\over 
     \int_0^\infty [6e^{-\tau(z')} + e^{-0.3\tau(z')}]\sqrt{1+z'}dz'},
\label{eq:ih}
\end{equation}
where $\tau(z)=c\sigma_T\int_{0}^{z}n_e(z')dz'(dt/dz')$ is the optical depth. 
Here, $\sigma_T$ is the Thompson cross section, $c$
is the speed of light, and $n_e$ is the free electron density.
The difference in brackets in the numerator sets the range in $z$
over which polarization is generated. For example, if
the decoupling epoch entailed an instantaneous transition from 
an extremely high optical depth ($\tau>>1$) to transparency ($\tau=0$),
there would be no polarization signal. 

To estimate the polarization fraction we compute
the optical depth using ordinary atomic physics 
and the thermal history of the universe 
\citep{peebles:1968, zeldovich/kurt/sunyaev:1969}. The result is shown
in Figure~1. From inserting $\tau(z)$ in Equation~\ref{eq:ih},
we find that the expected level of polarization anisotropy
is $\approx5$\% (in $E_{rms}/T_{rms}$) of the anisotropy. 

The polarization producing quadrupole is generated by different 
mechanisms at different epochs.
Near decoupling at $z_d=1088$  
\citep{page/etal:2003,spergel/etal:2003},
velocity gradients in the flow of the primordial plasma give rise to 
the quadrupole. More specifically, in the rest frame of an electron in 
such a flow, the radiation background has a quadrupolar pattern 
proportional to the velocity gradient, $\nabla\vec{\rm v}$, and the 
mean free path between scatterings, $\lambda$. Just before decoupling,
$z>z_d$, the photons are tightly coupled to the electrons 
and $\lambda$ is small. 
Thus, the polarization is small.
As decoupling proceeds $\lambda$ increases and the quadrupole magnitude 
increases. The process is cut off at lower redshift because the optical 
depth drops so rapidly. In the context of inflationary cosmology, 
\citet{harari/zaldarriaga:1993} show that in Fourier space the 
polarization signal is
$\propto kv\Delta $ where $k$ is the wavevector and 
$\Delta\approx\lambda$ is the width of the
last scattering surface. 

After decoupling there are no free electrons to scatter the CMB
until the first generation of stars ignite and reionize the
universe at $z_r$. The free electrons then scatter the intrinsic 
CMB quadrupole, $C_2(z_r)$, and produce a polarized signal
$\propto C_2(z_r)^{1/2} \tau(z_r)$. As this process occurs well 
after decoupling, the effects of the scattering are manifest
at comparatively lower values of $\ell$. We expect the maximum value of the 
signal to be at $\ell_{max} \approx \pi/\theta_H(z_r)$ where 
$\theta_H(z_r)$ is the current angular size of the horizon at 
reionization. For $6<z<30$ a simple fit 
gives $\theta_H(z)=4.8/z^{0.7}$, so that for $z_r=12$, $\ell_{max}\approx 4$.
Thus, the signature of 
reionization in polarization is cleanly separable from the
signature of decoupling. 
In the first data release the {\WMAP} team published a measurement of
the temperature-polarization (TE) cross spectrum for $2<\ell<450$
\citep{bennett/etal:2003,kogut/etal:2003} with distinctive anti-peak and
peak structure \citep{page/etal:2003}. 
The $\ell>16$ part of the spectrum was consistent with the prediction
from the temperature power spectrum,
while the $\ell<16$ part showed an excess that was 
interpreted as reionization at $11<z_r<30$ (95\% CL).

This paper builds on and extends these results. 
Not only are there three times as much data, but the analysis has 
improved significantly: 
1) The polarization mapmaking pipeline now self-consistently includes 
almost all known effects and correlations due to instrumental systematics, 
gain and offset drifts, unequal weighting, and masking 
\citep{jarosik/etal:prep}. For example, the noise matrix is no longer
taken to be diagonal in pixel space, leading to new estimates of the 
uncertainties. 2) The polarization power spectrum estimate now consistently
includes the temperature, E and B modes (defined below), 
and the coupling between them
\citep[see also ][]{hinshaw/etal:prep}. 3) The polarized foreground 
emission is now modeled and subtracted in pixel space 
(\S\ref{sec:fgr}). 
Potential residual contamination is examined $\ell$ by $\ell$
as a function of frequency. In addition to
enabling the production of full sky maps of the polarization and their
power spectra, the combination of these three improvements has led to a new
measure of the $\ell<16$ TE and EE spectra, and therefore a new evaluation 
of the optical depth based primarily on EE. The rest of the paper is 
organized as follows: we discuss the measurement
in \S\ref{sec:inst} and consider systematic errors and maps in
\S\ref{sec:syst}. In \S\ref{sec:foregrounds} we discuss 
foreground emission. We then consider, in \S\ref{sec:pspec} 
and \S\ref{sec:cosmo},
the polarization power spectra and their cosmological
implications. We conclude in \S\ref{sec:conclude}.   

\section{The Measurement}
\label{sec:inst}

\WMAP\ measures the difference in intensity 
between two beams separated by $\approx 140\degr$ 
in five frequency bands centered on 23, 33, 41, 61, and 94 GHz
\citep{bennett/etal:2003,page/etal:2003,jarosik/etal:2003}.
These are called K, Ka, Q, V, and W bands respectively.
Corrugated feeds \citep{barnes/etal:2002} couple radiation from 
back-to-back telescopes to the differential radiometers.
Each feed supports two orthogonal polarizations aligned so that 
the unit vectors along the direction of maximum electric field
for an A-side feed
follow $(x_s,y_s,z_s)\approx (\pm 1,-\sin 20^\circ,-\cos20^\circ )/\sqrt{2}$ 
in spacecraft coordinates
\citep{page/etal:2003}. For a B-side feed, the directions are
$(x_s,y_s,z_s)\approx (\pm 1, \sin 20^\circ,-\cos20^\circ )/\sqrt{2}$. 
The $z_s$ axis points toward the Sun along the 
spacecraft spin axis; the $y_s-z_s$ plane bisects the telescopes
and is perpendicular to the radiator panels 
\citep[Figure 2]{bennett/etal:2003} \citep[Figure 1]{page/etal:2003}.
The angle between the spacecraft spin axis and the optical axes
is $\approx 70^{\circ}$.
Thus the two polarization axes on one side are oriented 
roughly $\pm45^{\circ}$ with respect to the spin axis. 

The polarization maps are derived from the difference of two differential
measurements \citep{jarosik/etal:prep,kogut/etal:2003,hinshaw/etal:2003}. 
One half of one differencing 
assembly (DA) \citep{jarosik/etal:2003} measures the difference 
between two similarly oriented polarizations, $\Delta T_1$, from one 
feed on the A side and one feed on the B side
(e.g., W41: polarization 1 of the 4th W-band DA 
corresponding to $x_s=+1$ in both expressions above). The
other half of the DA measures the difference between the other polarizations
in the same pair of feeds, $\Delta T_2$ 
(e.g., W42: polarization 2 of the 4th W-band DA corresponding to $x_s=-1$
in both the expression above). 
The polarization signal is proportional to $\Delta T_1-\Delta T_2$. 
In other words, \WMAP\ measures a double difference in polarized 
intensity, not the intensity of the difference of electric 
fields as with interferometers and correlation receivers
\citep[e.g.,][]{leitch/etal:2002,keating/etal:2001,hedman/etal:2002}.

With these conventions,
the total intensity and polarization signals
as measured at the output of the detectors are
\citep[Eq. 3\&4]{kogut/etal:2003}:
\begin{eqnarray}
\Delta T_{I} &\equiv& \frac{1}{2}(\Delta T_1 + \Delta T_2)
         = I(\hat{n}_A) - I(\hat{n}_B)\\
\Delta T_{P} &\equiv& \frac{1}{2}(\Delta T_1 - \Delta T_2)\\
  &=& Q(\hat{n}_A) \cos 2\gamma_A + U(\hat{n}_A) \sin 2\gamma_A \\
  & & - Q(\hat{n}_B) \cos 2\gamma_B - U(\hat{n}_B) \sin 2\gamma_B.\nonumber
\label{eq:diff_defs}
\end{eqnarray}
where $n_A$ and $n_B$ are the unit vectors for the A and B sides;
$I$, $Q$, and $U$ are the Stokes parameters\footnote{Italics are used to
distinguish between the similarly notated Q
band and $Q$ Stokes parameter.}, and $\gamma$ is the angle
between the polarization direction of the electric field
and the Galactic meridian \citep{kogut/etal:2003}. 
In the mapmaking
algorithm \citep{wright/hinshaw/bennett:1996,hinshaw/etal:2003,
jarosik/etal:prep},
$I$, $Q$, and $U$
maps of the sky are produced from the time-ordered differential
measurements, $\Delta T_{I}$ and $\Delta T_{P}$. From these,
we form maps of polarization intensity, $P=\sqrt{Q^2+U^2}$, and direction,
$\gamma={1\over 2}\tan^{-1}(U/Q)$. This convention has 
$\gamma$ positive for North through West and
follows the convention in \citet{zaldarriaga/seljak:1997} and 
HEALPix \citep{gorski/hivon/wandelt:1998}. However, it differs 
from the standard astronomical position angle (PA) which has 
$\gamma_{PA}={1\over 2}\tan^{-1}(-U/Q)$ with $\gamma_{PA}$ positive for 
North through East. The choice of convention does not affect the 
plots.

For linear polarization in a given pixel,
the $Q$ and $U$ quantities are related to the $x$ and $y$ components
of the electric field, $E_x, E_y$, through the coherency matrix
\citep{born/wolf:POO:6e}:
\begin{eqnarray}
 \twobytwo{\langle E_xE_x^*\rangle}
{\langle E_xE_y^*\rangle}{\langle E_yE_x^*\rangle}
{\langle E_yE_y^*\rangle} &=&
\frac{1}{2} \twobytwo{I}{0}{0}{I} +
\frac{1}{2} \twobytwo{Q}{U}{U}{-Q} \cr
&=&
\frac{I}{2} \twobytwo{1}{0}{0}{1} +
\frac{P}{2} \twobytwo{\cos 2\gamma}{\sin 2\gamma}{\sin 2\gamma}
{-\cos 2\gamma}
\end{eqnarray}
where we have set Stokes $V=0$.
The polarized component of the coherency matrix is a spin-two field
on a sphere; the total power is the trace of the coherency matrix.
 
The Crab Nebula [Tau~A, 3C144, RA = ${\rm 05^h34^m31^s}$, 
Dec=$22^{\circ} 01^{\prime}$ (J2000)] is the brightest
polarized point source in the sky and provides a useful end-to-end
check of the sign conventions and mapmaking pipeline.
Figure~\ref{fig:crab} shows our measurement
of the Crab in Q band (41~GHz) in $I$, $Q$, $U$, $P$, and $\gamma$.
Note that its polarization direction ($U\approx0,~Q~$negative), 
is perpendicular 
to the polarization of the Galaxy ($U\approx0,~Q~$positive).
The \WMAP\ polarization direction and intensity are in general agreement
with previous measurements. Table~\ref{tab:crab} summarizes
the results in all five frequency bands and previous measurements in
our frequency range. A second check is needed to fully resolve
the sign convention because with $U=0$, $\gamma=\gamma_{PA}$. 
In Figure~\ref{fig:crab} we show that the polarization 
direction of the Centaurus A 
galaxy [Cen~A, NGC5128, RA=${\rm 13^h25^m27^s}$, 
Dec$= -43^{\circ}01^\prime 09^{\prime\prime}$ (J2000)] 
is consistent with that measured by \citet{junkes/etal:1993}. 

Figures~\ref{fig:pgmaps1} and \ref{fig:pgmaps2} show
the $P$ and $\gamma$ maps of the full sky for all five frequency bands
in Galactic coordinates.
Figure~\ref{fig:kstereo} shows a Lambert equal area projection
of the Galactic polar region in K band.
A number of features are immediately apparent to the eye.
K band is strongly polarized over a large fraction of the sky,
including the polar region.
The North Polar Spur and its southern extension are clearly 
evident.
The polarization has a coherent structure over large swaths of sky
which translates into significant emission at low $\ell$. The
polarization intensity decreases with increasing frequency but follows 
the same pattern. K band is a good monitor of polarized foreground emission
as discussed below.
Though not immediately apparent to the eye, there is somewhat more 
polarized emission at W band than V band.
The uneven weighting due to the scan strategy is also evident 
as increased noise in the ecliptic plane \citep[Figure 4]{bennett/etal:2003}.
Figure~\ref{fig:qumaps} shows the K and Ka bands in Stokes $Q$ and $U$.

While foreground emission is visible with
a high signal to noise ratio, the CMB polarization anisotropy is not,
a situation unlike that for the temperature anisotropy. 

\section{Systematic Errors}
\label{sec:syst}

Detection of the CMB polarization requires 
tight control of systematic errors, as small couplings to the
temperature field or instrument will dominate the 
polarization signal. \WMAP's differential nature and interlocked scan strategy
suppress potential polarization systematics in ways similar
to the suppression for temperature systematics.
The details are different however, and more complex
because of the tensorial nature of the polarization field 
and the double difference required to measure the polarization. 
Throughout our analyses, the overall level of systematic
contamination is assessed with null tests as described here and 
in \citet{jarosik/etal:prep} \& \citet{hinshaw/etal:prep}.

The mapmaking procedure is described
in \citet{jarosik/etal:prep}. End-to-end simulations of the 
instrument and scan strategy,
incorporating realistic models of the frequency response,
foreground emission, and detector noise characteristics, are used to
assess the possible levels of contamination.
Interactions between the slow $<1$~\% 
drifts in the gain, non-uniform weighting across the sky,
the 0.2\% correlation due to the oppositely directed beams, the time 
series masking of the planets, and the $1/f$ noise are accounted for 
in the map solution. 
In the following we discuss how the instrumental offset,
gain/calibration uncertainty, passband mismatch, main beam mismatch,
polarization isolation and cross polarization, 
loss imbalance, and sidelobes affect the polarization maps.

\textit{Offset and baseline drift----}
The instrumental offset is the output of the detector in the 
absence of celestial signal. The average polarization offset in 
the Q, V, and W bands is 250~mK. Changes in this offset on time scales
of minutes to hours arise from spacecraft temperature
changes and from $1/f$ drifts in the amplifier gain acting on the 250~mK. 
To measure polarization at the level of $0.1~\mu$K, we require that changes in 
the baseline be suppressed by roughly a factor of $10^6$. The first
step in achieving this is maintaining a stable instrument and environment.  
The physical temperature of the DAs averaged
over a spin period changes by less than 5 parts in $10^6$ 
\citep{jarosik/etal:prep}, suppressing changes in the baseline 
by a similar factor. The second step in achieving this is through 
the baseline removal in the mapmaking 
algorithm \citep{hinshaw/etal:2003,jarosik/etal:prep}.

If the precession of the satellite were stopped, the temperature 
data for $\ell>1$ would repeat in the time stream at the spin 
period (2.16~m).
The offset, though, would change sign relative to the celestial signal at 
half the spin period enabling the differentiation 
of celestial and instrumental signals. (Alternatively, one may imagine
observing a planet in which case the temperature data would change sign 
at half the spin period and an offset would be constant.)
By contrast, with our choice of polarization 
orientations, the polarization data $\Delta T_P$, would repeat
at half the spin period for some orientations of the satellite. 
Consequently, an instrumental offset would not change sign relative 
to a celestial signal upon a $180^\circ$ spacecraft rotation. 
Thus the polarization data are more sensitive to instrumental offsets 
than are the temperature data. In general, the polarization data 
enters the time stream in a more complex manner than does the temperature
data. 

{\it Calibration---} An incorrect calibration between channels
leads to a leakage of the temperature signal
into $\Delta T_P$, contaminating the polarization map.
Calibration drifts cause a leakage that varies across the sky.
\citet{jarosik/etal:2003} show that calibration drifts on $\approx 1$ day
time scales are the result of sub-Kelvin changes in 
the amplifier's physical temperature. The calibration can 
be faithfully modeled by fitting to the physical temperature of each DA
with a three parameter model. Here
again \WMAP's stability plays a key role. The residual
calibration errors are at the $\approx 0.2$\% level. These errors
do not limit the polarization maps because the bright Galactic plane 
is masked in the time ordered data when producing the high Galactic 
latitude maps \citep{jarosik/etal:prep}. The overall absolute calibration
uncertainty is still the first-year value, 0.5\% \citep{jarosik/etal:prep}.

{\it Passband mismatch---} The effective central frequencies 
\citep{jarosik/etal:2003b,page/etal:2003} for 
$\Delta T_1$ and $\Delta T_2$ are not the same. 
This affects both the beam patterns, treated below,
and the detected flux from a celestial source, treated in the following.
The passbands for the A and B sides of one polarization channel in a
DA may be treated as the same because the dominant contributions to the 
passband definition, the amplifiers and band defining filters, are
common to both sides.

Since \WMAP\ is calibrated on the CMB dipole, the presence of a 
passband mismatch means that the response to radiation with a non-thermal 
spectrum is different from the response to radiation with a CMB spectrum
\citep{kogut/etal:2003,hinshaw/etal:2003}. This would be true 
even if the sky were unpolarized, the polarization offset zero, 
and the beams identical. The effect produces a response in the
polarization data of the form:
\begin{eqnarray}
\Delta T_{P} &=& \Delta I_1 - \Delta I_2 + \\ \nonumber
  & & Q(\hat{n}_A) \cos 2\gamma_A + U(\hat{n}_A) \sin 2\gamma_A \\ \nonumber
  & & - Q(\hat{n}_B) \cos 2\gamma_B - U(\hat{n}_B) \sin 2\gamma_B.
\end{eqnarray}
where $\Delta I_1$ is the unpolarized temperature difference observed
in radiometer one, and similarly for $\Delta I_2$. If these differ due to 
passband differences, the polarization data will have an output
component that is independent of parallactic angle. Given sufficient
paralactic coverage, such a term can be separated from Stokes $Q$ and $U$
in the mapmaking process. We model the polarized signal as 
$Q\cos 2\gamma + U\sin 2\gamma + S$ where the constant, $S$,
absorbs the signal due to passband mismatch. 
We solve for the mismatch term simultaneously with $Q$ and $U$
as outlined in \citet{jarosik/etal:prep}. Note that we do not need
to know the magnitude of the passband mismatch, it is fit for in the 
mapmaking process. The $S$ map
resembles a temperature map of the Galaxy but at a reduced amplitude 
of $3.5\%$ in K band, 2.5\% in the V1 band, and on average $\approx 1$\% for 
the other bands. The maps of $S$ agree with the expectations based on the 
measured passband mismatch.

{\it Beamwidth mismatch---} The beamwidths of each polarization
on each of the A and B sides are different. The difference between
the A and B side beam shapes is due to the difference in shapes
of the primary mirrors and is self consistently 
treated in the window function \citep{page/etal:2003}.
The difference in beam shapes between $\Delta T_1$ and $\Delta T_2$
is due to the mismatch in central frequencies.\footnote{If the passbands
were the same, the beam solid angles for 
$\Delta T_1$ and $\Delta T_2$ would be the same to $<0.5$\% accuracy.} 

This effect is most easily seen in the K-band observations of 
Jupiter. We denote the brightness temperature and solid angle of 
Jupiter with $T_J$ and $\Omega_J$, and the measured quantities as
$\hat{T}_J$ and $\hat{\Omega}_J$.
Although the product $T_J\Omega_J=\hat{T}_J\hat{\Omega}_J$
is the same for the two polarizations (because
Jupiter is almost a thermal source in K band),
the beam solid angles differ by 8.1\% on the A-side and
6.5\% on the B-side \citep{page/etal:2003b}. 
The primary effect of the beamwidth mismatch is to complicate the
determination of the intrinsic polarization of point sources.

The difference in beams also leads to a small difference in window functions
between $\Delta T_1$ and $\Delta T_2$. The signature would be leakage of power 
from the temperature anisotropy into the polarization signal at high $\ell$. 
We have analyzed the data for evidence of this effect and found it to be 
negligible. Additionally, as most of the CMB and foreground polarization 
signal comes from angular scales much larger than the beam, the difference 
in window functions can safely be ignored in this data set.

\textit{Polarization isolation and cross polarization---}
Polarization isolation, $X_{cp}$, and cross polarization are measures 
of the leakage of electric field from one polarization into the measurement of 
the orthogonal polarization. For example, if a source were fully polarized 
in the vertical direction with 
intensity $I_v$ and was measured to have intensity $I_h=0.01I_v$ with a 
horizontally polarized detector, one would say that the cross polar response
(or isolation) is $|X_{cp}|^2=1\%$ or $-20$~dB. The term ``polarization 
isolation'' 
is usually applied to devices whereas ``cross polarization'' is applied to the
optical response of the telescope. We treat these together as a 
cross-polar response.
For \WMAP, the off-axis design and imperfections in the orthomode
transducers (OMT) lead to a small cross-polar response. 
The ratio of the maximum of the modeled crosspolar beam
to the maximum of the modeled copolar beam is $-25$, $-27$, $-30$, $-30$, 
\& $-35$~dB in K through W bands respectively. The determination of the 
feed and OMT polarization isolation is limited by component
measurement. The maximum
values we find are: $|X_{cp}|^2=$ $-40$, $-30$, $-30$, $-27$, \& $-25$~dB for 
K through W bands respectively \citep{page/etal:2003}. We consider the 
combination of beams plus components below.

Because \WMAP\ measures only the difference in power from two
polarizations, it measures only Stokes $Q$ in a reference
frame fixed to the radiometers,
$Q_{Rad}$. The sensitivity to celestial Stokes $Q$ and $U$ comes through
multiple observations of a single pixel with different orientations
of the satellite.
The formalism that describes how cross polarization interacts with the 
observations is given in Appendix~\ref{app:jones}.
To leading order,
the effect of a simple cross polarization of the 
form $X_{cp}=Xe^{iY}$ is to rotate some of the radiometer $U$ 
into a $Q$ component. The measured quantity becomes:
\begin{equation}
    \Delta T_P = Q^A_{Rad} + Q^B_{Rad} + 2X\cos(Y)(U^A_{Rad}+U^B_{Rad})
\end{equation}
where $Q^A_{Rad}$ and $Q^B_{Rad}$ are the Stokes $Q$ components for the A
and B sides in the radiometer frame, similarly 
with $U^A_{Rad}$ and $U^B_{Rad}$. Note that in the frame of the radiometers
$Q^B_{Rad}$ (Stokes $Q$ in the B-side coordinate system) is
$-Q^A_{Rad}$. This leads to the difference in sign conventions
between the above and Equation~\ref{eq:diff_defs}.  
System measurements limit the magnitude of
$|X_{cp}|^2$ but do not directly give the phase, $Y$.
Laboratory measurements of selected OMTs 
show $Y=90^\circ\pm5^\circ$, indicating
the effective cross polar contamination is negligible.

We limit the net effect of the reflectors and OMT
with measurements in the GEMAC antenna range \citep{page/etal:2003}. 
We find that for a linearly polarized input, the ratios of the
maximum to minimum responses of the OMTs are 1) $-25$, $-27$, $-25$,
$-25$, $-22$~dB for K through W band respectively; 2)
$90^\circ\pm2^\circ$ apart; and 3) within $\pm 1.5^{\circ}$ of the 
design orientation. Thus, we can limit any rotation of one
component into another to $<2^\circ$.
The comparison of $\gamma$ derived from \TauA\ to the measurement by 
\citet{flett/henderson:1979} in Table~\ref{tab:crab} gives 
further evidence that any possible rotation of the Stokes 
components is minimal. Based on these multiple checks, we
treat the effects of optical cross polarization and 
incomplete polarization isolation as negligible.

\textit{Loss imbalance---}
A certain amount of celestial radiation is lost to absorption by the optics
and waveguide components.
If the losses were equal for each of the four radiometer inputs
their effect would be indistinguishable from a change in the gain calibration.
However, small differences exist that produce a residual common-mode
signal that is separable from the gain drifts \citep{jarosik/etal:2003}.
The mean loss difference ($\bar{x}_{\rm im}$)
between the A- and B-sides is accounted for
in the mapmaking algorithm \citep{hinshaw/etal:2003b, jarosik/etal:2003}.
In addition, the imbalance between the two polarizations on a single side,
the ``loss imbalance imbalance,'' is also included \citep{jarosik/etal:prep}.
It contributes a term $2(L^A T^A+L^B T^B)$ to $\Delta T_P$.
Here $T^{A,B}$ is the sky temperature observed
by the A,B side, and $L^{A,B}$ is the loss imbalance between the two
polarizations on the $A,B$ side (see Appendix~\ref{app:jones}).
The magnitude of $L^{A,B}$ is $\la1\%$ \citep{jarosik/etal:2003b}.

A change in the loss across the 
bandpass due to, for example, the feed horns is a potential 
systematic error that we do not quantify with the radiometer passband 
measurements \citep{jarosik/etal:2003b}.
The magnitude of the effect is second order to the 
loss imbalance which is 1\%. We do not have a measurement of 
the effect. Nevertheless, as the effect mimics a passband
mismatch, it is accounted for in the map solution. 

\textit{Sidelobes---} When the sidelobes corresponding to $\Delta T_P$
are measured, there are two terms \citep{barnes/etal:2003}. 
The largest term is due to the passband mismatch and is consistently 
treated in the mapmaking process. The second smaller term is 
due to the intrinsic polarization. 
We assess the contribution of both terms by simulating the 
effects of scan pattern of the sidelobes on 
the $Q$ and $U$ polarization maps. The results are reported in 
\citet{barnes/etal:2003} for the 
first-year polarization maps. In K band, the net rms contamination 
is $1\,\uK$ outside of the Kp0 mask region \citep{bennett/etal:2003}.
The intrinsic polarized sidelobe pickup is $<1\,\uK$ and is
not accounted for in this three-year data release. The contamination 
is more than an order of magnitude smaller in the other bands.

\section{The Foreground Emission Model}
\label{sec:foregrounds}

The microwave sky is polarized at all frequencies 
measured by \WMAP. In K band the polarized flux exceeds 
the level of CMB polarization over the full sky. By contrast, 
unpolarized foreground emission dominates over the CMB
only over $\approx20\%$ of the sky. Near 60~GHz and $\ell\approx5$, 
the foreground emission temperature is roughly a factor of two 
larger than the CMB polarization signal.
Thus, the foreground emission must be subtracted before a
cosmological analysis is done. While it is possible to make significant 
progress working with angular power spectra, we find that due to the 
correlations between foreground components, a pixel space subtraction 
is required.
Table~\ref{tab:gcent} gives the foreground emission levels in a region around
the Galactic center.

The two dominant components of diffuse polarized foreground 
emission in the $23-94~$GHz range are synchrotron emission
and thermal dust emission \citep{weiss:1984,
bennett/etal:2003}. Free-free emission is 
unpolarized\footnote{There may be polarized emission at the edges of 
HII clouds as noted in \citet{keating/etal:1998}.}
and spinning dust grains are expected to have polarization 
fractions of ~1-2\% \citep{lazarian/draine:2000}.
The signal from polarized radio sources is negligible
\citep[Table 9,][]{hinshaw/etal:prep}.  
The detected polarized sources are all well known, and among the 
brightest objects in the temperature source catalog. They include  
3C273, 3C274 (M87, Vir A), 3C279, Fornax A, Pictor A, [HB93]2255-282, 
and [HB93] 0637-752 and are masked as discussed below.
The potential impact of polarized 
foreground emission on the 
detection of the CMB polarization has been discussed by many authors including
\citet{verde/peiris/jimenez:2006, ponthieu/etal:2005, 
deoliveira-costa/etal:2003a, 
giardino/etal:2002, tucci/etal:2002, baccigalupi/etal:2001, 
tegmark/etal:2000}. 

Synchrotron emission is produced
by cosmic-ray electrons orbiting in 
the $\approx6~\mu$G total Galactic magnetic 
field. The unpolarized synchrotron component has been well measured 
by \WMAP\ in the 23 to 94 GHz range \citep{bennett/etal:2003c}.
The brightness temperature of the radiation is characterized
by $T(\nu)\propto \nu^{\beta_s}$ where the index $-3.1<\beta_s<-2.5$
varies considerably across the sky \citep{reich/reich:1988,lawson/etal:1987}.
In the microwave range, the spectrum reddens ($\beta_s$ tends to more negative 
values) as the frequency increases \citep{banday/wolfendale:1991}. 

Synchrotron radiation can be strongly polarized in the direction 
perpendicular to the Galactic magnetic field \citep{rybicki/lightman:1979}. 
The polarization has been measured at 
a number of frequencies [from Leiden between 408 MHz to 1.4 
GHz \citep{brouw/spoelstra:1976,wolleben/etal:2005}, 
from Parkes at 2.4 GHz \citep{duncan/etal:1995,duncan/etal:1999}, and
by the Medium Galactic Latitude Survey at 
1.4 GHz \citep{uyaniker/etal:1999}].
At these low frequencies, Faraday rotation alters the 
polarization.
Electrons in the Galactic magnetic field rotate the plane of
polarization because the constituent left and right circular polarizations  
propagate with different velocities in the medium. 
In the interstellar medium,
the rotation is a function of electron density, $n_e$, and 
the component of the Galactic magnetic field along
the line of sight, $B_{||}$,
\begin{equation}
\Delta\theta = 420^{\circ}\biggl(\frac{1~{\rm GHz}}{\nu}\biggr)^2
\int_0^{L/1~{\rm kpc}}dr\, \biggl(  \frac{n_e}{ 0.1~{\rm cm^{-3}} } \biggr)
                      \biggl( \frac{B_{||}}{ 1~\mu{\rm G}}\biggr)
\end{equation}
where the integral is over the line of sight. With 
$n_e\sim 0.1\,{\rm cm}^{-3}$, $L\sim 1\,{\rm kpc}$,
and $B_{||}\sim 1\,\mu{\rm G}$, 
the net rotation is $\Delta\theta\sim 420\degr/\nu^2$, with $\nu$ in GHz.
At \WMAP\ frequencies the rotation is negligible, though
the extrapolation of low frequency polarization
measurements to \WMAP\ frequencies can be problematic.
In addition there may be both observational and 
astrophysical depolarization effects that are different at lower 
frequencies \citep{burn:1966,cioffi/jones:1980,cortiglioni/spoelstra:1995}.
Thus, our method for subtracting the foreground emission
is based, to the extent possible, on the polarization directions 
measured by \WMAP\ .
 
The other dominant component of polarized foreground emission comes 
from thermal dust.
Nonspherical dust grains 
align their long axes perpendicularly to the Galactic magnetic 
field through the
Davis-Greenstein mechanism \citep{davis/greenstein:1951}.
The aligned grains preferentially absorb
the component of starlight polarized along their longest axis. Thus,
when we observe starlight we see it polarized in the same direction
as the magnetic field. These same grains emit thermal 
radiation preferentially polarized along their longest axis, 
perpendicular to the 
Galactic magnetic field. Thus we expect to observe thermal dust emission
and synchrotron emission polarized in the same direction, while starlight is
polarized perpendicularly to both. 

In Section~\S\ref{sec:mag}, we describe a physical model of the polarized 
microwave emission from our Galaxy that explains the general features
of the \WMAP\ polarization maps. {\it However this model is not directly used 
to define the polarization mask or to clean the polarization maps.} 
We go on to define the polarization masks in \S\ref{sec:mask} and 
in \S\ref{sec:fgr} we describe how we subtract the polarized foreground 
emission.

\subsection{The Galaxy Magnetic Field and a Model of Foreground Emission.}
\label{sec:mag}

In the following, 
we present a general model of polarized foreground emission based on
\WMAP\ observations. We view this as a starting point aimed at understanding 
the gross features of the \WMAP\ data. A more detailed model
that includes the wide variety of external data sets that relate to
polarization is beyond the scope of this paper. 

For both synchrotron and dust emission, the Galactic magnetic field
breaks the spatial isotropy thereby leading to polarization. Thus, to 
physically model
the polarized foreground emission we
need a model of the Galactic magnetic field. As a first step, we note
that the K-band polarization maps suggest a 
large coherence scale for the Galactic magnetic field, as shown
in Figure~\ref{fig:pgmaps1}.

We can fit the large-scale field structure seen in
the K-band maps with a gas of cosmic ray electrons interacting
with a magnetic field that follows the spiral arms. The Galactic
magnetic field can be quite complicated
\citep{beck:2006,han/etal:2006,reich:2006, wielebinski:2005}: there are field
direction reversals in the Galactic plane; the field strength 
depends on length scale, appearing turbulent on scales $<80$~kpc
\citep[e.g.,][]{mitner/spangler:1996}; and the field
strength of the large-scale field depends on the Galactocentric 
radius \citep[e.g.,][]{beck:2001}.
Nevertheless, most external spiral galaxies have magnetic fields
that follow the spiral arm pattern  
\citep[e.g.,][]{wielebinski:2005,beck/etal:1996,sofue/etal:1986}.
Inspired by this, we model the field in cylindrical coordinates as:

\begin{eqnarray}
\label{eq:bfield} 
{\bf B}(r,\phi,z) = B_0 [&\cos& \psi(r) 
                        \cos \chi(z)\hat{r}+\\ \nonumber
   &\sin& \psi(r) \cos \chi(z)\hat{\phi} +\\ \nonumber
   &\sin& \chi(z)\hat{z} ]
\end{eqnarray}
where $\psi(r) = \psi_0 + \psi_1\ln(r/8~{\rm kpc})$, 
$\chi(z) = \chi_0 \tanh(z/1~{\rm kpc})$, $r$ and $z$ are measured in kpc
with respect to the center of the Galaxy,
$r$ ranges from $3~{\rm kpc}$ to $20~{\rm kpc}$, 
and the angles are in degrees. The coordinates follow those
in \citet{taylor/cordes:1993}. 
For a fixed radius, |{\bf B}| has the 
same value at all azimuths.
We term the expression the logarithmic spiral
arm (LSA) model to distinguish it from previous forms. We take 8~kpc as 
the distance to the center of the Galaxy \citep{eisenhauer/etal:2003,
reid/brunthaler:2005}. The values are determined by fitting to
the K-band field directions. While the tilt, $\chi(z)$ 
with $\chi_0=25^{\circ}$, and the 
radial dependence, $\psi(r)$ with $\psi_1=0\fdg9$, optimize the fit, 
the key parameter is $\psi_0$, the opening angle of the spiral arms. 
We find that the magnetic field is a loosely wound spiral with 
$\psi_0 \simeq 35^\circ$.

To model the cosmic ray electrons, we assume they have a power-law 
distribution with slope 
\footnote{\citet{bennett/etal:2003c} uses $\gamma$ in place of $p$.} 
$p=-(2\beta_s+3)=3$ \citep{rybicki/lightman:1979}
and are distributed in a exponential disk
with a scale height of $h_d=1$~kpc and a radial scale length of 
$h_r=5$~kpc \citep[e.g., ][]{drimmel/spergel:2001} as
\begin{equation}
n_e = n_0\exp(-r/h_r){\rm sech}^2(z/h_d).
\end{equation}
While the amplitude of the signal is sensitive to the details of the 
cosmic ray distribution and the magnetic field structure, we
may estimate its overall structure with the smooth field model 
(Eq.~\ref{eq:bfield}) and cosmic ray distribution.  
The model predictions are not very sensitive to the assumed scale 
height and scale length. We compute
the polarization direction in this simple model as:

\begin{eqnarray}
\label{eq:tanqu} 
\tan 2\gamma(\hat{n}) 
&=& \frac{U(\hat n)}{Q(\hat n)} \nonumber \\
&=& \frac{\int n_e(x,\hat n ) 2 B_s(x,\hat n) B_t(x,\hat n)~dx}
	{\int n_e(x,\hat n ) \left[B_s^2(x,\hat n) -B_t^2(x,\hat n)\right]dx}
\end{eqnarray}
where $\hat{n}$ is the line-of-sight direction, $x$ is the distance along 
that direction, $n_e$ is the electron 
distribution described above, and
$B_t$ and $B_s$ are orthogonal components of the field perpendicular to 
the line of sight, with $B_t$ the component
perpendicular to the $z$ axis of the Galactic plane.
The parameters of the LSA model are determined by fitting the
predicted directions, Equation~\ref{eq:tanqu}, to the measured
the K-band field directions.  

Figure~\ref{fig:coherence} shows the predicted magnetic field orientation
for the LSA model. The actual direction has a $180^\circ$ ambiguity. 
In the plane, the field lines are parallel
to the Galactic plane and the polarization projects into positive Stokes $Q$.  
Near the Galactic pole, the field lines point along the spiral 
arm direction.  
When projected into $Q$ and $U$, this leads to $\gamma$ rotating around 
the pole.  We assess the agreement between the model field orientation and 
the orientation inferred from the K-band polarization 
with the correlation
coefficient $r = \cos(2(\gamma_{model} - \gamma_{data}))$,
and take the rms average over 74.3\% of the sky (outside the P06 mask described
below). For our simple model the agreement is 
clear: $r=0.76$ for K band.

Using using rotation measures derived from pulsar observations in the Galactic
plane, one finds instead a spiral arm opening angle of
$\psi_0 \simeq 8^\circ$ as reviewed in \citet{beck:2001,han:2006b}. 
Our method is
more sensitive to the fields above and below the plane; and, unlike the 
case with pulsars, we have no depth information. It has been suggested
by R. Beck and others 
that the north polar spur may drive our best fit value
to $\psi_0 \simeq 35^\circ$. Though the agreement between our simple model
and the K-band polarization directions indicate that we understand 
the basic mechanism, more modeling is needed to connect the {\sl WMAP}
observations to other measures of the magnetic field.  

For a power law 
distribution of electrons moving in a homogeneous magnetic field, 
the polarization fraction is $\Pi_s = (p+1)/(p+7/3)\approx0.75$
\citep{rybicki/lightman:1979}. Because the field direction changes 
as one integrates along the line of sight, there is a geometric
suppression of the amplitude of the polarization signal.  We estimate 
this geometric suppression as
\begin{equation}
g_{sync}(\hat n) = \frac{P(\hat n)}{\Pi_s~~I(\hat n)},
\label{eq:geom}
\end{equation}
where all quantities are determined from the model:
$P(\hat{n}) = \sqrt{Q^2 + U^2}$ and $I$ is found by integrating the 
perpendicular component of the magnetic field, $(B_s^2+B_t^2)^{1/2}$, 
and cosmic ray distribution along 
the line of sight. The result is shown in Figure~\ref{fig:supgeom}.
Similar results have been found by \citet{ensslin/etal:2006}.
This geometric reduction factor ranges from unity to zero.

\subsubsection{Comparison to Low Frequency Observations}

The polarization of a number of spiral galaxies similar
to the Milky Way has been measured by \citet{dumke/etal:1995}.
The observations are at 10.55~GHz and thus probe primarily synchrotron 
emission. For one of the best measured edge-on spirals, NGC 891,
they find: (1) at distances $\approx 5~$kpc off the galactic
plane the polarization 
fraction can be $\geq$10\%; and (2) in the plane, at heights $<0.5~$kpc, 
the polarization fraction drops to $<$5\%.
Similar behavior is seen by \citet{sukumar/allen:1991} at 5~GHz.
In addition, 
\citet{hummel/dahlem/vanderhulst:1991} show that (3)
between 0.66~GHz and 1.5~GHz the spectral index ranges from
$\beta_s=-2.5$ in the plane to $\beta_s=-3.5$ well off the plane.
\WMAP\ observes qualitatively similar behavior in K band.

At 408~MHz, \citet{haslam/etal:1982} have surveyed the Galactic plane
in intensity. At this frequency, synchrotron emission dominates 
maps.  We test the magnetic field model by extrapolating 
the 408~MHz measurements to 22~GHz (an extrapolation of
40 in frequency and over 10,000 in amplitude):
\begin{eqnarray}
Q_{model}(\hat n) &=& qI_{Has}(\hat n) 
\left(\frac{22}{0.408}\right)^{\beta_s} \Pi_s g_{sync}(\hat n) 
\cos(2\gamma_{model}) \nonumber \\
U_{model}(\hat n) &=& qI_{Has}(\hat n) 
\left(\frac{22}{0.408}\right)^{\beta_s} \Pi_s g_{sync}(\hat n) 
\sin(2\gamma_{model}) 
\label{eq:qumod}
\end{eqnarray}
where $q$ is the ratio of the homogeneous field strength to the total 
field strength. Note that the model effectively 
has only one free parameter: an overall amplitude, which is described 
by a degenerate combination of the spectral index, $\beta_s$ and $q$.
For $\beta_s = -2.7$, the best fit value for $q$ is $0.7$.  This implies that 
the energy in the large scale field is roughly the same as the energy 
in small scale fields, consistent with other results for
the Milky Way \citep[][and references therein]{jones/klebe/dickey:1992,
beck:2001}.

Figure~\ref{fig:haslam} compares the K-band polarization signal to 
the extrapolated 408 MHz maps.  Given the simplicity of the model 
(uniform cosmic ray spectral index, $p$, and a uniform LSA field), 
the agreement is remarkably good.  The largest deviations are seen near 
spiral arms.  Recent observations \citep{enomoto/etal:2002} suggest
that cosmic rays are accelerated in star-forming regions.  If most 
cosmic rays are accelerated in spiral arms and then diffuse away 
from the arms, we would expect a flatter spectral index in the 
arms, consistent with the observations. In Figure~\ref{fig:loops}
we show that the radio loops \citep{berkhuijsen/haslam/salter:1971} 
seen at 408~MHz, probably from supernovae or ``blowouts,'' 
are also seen in the \WMAP\ data. 

\subsubsection{Starlight Polarization and Polarized Dust Emission}

Measurements of starlight polarization serve as a template for the analysis
of polarized microwave dust emission 
\citep{fosalba/etal:2002,bernardi/etal:2003}. We have combined several 
catalogs of optical dust polarization
measurements \citep{heiles:2000, berdyugin/etal:2001, 
berdyugin/teerikorpi:2002,berdyugin/piirola/teerikorpi:2004} to 
construct a template for the magnetic field direction in dusty environments.  
Since there are 
significant variations in the dust column density, we only use 
the measured direction to construct the dust template.  
The dust layer has a scale height 
of 100 pc \citep{berdyugin/teerikorpi:2001,drimmel/spergel:2001}.
Observations toward the 
Galactic poles suggest that most of the dust absorption occurs 
within 200 pc. To select stars outside the dust column
for $|b|>10^{\circ}$, we limit the sample to the 
1578 stars with heliocentric distances greater than 500 pc.  
For $|b|<10^{\circ}$, the model is problematic because 
there is ample dust emission from distances further away than
the stars sample.  

We represent the starlight polarization data, $(Q_\star,U_\star)$, 
in terms of a polarization amplitude, $P_\star$ and direction, $\gamma_\star$:
\begin{eqnarray}
Q_\star &=& P_\star \cos(2\gamma_\star) \nonumber \\
U_\star &=& P_\star \sin(2\gamma_\star) 
\end{eqnarray}
We then smooth the starlight data by convolving $(Q_\star/P_\star)$
and $(U_\star/P_\star)$ with a Gaussian window with a FWHM of $9\fdg2$.
The smoothing is required because the measurements are coarsely distributed.
As a result, this dust model is applicable only 
for $\ell\la 15$ and  $|b|>10^{\circ}$.
Above, $\gamma_{\star}$ describes the direction of this smoothed starlight 
polarization field.  We can quantify the agreement between the starlight
and \WMAP\ K-band polarization measurements by computing their 
correlation in each pixel, 
$Z = \cos(2 (\gamma_{\star} - \gamma_{K})+\pi)$
where $\gamma_{K}$ is the direction in K band.
Figure~\ref{fig:zzz} shows a plot of the correlation 
as a function of position.  
The median correlation coefficient is 0.72  implying that the dust and 
K-band directions typically agree to 20$^\circ$.  Because of noise in both 
the K-band and starlight maps, this is an underestimate
of the correlation. Nevertheless, the correlation tells us that the basic 
model relating the starlight, the dust, synchrotron emission, and 
the magnetic field agrees with observations. 

\subsubsection{Thermal Dust Emission}
\label{sec:thde}

Based on the detection of starlight polarization,
thermal dust emission is expected to be polarized at millimeter
and sub-millimeter wavelengths.
Archaeops has detected polarized thermal emission at 353 GHz
\citep{benoit/etal:2004}. An extrapolation from this high frequency
suggests that \WMAP\ should see polarized thermal dust emission at 94 GHz.
Here, we report on the \WMAP\ detection of dust polarization at 94 GHz.

We generate a template for the dust polarization by 
using the Maximum Entropy Method (MEM)\footnote{The dust, free-free,
synchrotron, and CMB MEM maps are derived from a maximum entropy solution 
to the five \WMAP\ bands, the FDS dust map
\citep{finkbeiner/davis/schlegel:1999}, the Haslam map, 
and a H$\alpha$ map \citep{finkbeiner:2003}}. 
dust intensity 
map \citep{bennett/etal:2003c},
the smoothed polarization direction from the starlight,
and the model geometric factor for the dust layer:
\begin{eqnarray}
\label{eq:dust_template}
Q_{dust}(\nu) &=& 
I_{dust}(\hat n)\Pi_d g_{dust}(\hat n) \cos(2\gamma_{dust}) \nonumber \\
U_{dust}(\nu) &=& 
I_{dust}(\hat n)\Pi_d g_{dust}(\hat n)  \sin(2\gamma_{dust})
\end{eqnarray}
where $\gamma_{dust}=\gamma_\star+\pi/2$ is the smoothed starlight 
polarization direction. The geometric suppression factor for the 
dust, $g_{dust}$, is 
computed along the same lines as $g_{sync}$ in Equation~\ref{eq:geom}
and is shown in Figure~\ref{fig:supgeom}. 
To compute $I(\hat n)$ for the analog to Equation~\ref{eq:geom}, 
we assume the dust has 
a scale height of 100~pc and a radial scale length of 3~kpc. To find
$P(\hat n)$ we use the LSA magnetic field model. The fractional
polarization, $\Pi_d=0.05$, is found with a best fit of the model to the data.
Similar results are found using the FDS dust map 
\citep{finkbeiner/davis/schlegel:1999} instead of the MEM dust map.
The uncertainty is estimated to be 50\%.

Figure~\ref{fig:dust_template} compares this predicted pattern of
polarization to the cleaned W-band observations.  We use the
K-band synchrotron template to clean Q, V and W bands and then
use the Q and V band maps to remove the CMB polarization signal
from the W-band maps; though removing the CMB component is not necessary.
The W-band map is then smoothed with a 10$^\circ$ 
beam for plotting. The appearance of the dust 
polarization signal pattern is similar
to that found by Archeops \citep[Figures 2 \& 3]{ponthieu/etal:2005}.
However, the signal to noise ratio is low due to the low level
of polarized dust emission at 94~GHz. 
The predominant feature is that the plane is dominated by positive
Stokes $Q$ emission. A visual comparison to the model is less robust.
One must keep in mind that since stars are
heavily obscured in the plane, the model is not expected to 
be accurate in the plane. Nevertheless, since Stokes $Q$ emission
corresponds to the dominant horizontal magnetic field, one does
not have to sample too deeply to pick it up. Similarly, we interpret 
the poor correlation between the model $U$ and the observed $U$ as due to 
the insufficient sampling of other magnetic field directions by
rather limited depth of the stars. Some common features between the 
model and W-band
data are seen for $|b|>10^{\circ}$. Fits of the data to the model 
are given in Section~\S\ref{sec:fgr}. Clearly, more integration time 
and more stellar polarization measurements are needed to fill out the 
model.

\subsubsection{Spinning Dust Emission?}

Electric dipole emission from rapidly spinning dust grains is potentially a 
significant source of emission at \WMAP\ frequencies 
\citep{erickson:1957,draine/lazarian:1998a}. Thermal fluctuations 
in the magnetization of magnetized grains may also be a potentially 
significant source of emission at microwave wavelengths
\citep{draine/lazarian:1999,prunet/lazarian:1999}.  Both
have been proposed as an explanation for the correlations seen between 
thermal dust emission at 140~$\mu$m  and microwave emission in many cosmic 
background experiments: COBE \citep{kogut/etal:1996a}, 
OVRO \citep{leitch/etal:1997}, Saskatoon \citep{deoliveira-costa/etal:1997}, 
the 19~GHz Survey \citep{deoliveira-costa/etal:1998a},
Tenerife \citep{deoliveira-costa/etal:1999,deoliveira-costa/etal:2004}, 
Python V \citep{mukherjee/etal:2003}, 
and COSMOSOMAS \citep{fernandez-cerezo/etal:2006}.  

The spectral shape of spinning dust emission can be similar to 
synchrotron emission in the 20-40 GHz range. Thus models with either variable 
synchrotron spectral index \citep{bennett/etal:2003} or
with a spinning dust spectrum with a suitably fit cutoff frequency
\citep{lagache:2003,finkbeiner:2004} can give reasonable fits to the data.
However, at $\nu<20~$GHz there is a considerable body of evidence,
reviewed in \citet{bennett/etal:2003} and \citet{hinshaw/etal:prep},
that shows (1) that the synchrotron index varies across the sky
steepening with increasing galactic latitude (as is also seen in \WMAP\ ) 
and (2) that in other galaxies and our galaxy there is a strong correlation
between 5~GHz synchrotron emission and 100~$\mu$m (3000~GHz) dust emission.
The combination of these two observations imply that the 
$\nu<40$~GHz \WMAP\ foreground emission is dominated by synchrotron 
emission as discussed in \citet{hinshaw/etal:prep}.  
Nevertheless, we must consider spinning dust as a possible emission
source. While on a Galactic scale it appears to be sub-dominant, 
it may be dominant or a significant fraction of the emission in some 
regions or clouds.

Spinning dust models predict an unambiguous signature in intensity maps: at
5-15~GHz, the dust emission should be significantly less than the 
synchrotron emission.  \citet{finkbeiner:2004} and 
\citet{deoliveira-costa/etal:2004} 
argue that the Tenerife and Green Bank data show evidence for a rising 
spectrum between 10 and 15 GHz, suggesting the presence of spinning dust. 
Observations of individual compact clouds also show evidence for spinning dust 
emission \citep{finkbeiner/etal:2002,finkbeiner/langston/minter:2004,
watson/etal:2005} though the signature is not ubiquitous. 
The status of the observations is discussed further 
in \citet{hinshaw/etal:prep}.

The \WMAP\  polarization measurements potentially
give us a new way to distinguish between 
synchrotron and dust emission at microwave frequencies.  While synchrotron 
emission is expected to be highly polarized, emission from spinning dust 
grains is thought to be weakly polarized.
While promising, the signature is not unique as a tangle of magnetic
field lines can also lead to a low polarization component 
\citep{sukumar/allen:1991} as seen at 5~GHz where spinning dust emission is
expected to be negligible.  
Using a model for the polarization fraction of the synchrotron emission 
based on the LSA structure, we separate the  microwave intensity emission into
a high and low polarization component:
\begin{eqnarray}
\label{eq:low_high}
I_{high}^{\nu}(\hat n) &=& P^{\nu}(\hat n)/q \Pi_s g_{sync}(\hat n)\nonumber \\
I_{low}^{\nu}(\hat n) &=& I^{\nu}(\hat n) - I^{\nu}_{high}(\hat n) -
        I_{CMB}^{ILC} - I_{FF}^{MEM, \nu}
\end{eqnarray}
where $I^{\nu}$ and $P^{\nu}$ are the intensity and polarization maps at
frequency $\nu$. For notational convenience, we use $\nu={\rm K,Ka,Q,V,W}$.
$I_{CMB}^{ILC}$ is the foreground-free  CMB map made with a linear combination
of \WMAP\ bands, and
$I_{FF}^{MEM,\nu}$ is the MEM free-free map for band $\nu$
\citep{bennett/etal:2003}. In effect, we use the \WMAP\ polarization 
maps to extract the intensity map of the low-polarization component 
in the data.

Figure \ref{fig:lowpol} compares the morphology of the 
low polarization K-band map to
the W-band MEM dust map \citep{bennett/etal:2003c}. Even in this simple
model based on a number of assumptions, the agreement in 
morphology is striking.  
We quantify this by computing the rms deviation between the two scaled maps,
\begin{equation}
d^2 = \frac{\sum (I_{\rm W}(\hat n) - \alpha I_{low}^{K}(\hat n))^2}
{\sum I_{\rm W}(\hat n)^2}
\end{equation}
where W is the W-band map, the scale factor is $\alpha=0.105$, 
and the sum is taken over pixels. 
We find $d = 0.05$. In other words, we can ``predict'' the distribution 
of dust in W band from just the K band intensity and
polarization maps. The low polarization fraction component
has a spectral index of $\beta = -2.6$ between K and Q bands. 
This correlation between the low polarization emission regions
at 22-45~GHz and the thermal emission at 90~GHz and higher 
may be interpreted as either a very tight correlation
between tangled field lines in star forming (dusty) regions
or as evidence for spinning dust emission. 
More polarization data, $\nu<22$~GHz observations, and extensive
modeling are needed to conclusively
delineate the magnitude and morphology of the various components. 

\subsection{Masking Polarized Foreground Emission} 
\label{sec:mask}

To compute the CMB power spectrum, we must mask the regions
with the brightest foreground emission. For polarization we create 
a set of masks with a process that is somewhat analogous to the 
creation of the temperature masks \citep{bennett/etal:2003}. 
First, the K-band $Q$ and $U$ polarization 
maps are used to compute a positive-definite HEALpix r4 
\footnote{The number of pixels is $12N_{side}^2$ where
$N_{side}=2^4$ for r4, or resolution 4 
\citep{gorski/hivon/wandelt:1998}. See notation 
in \citet{bennett/etal:2003}.}
P map. From this a noise-bias variance map \citep{jarosik/etal:prep} 
is subtracted.  The rectified noise-bias 
correction is small because of the coarse resolution at r4.  
A histogram of pixel polarization amplitudes in 
this noise-bias-corrected map approximates a power law.  
The peak is near the zero pixel value, there are just 
a few negative pixels (due to the noise bias correction), and there is a long 
positive tail.  

Unlike the process in which the temperature masks 
were created, there is no natural cut level based on the histogram peak.  
Instead, the cuts are given in terms of the mean of the noise bias 
corrected map of P at K band. The cut level at the mean is denoted ``P10''.  
The cut level at 0.2 times the 
mean is ``P02'', etc. For each cut level, a preliminary mask is made by 
setting r4 pixels greater than the cut level to 1, and all others to 0.  
This mask is expanded to r9 and smoothed by a $7\fdg5$ FWHM Gaussian.  
This mask map is set back to all 0s and 1s using the 0.5 level as a 
cut-off and the sense of the mask is reversed, so that the masked-out 
parts of the sky have zeros (the \WMAP\ convention).  
The above process results in a synchrotron polarization 
mask. 

In the case of temperature masks, we found that additional masking 
based on the higher frequency bands was redundant.  
This is not the case with polarization. 
Thus we make a dust polarization mask in a similar manner. We begin 
with the first-year MEM 
dust model box-averaged to r7. Half the maximum value found in a subset of 
pixels in the polar caps ($\vert b\vert > 60^\circ$) is adopted as the 
cut-off level.  A preliminary mask is made by setting r7 pixels greater 
than the cut-off level to unity, and all others to zero.  This mask is 
then resolution expanded to a r9 map,  smoothed by a $4\fdg0$ Gaussian, 
and set back to digital levels with a 0.5 cut-off.  The sense of the mask 
is then reversed to fit the \WMAP\ convention. Each synchrotron 
polarization mask is ANDed with the (constant) dust polarization mask and a 
constant polarized source mask. 

We find, in general, that the extragalactic 
point sources are minimally polarized in the \WMAP\ bands, as discussed 
in \citet{hinshaw/etal:prep}.  We construct a source mask based on the 
exceptions.  The most significant exception (not already covered by 
the synchrotron or dust polarization masks) is Centaurus A, an extended 
and polarized source.  We found excellent agreement between \WMAP\ and 
previously published maps of Cen A (Figure~\ref{fig:crab}). 
Based on this information, we custom-masked 
the full extent of Cen A.  Six other bright polarized sources that we masked 
are Fornax A, Pictor A, 3C273, 3C274, 3C279, PKS 1209-52. 
(Some bright polarized sources already covered by the synchrotron and 
dust mask regions include: 3C58, Orion A, Taurus A, IC443, 1209-52, W51, 
W63, HB21, CTB104A). We have determined that, for most 
applications, the mask that we call ``P06'' 
is the best compromise between maximizing usable sky area while minimizing 
foreground contamination.  With the above considerations, the P06
mask masks 25.7\% of the sky, mostly near the Galactic plane.
We use the terminology ``outside the P06 mask''
to refer to data in the 74.3\% of the sky left for cosmological analysis.
Various masks are shown in Figure~\ref{fig:polmasks}. 

\subsection{Removing the Polarized Foreground Emission from the Maps} 
\label{sec:fgr}

Based on our analysis of the Galactic foreground emission, we have generated
synchrotron and dust template maps for the purposes of foreground removal.  
The template maps are fit and subtracted from the Ka through W band data 
to generate cleaned maps that are used for CMB analysis.  We assess 
the efficacy of the subtraction with $\chi^2$ and 
by examining the residuals as a function 
of frequency and multipole $\ell$, as described in \S\ref{sec:psfgm}. 

We use the K-band data to trace the synchrotron emission, taking care to 
account for the (relatively weak) CMB signal in the K-band map when fitting 
and subtracting the template.  For dust emission, we construct a template 
following Equation (\ref{eq:dust_template}) that is
based on the starlight-derived polarization directions
and the FDS dust model
eight \citep{finkbeiner/davis/schlegel:1999} evaluated at
94~GHz to trace the dust intensity, $I_{\rm dust}$.  We call this 
combination of foreground templates ``KD3Pol''.  

The synchrotron and dust templates are fit simultaneously in 
Stokes $Q$ and $U$ to three-years maps in Ka through W bands.  
The three-year maps are constructed by optimally combining the 
single-year maps for each DA in a frequency band. Specifically
\begin{equation}
[Q_{\nu},U_{\nu}] = \big( \sum_i {\bf N}_i^{-1} \big)^{-1} \sum_i 
{\bf N}_i^{-1} [Q_i,U_i]
\end{equation}
where $i$ is a combined year and DA index, $[Q_i,U_i]$ is a polarization map
degraded to r4 \citep{jarosik/etal:prep}, and ${\bf N}_i^{-1}$ is the inverse
noise matrix for polarization map $i$.  The fit coefficients, $\alpha_s$ and
$\alpha_d$ are obtained by minimizing $\chi^2$, defined as
\begin{equation}
\chi^2 = \sum_p \frac{([Q_{\nu},U_{\nu}] - \alpha_{s,\nu}[Q_s,U_s]
                - \alpha_{d,\nu}[Q_d,U_d])^2}{[\sigma_Q^2,\sigma_U^2]},
\end{equation}
where $[Q_s,U_s]$ is the K-band polarization map (the synchrotron template),
$[Q_d,U_d]$ is the dust template, and $[\sigma_Q^2,\sigma_U^2]$ is 
the noise per pixel per Stokes parameter in the three-year combined maps.  
We have tried using optimal (${\bf N}^{-1}$) weighting for the fits as 
well, and found similar results for the coefficients.  
The results reported here are based on the simpler diagonal weighting.  
The fit is evaluated for all pixels outside the processing 
mask \citep{jarosik/etal:prep}.

The fit coefficients are given in the top half of Table~\ref{tab:fgfit}.  
For each emission
component we also report the effective spectral index derived from the fit:
$\beta_s(\nu_K,\nu)$ for synchrotron emission, and $\beta_d(\nu,\nu_W)$ for
dust.  These results indicate that the spectrum of the component traced by
K-band is systematically flattening with increasing frequency, which is
unexpected for synchrotron emission.  This behavior is statistically 
significant, and is robust to
variations in the dust model and the data weighting.  We do not have a
definitive explanation for this behavior.

To guard against the possibility of subtracting CMB signal, we
modified the template model as follows.  We take the four synchrotron 
coefficients in Table~\ref{tab:fgfit} and fit them to a spectrum model 
of the form
\begin{equation}
\alpha_s(\nu) = \alpha_{s,0} \cdot g(\nu)(\nu/\nu_K)^{\beta_s} + \alpha_c,
\end{equation}
where $\alpha_{s,0}$, $\beta_s$, and $\alpha_c$ are model parameters that are
fit to the $\alpha_s(\nu)$, and $g(\nu)$ is the conversion from antenna
temperature to thermodynamic temperature at frequency $\nu$.  
This results in a modified set of synchrotron coefficients that are 
forced to follow a power-law
that is largely determined by the Ka and Q-band results.  Specifically, the
modified coefficients are given in Table~\ref{tab:fgfit}.  
The implied synchrotron spectrum is $\beta_s = -3.33$.  
This results in a 12\%
reduction in the synchrotron coefficients at Q-band, and a 33\% reduction at
V-band.  However, because the K-band template is dominated by an 
$\ell=2$ E-mode
signal (see \S\ref{sec:psfgp06}), this change has a negligible effect 
on our cosmological
conclusions, which are dominated by E-mode signal at $\ell>2$. 
A comparison of selected ``before and after'' cleaning maps is 
shown in Figure~\ref{fig:inout}.

We also account for the cleaning in the map error bars. 
Since the K-band data are a combination
of synchrotron and CMB emission, subtracting a scaled version of K band
from a higher frequency channel also subtracts some CMB signal.
If the fit coefficient to the higher frequency channel is $a_0$, 
then the cleaned
map is $M^\prime(\nu ) = (M(\nu )-a_0M(\nu = K))/(1-a_0)$, where $M$
is the map and $\nu$ denotes the frequency band.  
The maps we use for cosmological analysis were cleaned using the 
coefficients in the bottom half of Table~\ref{tab:fgfit}.  The factor 
of $1/(1-a_0)$ 
dilates the noise in the cleaned map.  To account for this in the error 
budget we scale the covariance matrix of the cleaned map by a factor 
of $1/(1-a_0)^2$.  Additionally, we modify the form of the inverse covariance 
matrix by projecting from it any mode that has the K-band polarization 
pattern: ${\bf N}^{-1} \rightarrow {\bf N'}^{-1}$, where 
${\bf N'}^{-1} [Q_K,U_K] = 0$.  This ensures that any residual signal 
traced by K-band (due, for example, to errors in the form of the spectrum) 
will not contribute to cosmological parameter constraints.

One measure of the efficacy of the foreground removal is the change in 
$\chi^2$, relative to a null signal, between pre-cleaned and cleaned maps.
Table~\ref{tab:fgchi}
gives the values for the full sky and the P06 cut. In both cases the 
full pixel covariance matrix was used to compute $\chi^2$ for Stokes
$Q$ and $U$ simultaneously. For the full sky the number of degrees of 
freedom, $\nu$, is 6144 (twice the number of pixels in an r4 map) and 
outside the P06 mask $\nu=4534$. Note the large $\Delta\chi^2$ achieved 
with just a two parameter fit. By comparing the full sky to the 
P06 $\chi^2$, we 
find that the starlight-based dust template is insufficient in the 
plane as discussed in \S\ref{sec:thde}. We also see that outside the P06 mask, 
that Q and V bands are the cleanest maps and that they are
cleaned to similar levels. Since $\chi^2/\nu$ 
for Q and V bands is so close to unity for the cleaned maps, it is no 
longer an effective measure of cleaning. Instead, we examine the power
spectra $\ell$ by $\ell$ to assess the cleaning, and then test the
sensitivity of the cosmological conclusions to cleaning 
by including Ka and W band data.  

We have tried a number of variants on the KD3Pol cleaning. We find, 
for example, that setting $g_{dust}=1$ across the sky has negligible 
effect on the fits or the derived optical depth. Alternatively,
when one uses  the K-band polarization 
direction to trace the dust direction, $\gamma_{\rm dust} = \gamma_{\rm K}$
in Equation~\ref{eq:dust_template}, the cleaning is not as effective. Outside
the P06 mask, the reduced $\chi^2$ in the Q and V band maps is
1.022 and 1.016 as compared to 1.014 and 1.018 for the starlight-based
directions. Thus the latter are used. Regardless of
template, we find that our cosmological conclusions are relatively
insensitive to the details as indicated in Figure~\ref{fig:tau_likes}.

\section{Power Spectra}
\label{sec:pspec}

The $Q$ \& $U$ maps are well suited 
to analyzing foreground emission, are useful for comparing
to other polarization maps, and have straightforward noise properties.
However, they are not well suited to quantifying the 
CMB polarization anisotropy because their definition is coordinate dependent.
The $Q$ and $U$ maps may be transformed into
scalar and pseudo-scalar quantities called E and B modes
\citep{seljak:1997,
kamionkowski/kosowsky/stebbins:1997,zaldarriaga/seljak:1997}.
E and B are so named because they comprise a curl-free and divergence-free
decomposition of the spin-2 polarization field, analogous to static electric 
and magnetic fields.
The problem of separating E and B modes
with an unevenly sampled and cut sky has been
considered by a number of authors 
\citep[e.g.,][]{tegmark/etal:2000,lewis/challinor/turok:2002,
bunn/etal:2003}. In our analysis, we work 
directly with $Q$ and $U$ maps to produce the E and B angular power spectra. 
The conventions follow Appendix A of
\citet{kogut/etal:2003}.\footnote{In this 
paper we do not use the rotationally invariant $Q^\prime$ and $U^\prime$
of \citet{kogut/etal:2003}.}

Fundamental symmetries in the production and growth of the polarized signal
select the possible configurations for the CMB polarization.
Scalar (density) perturbations to the matter power spectrum
give rise to T and E modes. Tensor perturbations
(gravitational waves) give rise to T, E, and B modes primarily at
$\ell\la 200$\footnote{For $r<0.03$ and $\ell \ga 70$, primordial B modes are 
dominated by the gravitational lensing of E modes.}.
Both scalar and tensor perturbations can produce polarization patterns
in both the decoupling and reionization epochs. 
Vector perturbations 
\footnote{Vector modes are produced by purely
rotational fluid flow.
Based on the fit of the adiabatic $\Lambda$CDM model to \WMAP\ TT data, 
the contribution of such modes is not large \citep{spergel/etal:2003}.
However, a formal search for them has not been done.}
(both inside and outside the horizon) are redshifted away with 
the expansion of the universe, unless there are active sources creating
the vector modes, such as topological defects. We do not consider these
modes here.

At the noise levels achievable with \WMAP\ ,
the standard cosmological model predicts that 
only the E mode of the CMB polarization and its correlation with T
will be detected. The B-mode polarization signal is expected to 
be too weak for \WMAP\ to detect, while the correlations of T and E with
B is zero by parity. Thus the TB and EB signals serve as a useful 
null check for systematic effects.
The polarization of foreground emission is produced by different mechanisms.
Foreground emission can have any mixture of
E and B modes, it can be circularly polarized (unlike the CMB),
and E and B can be correlated with T.

We quantify the CMB polarization anisotropy with the $C^{TE}_\ell$, 
$C^{EE}_\ell$,
and $C^{BB}_\ell$ angular power spectra, where
\begin{eqnarray}
C^{XY}_\ell &=& \langle a^X_{\ell m}
a^{Y*}_{\ell m}\rangle.
\end{eqnarray}
Here the ``$\langle\rangle$'' denote an ensemble average, 
$a^T_{\ell m}$ are the multipoles of the temperature map,
and $a^E_{\ell m},a^B_{\ell m}$ are related to the spin-2 decomposition of the
polarization maps
\begin{equation}
[Q\pm iU](\hat{x}) = \sum_{\ell>0}
\sum_{m=-\ell}^{\ell}{{}_{\mp2}a_{\ell m}{}_{\mp2}Y_{\ell m}(\hat{x})}
\end{equation}
via
\begin{eqnarray}
{}_{\pm2}a_{\ell m} = a^E_{\ell m} \pm i a^B_{\ell m} 
\end{eqnarray}
\citep{zaldarriaga/seljak:1997}.
The remaining polarization spectrum combinations (TB, EB) have no expected
cosmological signal because of the statistical isotropy of the universe.

We compute the angular power spectrum after applying the P06
polarization mask using two methods depending on the $\ell$ range.
All power spectra are initially based on the 
single-year r9  
$Q$ and $U$ maps \citep{jarosik/etal:prep}.
For $\ell>23$
\footnote{$\ell=23=3N_{side}-1$ is the 
Nyquist limit on $\ell$. For some analysis methods
(\S\ref{app:ektau}) we use HEALPix r3 for which $n_{side}=2^{3}=8$}, 
we compute the power spectrum following the 
method outlined in \citet{hivon/etal:2002}, and
\citet[Appendix A]{kogut/etal:2003} as updated in \citet{hinshaw/etal:prep}
and Appendix~\ref{sec:masterP}. The statistical weight per pixel is 
$N_{obs}/\sigma_0^2$ where $\sigma_0$ is the 
noise per observation \citep{jarosik/etal:prep,hinshaw/etal:prep}. 
Here $N_{obs}$ is a 2x2 weight matrix that multiplies the vector $[Q,U]$
in each pixel
\begin{equation}
N_{obs} = \twobytwo{N_Q}{N_{QU}}{N_{QU}}{N_{U}},
\end{equation}
where $N_Q$, $N_U$, and $N_{QU}$
are the elements of the weight arrays provided with the sky map data.
Note that the correlation
between $Q$ and $U$ within each pixel is accounted for.
We refer to this as ``$N_{obs}$ weighting.''
From these maps, only cross power spectra between DAs 
and years are used.
The cross spectra have the advantage that only signals common to 
two independent maps contribute and there are no noise biases 
to subtract as there are for the auto power spectra.
The covariance matrices for the various $C_\ell$ are given in 
Appendix~\ref{app:covar}. 

For $\ell<23$ we mask and degrade the r9 maps to 
r4 \citep[see the last paragraph of Appendix~D and][]{jarosik/etal:prep}
so that we may use the full r4 inverse pixel noise matrix, ${\bf N^{-1}}$,
to optimally weight the maps prior to evaluating the pseudo-$C_\ell$.
This is necessary because the maps have
correlated noise that is significant compared to the faint CMB signal.
By ``${\bf N^{-1}}$ weighting'' the maps, we efficiently suppress modes in 
the sky that are poorly measured given the \WMAP\ beam separation 
and scan strategy (mostly modes with structure in the ecliptic plane). 
We propagate the full noise errors through to the Fisher matrix of the 
power spectrum. For the spectrum plots in this section, the errors are 
based on the diagonal elements of the covariance matrix which is 
evaluated in Appendix~\ref{app:master}. 

Figure~\ref{fig:nobsninv} shows the effect that correlated noise 
has on the low $\ell$ errors in the EE and BB spectra.
The curves show the diagonal elements of the inverse Fisher matrix
(the $C_\ell$ errors) computed in two ways: (1) assuming the noise 
is uncorrelated in pixel space and described by $N_{obs}$ (red)
and (2) assuming it is correlated and correctly described by
${\bf N^{-1}}$ (black).The smooth rise in both curves toward low $\ell$
is due to the effects of $1/f$ noise and is most pronounced in the W4 DA, 
which has the highest $1/f$ noise. The structure in the black trace 
is primarily due to the scan strategy. Note in particular,
that we expect relatively larger error bars on $\ell=2,5,7$ in EE 
and on $\ell=3$ in BB. {\it We caution those analyzing maps that to obtain accurate results, the ${\bf N^{-1}}$ 
weighting must be used when working with the $\ell<23$ power spectra.}
For the Monte Carlo Markov Chains (MCMC) and cosmological parameter 
evaluation, we do not use the 
power spectrum but find the exact
likelihood of the temperature and polarization maps given the cosmological 
parameters \citep[Appendix~\ref{app:ektau} \&][]{hinshaw/etal:prep}. 

For both r4 and r9 maps there are 15 MASTER cross power spectra
(see Table~\ref{tab:eebb}). For the full three-year result, we form
${\rm \sum_{i,j=1}^{3}ui\times vj/6}$ omitting the
${\rm u1\times u1}$, ${\rm u2\times u2}$, and ${\rm u3\times u3}$ 
auto power spectra. In this expression, $u$ and $v$ denote the frequency
band (K-W) and $i$ and $j$ denote the  year. 
The noise per $\ell$ in the limit of 
no celestial signal, $N_\ell$, is determined from analytical models
that are informed by full simulations for r9
(including $1/f$ noise),
and from the full map solution for r4. 

\subsection{Power Spectrum of Foreground Emission Outside the P06 Mask.}
\label{sec:psfgp06}

Figure~\ref{fig:rawps_lspace} shows the EE and BB power spectra for the 
region outside the P06 mask, 74.3\% of the sky, before any cleaning. 
The 15 cross spectra have 
been frequency averaged into four groups (Table~\ref{tab:eebb})
by weighting with the diagonal
elements of the covariance matrix. Data are similarly binned over
the indicated ranges of $\ell$. 
It is clear that even on the cut 
sky the foreground emission is non negligible. 
In K band, we find
$\ell(\ell+1)C^{EE}_{\ell=<2-6>}/2\pi =  66~(\mu K)^2$
and $\ell(\ell+1)C^{BB}_{\ell=<2-6>}/2\pi =  48~(\mu K)^2$,
where $\ell=<2-6>$ denotes the weighted
average over multipoles two through six.
The emission drops by roughly a factor of 200 in $C_\ell$ by 61~GHz resulting
in $\la 0.3~(\mu K)^2$ for both EE and BB.
There is a ``window'' between 
$\ell=4$ and $\ell=8$ in the EE where the emission is comparable to,
though larger than, the detector noise.  
Unfortunately, BB foreground emission 
dominates a fiducial $r=0.3$,  $\tau=0.09$ model 
by roughly an order of magnitude
at $\ell<30$. 
In general, the power spectrum of the foreground emission
scales approximately as $\ell^{-1/2}$ in $\ell(\ell+1)C_\ell$.

Figure~\ref{fig:rawps_fspace} shows the power spectra 
as a function of frequency for a few $\ell$ bands. 
The spectrum of the emission follows that of synchrotron with
$T\propto\nu^{\beta_s}$ with $\beta_s=-2.9$ for both EE and 
BB\footnote{The fits to the power law index were done with 
the power spectra in CMB temperature units. The ratio
of these and the indicies corresponding to antenna are
approximately 1.03 and 0.99 at 90 GHz (where the difference is largest)
for dust and synchrotron respectively. The difference is negligible.}. 
There is some evidence for another component
at $\nu>60$ as seen in the flattening of the EE $\ell=2$ term. We interpret 
this as due to dust emission. In the foreground 
model, we explicitly fit to a dust template and detect
polarized dust emission. However, there is not yet a sufficiently
high signal to noise ratio to strongly constrain the dust index or amplitude 
outside the P06 mask.

A simple parameterization of the foreground emission outside the P06
mask region is given by
\begin{equation}
\ell(\ell+1)C_{\ell}^{fore}/2\pi =
({\cal B}_{s}(\nu/65)^{2\beta_s}+{\cal B}_{d}(\nu/65)^{2\beta_d})
\ell^{m}.
\label{eq:fgmod}
\end{equation}
We have introduced the notation 
${\cal B}^{XX}\equiv\ell(\ell+1)C_\ell^{XX}/2\pi$ to simplify
the expression.
The ``d'' and ``s'' subscripts stand for ``dust'' and 
``synchrotron.''
From an unweighted fit to all the raw $\ell<100$ data with
the dust index fixed at $\beta_d=1.5$, we find for EE
${\cal B}_{s}=0.36~(\mu{\rm K})^2$, $\beta_s=-3.0$, 
${\cal B}_{d}=1.0~(\mu{\rm K})^2$ and $m=-0.6$; and for BB
${\cal B}_{s}=0.30~(\mu{\rm K})^2$, $\beta_s=-2.8$, 
${\cal B}_{d}=0.50~(\mu{\rm K})^2$ and $m=-0.6$.
This model is given as an 
approximate guide. Its $\ell$ dependence is shown in 
Figure~\ref{fig:rawps_lspace} for $\nu=65~$GHz and its frequency dependence
is shown in Figure~\ref{fig:rawps_fspace} for BB $\ell=2$. One can see that
this scaling model picks up the general trends but not the details of 
the foreground emission. For example, it ignores correlations between
dust and synchrotron emission. It predicts an average foreground emission
of $\approx 1$~($\mu$K)$^2$ at 30~GHz and $\ell=300$. \citet{leitch/etal:2005}
give an upper bound of $\approx 1~(\mu{\rm K})^2$ for synchrotron emission
in this range. As DASI observes a relatively synchrotron-free region
and at $\ell$s beyond where this simple parametrization
can be tested, there is not a conflict with their results.
The same is true for the CBI experiment \citep{readhead/etal:2004}
which also observed at 30~GHz but at a predominantly higher $\ell$
and in a predetermined clean region of sky. The dust amplitude in the 
model is especially uncertain. Depending on the region of sky within the 
P06 cut, and the $\ell$ of interest, it may be an order of magnitude off.

For a more complete model of the power spectra of foreground emission, one
must take into account the correlations or anticorrelations
between various foreground components and between the foreground 
components and the CMB. For example, a reasonable fit to the
$\ell=2$ EE spectrum, which is dominated by foreground emission, 
is given by
\begin{equation}
{\cal B}^{EE}(\nu) = a_{s}(\nu_1\nu_2)^{\beta_s} + 
         \rho_{sd}a_{s}a_{d}(\nu_1^{\beta_s}\nu_2^{\beta_d}+ 
                                  \nu_1^{\beta_d}\nu_2^{\beta_s} )+ 
         a_{d}(\nu_1\nu_2)^{\beta_d}
\label{eq:dscorr} 
\end{equation} where
$\rho_{sd}$ is the dust synchrotron correlation coefficient,
$\nu_1$ and $\nu_2$ are the frequencies of the two spectra
that are correlated, the $\beta_d$ and $\beta_s$ are the dust and synchrotron
spectral indices, and $\nu=\sqrt{\nu_1\nu_2}$.
This fit is shown in Figure~\ref{fig:rawps_fspace}. After normalizing
the frequency to 65~GHz, the following coefficients
were found to reasonably represent the data: $a_{s}=0.64$, 
$\beta_s=-2.9$, $a_{d}=0.65$, $\beta_d=1.5$,
and $\rho_{sd}=0.46$. In order to produce the KV, KW, and KaW
features, there must be significant correlations between dust
and synchrotron emission.
For the $\ell=4$ EE spectrum a similar expression fits the data
if $\rho_{sd}$ is negative. 
   
Some care is needed in interpreting the statistical significance
of power spectra that include foreground emission and a cut sky.
The lack of statistical 
isotropy of the foreground emission means that it must be treated 
separately from the CMB when assessing the net noise. In the presence 
of foregrounds, the random uncertainty becomes
\begin{equation}
\Delta C_\ell^2 = {2\over (2\ell+1)f_{\rm sky}^2}[N_\ell^2 
+ 2N_\ell {\cal F}_\ell]
\end{equation}
where ${\cal F}_\ell$ is the foreground emission at each $\ell$.
We plot only the first term in 
Figures~\ref{fig:rawps_lspace} \& \ref{fig:rawps_fspace} 
to indicate the size of the statistical error. Additionally, with the sky cut
there is a noise-foreground 
coupling between $N_\ell^{E,B}$ and ${\cal F}_{\ell\pm 2}^{E,B}$, and
between $N_\ell^{E,B}$ and ${\cal F}_{\ell\pm 1}^{B,E}$. This is analogous
to the noise coupling shown in Appendix~\ref{app:fisher}.

\subsection{Power Spectrum of Foreground-Cleaned Maps Outside the P06 Mask}
\label{sec:psfgm}

We next discuss the power spectrum after removing the foreground 
emission from the {\it maps}.
Cleaning foregrounds not only changes the mean of 
$C_\ell$, but it reduces $\Delta C_\ell$ because of the couplings. 
The choice of model makes 
little difference to the conclusions. For all the following we
have subtracted the best fit KD3Pol $Q$ and $U$ templates from the 
Ka through W maps (both r4 and r9 versions) as described
in Section~\ref{sec:foregrounds}. Table~\ref{tab:eebb}
shows the EE $\ell=2$ and BB $\ell=5$, the multipoles
with the largest foreground contributions, for both before and 
after the subtraction. Where the foreground signal is dominant, 
the subtraction can reduce its level by a factor of 6-10 in temperature. 

When we fit and subtract the foreground templates, we use essentially 
all of the available data on polarized foreground emission. 
The error bar on the power spectrum of the cleaned maps is dilated 
in the cleaning process as discussed above. We do not include
an additional error for systematic uncertainty in the model. Rather, by 
comparing spectra of pre-cleaned to cleaned maps, we estimate that the
model removes at least 85\% of the synchrotron. This is demonstrated,
for example, in the KKa and KaKa combinations for $\ell=2$ EE in 
Table~\ref{tab:eebb}, in the subtraction shown in Figure~\ref{fig:inout},
and to a lesser degree by the null EB and BB power spectra.
We also note that to a good approximation foreground emission adds only in 
quadrature to CMB emission.

Figure~\ref{fig:lbylEEBB} shows the power spectra of the foreground cleaned
maps as a function of frequency for $\ell=2-9$. It also shows what 
we estimate to be the maximum levels of residual foreground contamination
in the power spectrum. In the figure, we plot the synchrotron spectrum scaled 
to 0.15 of the pre-cleaned Ka band value (in temperature). 
This shows that there is negligible residual synchrotron from 
40 to 60 GHz with the possible exception of $\ell=2$ at 40~GHz.
Given the size of the $\ell=2$ error bar, this potential contribution
to the determination of the optical depth is negligible as discussed 
in Section~\S\ref{sec:tau}. 
Constraining the residual 
dust contamination is more difficult. In Figure~\ref{fig:lbylEEBB},
we also show the MEM temperature dust model scaled by 5\%,
a typical dust polarization value. A similarly scaled FDS model
is almost identical. This shows that even if we did not model and subtract 
dust, the contamination from it would not be large in Q and V bands. 
A more detailed model might have to take into account the possibility 
that the electrons and 
dust grains are in regions at different line of sight distances with
different magnetic fields or that variations in the 
magnetic field could alias power from low multipoles to higher ones. 

The cross power spectra of the cleaned maps are combined by frequency
band for testing cosmological models. The 10 cross spectra (since K-band 
is used in the model,
there is no K-band cleaned spectrum) are assessed $\ell$ by $\ell$ 
with a least squares fit to a flat line in Figure~\ref{fig:lbylEEBB}.
The results are shown 
for the QQ+QV+VV (denoted ``QV combination'') and QV+KaV combinations in 
EE, we find $0.1<{\rm PTE}<1$ for all $\ell<16$, where PTE is for
``Probability to Exceed'' and is the probability that a 
random variable drawn from the same distribution exceeds the measured 
value of $\chi^2$. When W band is added to the mix, we 
find ${\rm PTE}<0.03$ for 
$\ell=5,7,9$, though all other values of $\ell$ give reasonable values.
For BB, all frequency combinations yield reasonable PTEs for all $\ell$. 
Thus, there is a residual signal in our power spectra that we do not
yet understand. It is evident in W band in EE at $\ell=7$
and to a lesser degree at $\ell=5$ and $\ell=9$. We see no clear evidence
of it anywhere else. 

\subsection{Null Tests and Systematic Checks Outside the P06 Mask}

Null tests are critical for assessing the quality of the data.
We have examined the data in a wide variety of ways based
on differencing assembly, frequency band, $\ell$ range, and year. 
We present selected, though typical, results in the following.
A particularly important test is
the null measurement of the BB, TB, and EB signals
as shown in Table~\ref{tab:chisq} and Figure~\ref{fig:chisq}.
These data combinations are derived from the 
same processing as the EE, TE, and TT combinations, where a signal is 
detected. Thus, the null result highlights the stability of the \WMAP\ data,
the mapmaking, the foreground cleaning, and the power spectrum estimation.

The power spectrum of the difference of the individual yearly maps 
is another significant test. Table~\ref{tab:chisq2} shows 
the results for all the yearly differences for $\ell=2-16$, the critical
region of $\ell$-space for the cosmological analysis.
We have also used
the ${\rm (u1\times v1+u2\times v2-u1\times v2-u2\times v1)/4}$
cross spectra to similar effect. Here again the $u$ and $v$ denote
different frequency bands.
This combination is equivalent to forming the power spectrum of 
the difference between year one and year two maps. 
In principle it does not contain any signal. The cross-spectrum method 
treats the noise in a slightly different way from the straight map 
method,  where one must use the error bars from one of the maps.
It has been checked with simulations. Similar combinations are used 
for the other years.

Using a variant of cross-spectrum method, we have also tested combinations
of DAs for multiple ranges in $\ell$ within each frequency band.
For all null tests, we find the expected null measurements, apart 
from the previously mentioned residuals at $\ell=5\&7$
in W band. Table~\ref{tab:chisq} gives the reduced $\chi^2$ for all
combinations of T, E, and B data for a number of data combinations. 

From Figure~\ref{fig:lbylEEBB}, it is clear that the large signal in W band 
is not residual
dust contamination because the dust would not fit measurements in VW.
Additionally, if one assumes that the polarized emission at a particular
$\ell$ is a fraction times the intensity at the same $\ell$, it would 
require $>40$\% dust polarization, which is unreasonable. Though this simple 
picture does not take into account the aliasing of intensity from a lower
$\ell$, we do not observe a similar effect with the synchrotron emission,
which in the simplest case is polarized by the same magnetic fields.
The W-band EE $\ell=7$ value is essentially unchanged 
by cleaning, removing a $10^{\circ}$ radius around the Galactic caps, or
by additionally masking $\pm 10^{\circ}$ in the ecliptic plane.

A number of tests have been done to identify this artifact of the data.
We are not yet certain if it is due to an ersatz signal or 
an incorrect noise term. The error bars on the individual year 
differences are too 
large to clearly see if the effect is the same from year to year.
Simulations show that $1/f$ alone cannot explain the signal.
The scan pattern in combination with the change in polarization is 
directly related to the large  
error bars at $\ell=5,7,9$ and is well understood. We have not identified
a mechanism that leads to a further increase in these uncertainties. 
We know that different treatments of the noise, for example using 
$N_{obs}$ weighting, decreases the magnitude of the discrepancy, though
we are confident that the ${\bf N^{-1}}$ treatment of the pixel noise is 
the correct approach. The discrepancy can be made smaller by eliminating
the W1 data simply because the error bars increase. The W1 radiometer
has the lowest noise but also the largest number of ``glitches'' 
\citep[13, 4, 1 in years 1, 2, \& 3 respectively,][]
{limon/etal:2003,hinshaw/etal:2003}. 
However, since we could not identify any correlation between the glitch
rate (assuming that unmasked glitches are responsible) and the magnitude 
of the signal, we do not have a basis for eliminating this channel.

We believe there is an as yet unknown coupling in the W-band
data that is driving the signal but more simulations and
more sensitivity are needed to understand it. We cannot rule
out similar lower-level problems in other bands, but we see no 
evidence of systematic effects
in BB, EB, or other values of $\ell$ and other frequencies in EE.
To avoid biasing the result by this residual artifact which also
possibly masks some unmodeled dust and synchrotron contamination, we limit the 
cosmological analysis to the QV combination. We also show that 
including W band EE does not alter our conclusions.   

\subsection{Analysis of Foreground-Cleaned Power Spectra Outside the P06 Mask.}

A comparison of the raw spectra and foreground cleaned spectra 
is shown in 
Figure~\ref{fig:correebb}. We start with the weighted sum of the 
8 cross spectra with $\nu>40$~GHz (without KW). This is the upper-level 
line (green) in the figure.
The individual maps are then cleaned and the power spectra remade
and coadded. This is shown in violet. Similar comparisons are repeated
for the QVW and QV combinations. A simple visual inspection shows that even
at the $\ell$s with the highest foreground contamination, the cleaning 
is effective.

From the bottom left panel in Figure~\ref{fig:correebb} one sees that 
there is a clear signal
above the noise in EE at $\ell<7$. For the QV combination,
${\cal B}^{EE}_{\ell=<2-6>}= 0.086\pm0.029~(\mu K)^2$.
The signal has persisted through a number of different analyses. 
We cannot rule out that this signal might find explanation in an 
unmodeled foreground component;
however, we find this explanation unlikely since the emission
would have to be strikingly different from the measured spatial and frequency
characteristics of the polarized foreground emission. Additionally, 
when different bands are coadded, the signal level is consistent:
for QVW ${\cal B}^{EE}_{\ell=<2-6>}= 0.098\pm0.022~(\mu K)^2$
and for all channels with $\nu>40~$GHz except KW, 
${\cal B}^{EE}_{\ell=<2-6>}= 0.095\pm0.019~(\mu K)^2$.
We have searched for systematic effects in the 
EE $\ell=2-8$ range and have not been able to identify any, other than the 
one discussed above. We cannot find a more plausible explanation than that
the signal is in the sky. We are thus led to interpret it cosmologically.
This is done in the next section. 

We show the EE signal for 
$\ell>20$ in Table~\ref{tab:highee} and in Figure~\ref{fig:highee} 
along with a comparison to other 
recent measurements \citep{leitch/etal:2005, sievers/etal:2005, 
barkats/etal:2005, montroy/etal:2005}. 
Based on the best fit to the TT spectra, we produce a template for the 
predicted EE spectrum $C_\ell^{EE,T}$ and form:
\begin{equation}
\chi^2(A^{EE}) = \sum_{\ell=50}^{800}
\delta C_\ell 
Q_{\ell \ell}^{EE} 
\delta C_{\ell} 
\label{eq:aee}
\end{equation}
where $\delta C_\ell = C_\ell^{EE}-A^{EE}C_\ell^{EE,T}$,
$A^{EE}$ is the fit amplitude,
and $Q^{EE}_{\ell \ell}$ is the diagonal Fisher matrix in 
Appendix~\ref{app:covar}. Off diagonal elements in $Q^{EE}_{\ell \ell'}$ 
have a negligible effect on the results. 

The results of the fit are plotted in Figure~\ref{fig:eeteampphase}
for various frequency combinations. We plot $\Delta\chi^2$ from the 
minimum value and find that $A^{EE}=0.95\pm0.35$ for the pre-cleaned 
QVW combination, where the uncertainty is determined from the bounds 
at $\Delta\chi^2=1$. 
The reduced $\chi^2$ at the minima are 1.34, 1.34, 1.24, and 1.30
for QV, VW, QVW, and KaQVW combinations respectively, most likely indicating
residual foreground contamination. At a relative amplitude
of zero, $\Delta\chi^2 = 1.0,~3.3,~6.2, ~\&~16$ respectively for the same 
frequency combinations. It is clear that the
noise is not yet low enough to use just QV as was done at low multipoles.
In addition, cleaning the maps with the KD3Pol is problematic because
the K-band window function is reduced to 0.1 by $\ell=250$.
When the same code is used to analyze EB and BB
data, the fitted amplitude is always consistent with zero.
To summarize, the \WMAP\ EE data are
consistent with a model of adiabatic fluctuations based on the temperature
maps at greater than the $2\sigma$ level for the QVW and KaQVW combinations.  

Figure~\ref{fig:tespec} (left panel) shows the TE spectrum for $\ell<16$. 
We use V band for temperature and the QV combination for polarization.  
Several aspects of the new processing led to increased errors and a 
reduced low-$\ell$ signal estimate relative to the first-year result 
\citep[Figure~8,][]{kogut/etal:2003}.  These include: improvements in 
mapmaking and power spectrum estimation (especially accounting for 
correlated noise and applying ${\bf N^{-1}}$ weighting); limiting the 
bands to just Q and V instead of Ka-W; increasing the cut from KP0 to 
P06; and improvements in foreground modeling, including a new estimate 
of dust polarization. 
Recall also that the first-year result was based on the combinations 
of Ka, Q, V, 
and W bands and did not include a dust polarization template in
contrast to the new prescription. Furthermore, 
if the year-two data are processed in the same way as the first-year data,
we obtain a spectrum similar to that in \citet{kogut/etal:2003} 
indicating that the major difference between first-year and three-year
results rests on 
new knowledge of how to make and clean polarization maps.
The new spectrum is fully consistent with the first-year results
and prefers a model based just on TT and EE data to a null signal
at the $2\sigma$ level. However, the new spectrum is also consistent with 
the absence of a TE signal. Thus, it will take greater signal-to-noise
to clearly identify the TE signal with our new analysis methods.

Figure~\ref{fig:tespec} (right panel) shows the TE signal over the 
full range in $\ell$.
Other detections of TE at $\ell>100$ have been reported by 
DASI ($2.9\sigma$) \citep{leitch/etal:2005}, Boomerang ($3.5\sigma$)
\citep{piacentini/etal:2005}, and CBI ($3.3\sigma$)
\citep{sievers/etal:2005}.
The \WMAP\ data have had foreground models subtracted from both 
the temperature and polarization maps prior to forming the cross correlation. 
The expected anticorrelation
between the polarization and temperature is clearly evident.
To quantify the consistency with the TT data we make a TE template
based on the model fit to TT. Next, a fit is made to the TE data
for $20<\ell<500$ with the following: 
\begin{equation}
\chi^2(A^{TE},\Delta\ell) = \sum_{\ell \ell'}
\delta C_\ell 
Q_{\ell \ell'}^{TE} 
\delta C_{\ell'} 
\label{eq:ate}
\end{equation}
where $\delta C_\ell = C_\ell^{TE}-A^{TE}C_\ell^{TE,T}(\Delta\ell)$,
$C_l^{TE,T}(\Delta\ell)$ is the predicted power spectrum shifted
by $\Delta\ell$, $A^{TE}$ is the fit amplitude,
and $Q^{TE}_{\ell \ell'}$ is the diagonal Fisher matrix in 
Appendix~\ref{app:covar}. Off diagonal elements in $Q^{TE}_{\ell \ell'}$ 
have a negligible effect on the results. Similar 2D fits were done in
\citet{readhead/etal:2004b}. We show the combination that uses V and W bands 
for T and Q and V bands for E. The result,
shown in Figure~\ref{fig:eeteampphase},
is $A^{TE}=0.93\pm0.12$ and $\Delta\ell=0\pm 8$
with $\chi^2/\nu=468/482$ (PTE=0.66). Similar results are obtained
with other band combinations. Thus the TE data are consistent 
with the TT data to within the limits of measurement. 

Figure~\ref{fig:grandspec} shows a summary of the various components 
of the CMB anisotropy.

\section{Cosmological Analysis}
\label{sec:cosmo}

The $\ell<100$ region of the CMB polarization spectra is rich
with new tests of cosmology. The EE spectrum gives us a new measure of 
the optical depth. The same free electrons from reionization
that lead to the $\ell<10$
EE signal act as test particles that scatter the quadrupolar temperature
anisotropy produced by gravitational waves (tensor modes) originating
at the birth of the universe. The scatter results in polarization B
modes. Tensor modes 
also affect the TT spectrum in this region. A combination of these 
and related observations leads to direct tests of models of inflation.

The detection of the TE anticorrelation near $\ell\approx30$
is a fundamental measurement of the physics of the formation
of cosmological perturbations \citep{peiris/etal:2003}. 
It requires some mechanism like inflation to produce and 
shows that superhorizon fluctuations must exist.
\citet{turok:1996} showed that with enough free parameters one 
could in principle make a model based on post-inflation causal physics that 
reproduced the TT spectrum. \citet{spergel/zaldarriaga:1997} show 
that the TE anticorrelation is characteristic of models with 
superhorizon fluctuations. The reason is that the anticorrelation 
is observed on angular scales larger than the acoustic horizon
at decoupling. Thus, the observed velocity-density correlations
implied by the TE data must have existed on scales larger than the 
horizon and were not produced by post-inflation causal processes.

Although multiple distinct physical mechanisms affect the 
$\ell<100$ spectra, their effects can be disentangled
through an analysis of the full data complement \citep{spergel/etal:prep}.
The separation, though, is not perfect and there remain
degeneracies. In particular, to some degree, the values 
of the scalar spectral index, $n_s$, optical depth, and the tensor to scalar 
ratio, $r$, may be traded against each other, although
far less than in the first-year \WMAP\, results.
As the data improve, or as more data sets are added, the 
degeneracy is broken further.
In the following we take a step back from the full MCMC
analysis \citep{spergel/etal:prep} and estimate $\tau$ and
$r$ from analyses of just the $\ell<10$ polarization spectra. 
This approach 
aids our intuition in understanding what it is in the data 
that constrains the cosmological parameters.

\subsection{The Optical Depth of Reionization}
\label{sec:tau}

Our knowledge of the optical depth ripples through the 
assessment of all the cosmic parameters.
Free electrons scatter the CMB photons thereby reducing the amplitude 
of the CMB spectrum. This in turn directly impacts the determination of 
other parameters. 

The distinctive signature of reionization is at $\ell<10$ in EE.
The only known contamination is from foreground emission 
which has been modeled and subtracted. 
The amplitude of the reionization signal is proportional to $\tau$
in TE and is proportional to  $\tau^2$ in EE and BB.
In the first year analysis, we imposed a prior that
$\tau<0.3$ \citep{spergel/etal:2003}. Such a high value
would produce a signal $>6$ times the model in 
Figure~\ref{fig:correebb} and is clearly inconsistent with the EE
data. Thus this new analysis is a significant improvement over the 
previously assumed prior.

We assess $\tau$ using three methods: (1) with template fits to the 
EE power spectra; (2) with an exact likelihood 
technique based directly on the maps as described
in Appendix~\ref{app:ektau};
and (3) with a multiparameter MCMC fit to all the data as reported
in \citet{spergel/etal:prep}. The first method is based directly on the MASTER 
spectrum \citep[and Appendix~\ref{app:master}]{hivon/etal:2002}
of EE data and serves as a simple check of the other two.
Additionally, the simplicity allows us to 
examine the robustness of the EE and TE detections to
cuts of the data. The second method is robust and takes into account 
the phases of the EE and TE signals. It is run either as a stand alone
method, as reported here, or as part of the full MCMC chain
as reported in \citet{spergel/etal:prep}. The best estimate of the optical
depth comes from the full chains.

For the template fits,
$\Lambda$CDM power spectra were generated for $0\le\tau\le0.3$,
with the remaining parameters fixed to $n_s=0.96$, $\omega_b=0.0226$,
$\omega_m=0.133$, and $h=0.72$.
For each spectrum, the scalar amplitude $A$ is fixed by 
requiring that ${\cal B}^{TT}_{\ell=200} = 5589\,\mu{\rm K}^2$.
We then form:
\begin{equation}
{\cal L}(\tilde\tau)=
\frac{1}{(2\pi)^n\sqrt{{\rm det}(D)}}
\exp[-\sum_{\ell}(\vec{x}_\ell-\vec{x}^{th}_\ell)D^{-1}
(\vec{x}_\ell-\vec{x}^{th}_\ell)/2]
\end{equation}
where $\vec{x}_\ell$ is the data as shown in Figure~\ref{fig:correebb}, 
$\vec{x}^{th}_\ell={\cal B}^{EE}_\ell (\tilde\tau)$ is the model 
$\Lambda$CDM spectrum,
\begin{equation}
D_\ell = \frac{2}{2\ell + 1}\frac{1}{f_{\rm sky}^{EE}(\ell)^2}
({\cal B}^{EE}_\ell (\tilde\tau) + N^{EE}_\ell)^2 
\end{equation}
as in C.14, and $N^{EE}$ is the uncertainty 
shown in Figure~\ref{fig:correebb} and is
derived from the MASTER spectrum determination.
We use the symbol $\tilde\tau$ in this context
because the likelihood function we obtain is not the 
full likelihood for $\tau$. 
Uncertainties in other parameters, especially $n_s$, have been
ignored and the $C_\ell$ distribution is taken to be Gaussian.
Thus ${\cal L}(\tilde\tau)$ does not give a good estimate of the 
uncertainty. Its primary use is as a simple parametrization of the data. 
We call this method ``simple tau.''
Table~\ref{tab:taucomp} shows that simple tau is stable
with data selection. One can also see that if the QQ component
is removed from the QV combination, $\tau$ increases slightly. This is
another indication that foreground emission is not biasing the result.
Additionally, one can see that removing $\ell=5,7$ for all band combinations
does not greatly affect $\tau$.

The optimal method for computing the optical depth is 
with the exact likelihood (as in Appendix~\ref{app:ektau}). 
The primary benefits are: it 
makes no assumptions about the distribution of $C_\ell$ at each $\ell$
but does assume that the polarization signal and noise in the maps
are normally distributed;
it works directly in pixel space, taking advantage of
the phase relations between the T and E modes both together
and separately; and  it is unbiased. The only disadvantages 
are that it is computationally intensive and that it is not easy to 
excise individual values of $\ell$ such as $\ell=5,7$. 

In the exact likelihood method we take into account errors in our 
foreground model by marginalizing the likelihood function given in 
(Appendix~\ref{app:ektau}) over the errors in the fitting coefficient 
for synchrotron emission, $\alpha_{s}$. We ignore foreground errors in 
dust emission, as polarized dust emission is negligible in any of 
combinations of frequencies reported in Table~9. When errors in $\alpha_{s}$ 
are Gaussian, the marginalization simply yields an additional term in 
the covariance matrix,
\begin{equation}
C_{ij} = S_{ij} + N_{ij} \rightarrow C_{ij} = S_{ij} + N_{ij} + F_{ij},
\end{equation}
where 
\begin{equation} 
F_{ij} = \sigma_s^2 f_i f_j,
\end{equation}
and $\sigma_s^2=\langle\alpha_{s}^2\rangle-\langle\alpha_{s}\rangle^2$ and 
$f_i$ is a template map of polarized foreground (i.e., Q and U maps in K 
band). Here, the mean 
values, $\langle\alpha_{s}\rangle$, are given in the second column 
of Table~3. We find $\sigma_s=0.007$ in Q and V band. To be conservative, 
we adopt $\sigma_s=0.01$ as our foreground error, which is the 2-$\sigma$ 
bound on the foreground error in QV combination. As the foreground 
marginalization yields a new positive term in the covariance matrix, 
a fraction of the signal that was attributed to CMB before is now 
attributed to foreground, when the spatial distribution of the signal 
is the same as that of K-band maps. The values of $\tau$ with 
the foreground marginalization are tabulated Table~9. The marginalization 
reduces $\tau$ by 0.0017 in QV. The largest effect is seen in Q band, 
for which $\tau$ drops by 0.0027. Thus, the foreground error does not 
significantly affect our determination of $\tau$.

Table~\ref{tab:taucomp} shows that similar vales of $\tau$ are obtained
for a wide variety of band combinations. This is another indication
that foreground emission is not significant. We conservatively select
the QV combination. Table~\ref{tab:taucomp} also compares the exact 
likelihood for the 
EE QV combination to the simple tau method. One can see that
simple tau is slightly biased high when compared to the exact likelihood
and underestimates the likelihood at 
$\tau=0$. One source of the bias is the assumption of a Gaussian
likelihood. Nevertheless, it is reassuring that a variety of combinations 
of data give consistent values of $\tau$.

The values given here are just for the EE and TE data 
considered alone, with the first peak TT amplitude fixed. 
When the exact likelihood is used in the full MCMC analysis
\citep{spergel/etal:prep} yielding the best estimate, 
we find $\tau=\ensuremath{0.089 \pm 0.030}$, slightly lower
than the values reported here but with the same uncertainty, indicating
that the simple analysis has exhausted most of the information on the optical
depth contained in the polarization data. 

As discussed in 
\citep{spergel/etal:prep} there is a degeneracy between the scalar 
spectral index, $n_s$ and $\tau$. If we had instead selected the 
K-band directions for the dust polarization template, we would
have found $\tau=\ensuremath{0.107}$ and an increase
in $n_s$ of 0.004. A similar shift would have been found using the
KaQVW combination shown in  
Table~\ref{tab:taucomp} and Figure~\ref{fig:tau_likes}. This is another
indication of the relative insensitivity of the results to the cleaning
method.

\subsection{Gravitational Waves}

The $C_\ell^{BB}$ spectrum directly probes
the primordial gravitational wave background produced by tensor 
fluctuations in the early universe.
The existence of these gravitational waves was proposed by
\citet{starobinsky:1979}. Modern treatments may be found
in, for example, \citet{liddle/lyth:CIALSS, dodelson:MC, mukhanov:PFC}.
While scalar and tensor fluctuations both contribute to the
TT and EE spectra, only tensors produce B modes.
Inflation models generally predict similar scalar spectra,
but differ in their prediction of the tensor component. For example,
ekpyrotic/cyclic models 
\citep{khoury/etal:2002,steinhardt/turok:2002} predict 
no observable tensor modes.  

The tensor contribution is quantified with the 
tensor to scalar ratio $r$. We follow the convention in
the CAMB code \citep[Version, June 2004]{lewis/challinor/lasenby:2000},
in CMBFAST v4.5.1 \citep{seljak/zaldarriaga:1996}  
and in \citet{peiris/etal:2003,verde/etal:2003}:
\begin{equation}
r\equiv { \Delta_h^2(k_0)\over \Delta_{\cal R}^2(k_0) }.
\end{equation}
Here, $\Delta^2_{\cal R}$ and $\Delta^2_h$ 
are the variance due to 
scalar and tensor modes respectively. They are defined through
\begin{eqnarray}
\langle {\cal R}^2\rangle &=&\int \frac{dk}{k} \Delta_{\cal R}^2(k) \\ 
{\rm and~~~}\langle h_{ij}^{prim}h^{prim,ij} \rangle 
&=&\int \frac{dk}{k}\Delta_{h}^2(k),
\end{eqnarray}
where $h_{ij}^{prim}$ is the primordial tensor metric perturbation in
real space that was generated during inflation and stretched to 
outside the horizon\footnote{Note that our convention yields $r=16\epsilon$
for slow-roll inflationary models with a single scalar field.
Here, $\epsilon$ is the slow roll parameter related to the square of the 
slope of the inflaton potential.}. \citet{peiris/etal:2003} shows 
the $k$-dependence of these expressions.

The expression for $r$ is evaluated at $k_0=0.002~{\rm Mpc}^{-1}$ 
corresponding to $l\approx \eta_0k=30$ with the distance to 
the decoupling surface $\eta_0\approx 14,400~$Mpc.
Following \citet{verde/etal:2003}, we use
$\Delta^2_{\cal R}(k_0)=2\times10^4\pi^2A(k_0)/9T^2_0
\approx 2.95\times10^{-9}A(k_0)$ with $T_0$ in microkelvins.
Some of the simple models of inflation in a $\Lambda$CDM cosmology
predict $r\simeq 0.3$ \citep[e.g., ][]{liddle/lyth:CIALSS,
boyle/steinhardt/turok:2005}. For example, 
near this range inflationary models with a massive scalar field,
$V(\phi)=m^2\phi^2/2$, predict $r=8/N_e=4(1-n_s)=0.13-0.16$  
\citep{linde:1983} and models with a self coupling, 
$V(\phi)=\lambda\phi^4/4$, predict $r=16/N_e=16(1-n_s)/3=0.27-0.32$
for $N_e=60-50$. Here, $N_e$ is the number of e-foldings before the 
end of inflation.  However, some variants produce $r>0.32$
\citep[e.g.][]{mukhanov/vikman:2005} while many other have $r\ll 0.1$. 

For the best fit {\WMAP}-only $\Lambda$CDM plus tensor model,
the optical depth is $\tau=\ensuremath{0.091}$. 
If we add to this model
a tensor component with $r=0.3$, then
${\cal B}^{BB}_{\ell=<2-6>}=0.001\,\mu{\rm K}^2$.
A simple average of the $C_\ell^{BB}$ data gives  
${\cal B}^{BB}_{\ell=<2-6>} = -0.044\pm0.030\,\mu{\rm K}^2$,
${\cal B}^{BB}_{\ell=<2-6>} = -0.018\pm0.023\,\mu{\rm K}^2$,
${\cal B}^{BB}_{\ell=<2-6>} = 0.003\pm0.020\,\mu{\rm K}^2$,
for QV, QVW, and $\nu>40~$GHz (no KW) combinations respectively.
To detect a signal at the upper range of the predictions would require
maps with $\approx 5 $ times smaller error bars.

We constrain $r$ by directly fitting a template 
of $C^{BB}_l$ to the BB data. 
With the above definition, $r$ directly scales the $C_l^{BB}$ power 
spectrum. Additionally, the amplitude of $C_l^{BB}$ 
for $\ell<16$ scales as $\tau^2$. 
We set the template to be the standard $\Lambda$CDM model 
\citep{spergel/etal:prep} and use 
the  single field inflation consistency relation, $n_t=-r/8$, 
to fix the tensor spectral index. We assume the spectral index does not run
and set $n_s=0.96$. We distinguish the $r$ in the template fit
by the $\tilde r$ notation. The sum is over $2\le \ell \le 11$.

The results of the fit are plotted in Figure~\ref{fig:simple_rtau}.
When we consider just the limit on $\tilde r$ from the polarization
spectra, ignoring the tensor contribution to TT, we find
$\tilde r < 2.2$ (95\% CL) after marginalizing over $\tilde\tau$.
It is clear that the BB spectrum is not driving the limit on
$r$. After including the TT data, the limit drops 
to $\tilde r < 0.27$ (95\% CL). This shows that the TT data 
in combination with the limits on $\tau$ from EE and TE are
leading to the limit on $r$. The full MCMC analysis
gives  $r< \ensuremath{0.55}$  (95\% CL) 
with just the \WMAP\ data.
The increase in the error over the simple method given above 
is the result of the marginalization over
the other parameters, particularly $n_s$. Additionally, when $n_s$ is allowed
to depend on $k$, the error in $r$ increases dramatically, allowing
$r< 1.3$ (95\% CL). 

We can relate $r$ to the current energy density in primordial gravitational
radiation \citep{krauss/white:1992,peiris:phd},
\begin{equation}
\Omega_{GW} = \frac{1}{12H^2_0}\int{\frac{dk}{k}
	\Delta^2_h(k) \dot{T}^2(k,\eta)},
\label{eqn:omegagw}
\end{equation}
where $\eta$ is conformal time and the transfer function,
$T(k,\eta)$, is given in Equation~\ref{eq:tffinal}.
The approximation given in  
Equation~\ref{eq:ogw} evaluated for $A=0.838$ and $\tilde r<2.2$
yields $\Omega_{GW}<9.6\times 10^{-12}$ (95\% CL) and for 
$r<\ensuremath{0.55}$, 
$\Omega_{GW}<2.0\times 10^{-12}$ (95\% CL).

\section{Discussion and Conclusions}
\label{sec:conclude}

\WMAP\ detects significant levels of polarized
foreground emission over much of the sky. The minimum
in contamination is near 60 GHz outside the P06 mask. 
To detect the polarization
in the CMB at $\ell<10$ a model of the foreground emission must be
subtracted from the data. This situation differs
from that of the analysis of the temperature anisotropy 
for which the foreground emission may be simply masked
as a first approximation.\footnote{Of course our full analysis
\citep{hinshaw/etal:2003} involved extensive modeling of 
the foregrounds \citep{bennett/etal:2003}.}

\WMAP\ has detected the primary temperature anisotropy, the
temperature polarization cross correlation, and the E-mode polarization
of the CMB. We detect the optical depth with 
$\tau=\ensuremath{0.089 \pm 0.030}$ in a full fit to all
\WMAP\  data. This result is supported by stand-alone analyses of the 
polarization data.
Using primarily the TT spectrum, along with the optical depth established
with the TE and EE spectra, the tensor to scalar ratio is limited
to $\ensuremath{r_{0.002} < 0.55 \mbox{ } (95\%\mbox{\ CL})}$. 
When the large scale structure power spectrum is added to 
the mix \citep{spergel/etal:prep}, 
the limit tightens to $\ensuremath{r_{0.002} < 0.28 \mbox{ } (95\%\mbox{\ CL})}$. 
These values are approaching the predictions of the simplest inflation models.

A clear detection of the B modes at $\ell<100$ would give a direct
handle on the physics of the early universe at energy scales
of $10^{15}-10^{16}~$GeV. This paper shows that care
will be required to
unambiguously separate the intrinsic signal from the foreground emission.
However, the BB spectrum is particularly clean in \WMAP\ and, at least  
for $\ell=2,3$, the foreground contamination is relatively low. 
In the noise dominated regime,
the error bar on $C_\ell^{BB}$ decreases in proportion to time.
Continued \WMAP\ operations combined with other experimental
efforts are nearing a range of great interest.

These new results involve a complete reevaluation of all
the components of our previous analyses, from the beams and gain models through
to the mapmaking and foreground modeling.
The data and most of the derived data products 
are available through the LAMBDA
website, http://lambda.gsfc.nasa.gov/.
\WMAP\ continues to operate nominally. In the future we will
address a number of the open issues raised above. 
In particular, we can anticipate a better understanding of systematic
errors and foreground emission, and therefore improved constraints 
on $\tau$ and $r$. It is remarkable that our understanding of the cosmos 
has reached the point where we have begun to quantitatively
distinguish between different models of the birth of the universe.

\section{Acknowledgments}

We wish to thank Uros Seljak, Suzanne Staggs, and Paul Steinhardt
for enlightening conversations on polarization and inflation, and 
Jim Peebles for his explanation of reionization. We thank Bruce Draine
for helpful discussions on dust polarization and Rainer Beck for 
many suggestions on modeling the Galactic magnetic field.
We thank Andr{\'e} Waelkens and Torsten En{\ss}lin for pointing 
out a mistake in the original calculation of $g_{sync}$. (The fix 
affected Figures 8 and 9 but not the conclusions.) We thank Bruce 
Winstein for a detailed set of comments on the submitted draft. 
EK acknowledges support
from an Alfred P. Sloan Research Fellowship. HVP wishes to acknowledge
useful discussions with R. Easther, S. Larson, D. Mortlock, and A. Lewis.
The {\WMAP} mission is made possible by the support of the Office of Space 
Sciences at NASA Headquarters and by the hard and capable work of scores of 
scientists, engineers, technicians, machinists, data analysts, 
budget analysts, managers, administrative staff, and reviewers.
HVP is supported by NASA through Hubble Fellowship grant \#HF-01177.01-A
awarded by the Space Telescope Science Institute which is operated
by the Association of Universities for Research in Astronomy, Inc., for 
NASA under contract NAS 5-26555.
This research has made use of NASA's Astrophysics Data System Bibliographic
Services, the HEALPix software, CAMB software, and the CMBFAST software.
This research was additionally supported by NASA LTSA03-000-0090,
NASA ATPNNG04GK55G, and NASA ADP03-0000-092 awards.

\appendix

\newcommand{\imag}{i}
\newcommand{\Mfunction}[1]{#1}

\newcommand{\identity}{\mathbf{I}}
\newcommand{\pauli}{{\bm\sigma}}
\newcommand{\Jones}{\mathbf{J}}
\newcommand{\hatJones}{\mathbf{\hat J}}
\newcommand{\Pol}{\mathbf{P}}
\newcommand{\Polin}{\Pol_{\rm in}}
\newcommand{\Polout}{\Pol_{\rm out}}
\newcommand{\hatPol}{\mathbf{\hat P}}
\newcommand{\hatPolin}{\hatPol_{\rm in}}
\newcommand{\Ein}{E_{\rm in}}
\newcommand{\Eout}{E_{\rm out}}
\newcommand{\D}{\mathbf{D}}
\newcommand{\zero}{{\bf 0}}

\newcommand{\M}{\mathbf{M}}
\newcommand{\DA}{{\rm DA}}
\newcommand{\connect}{{\rm connect}}
\newcommand{\radiometer}{{\rm radiometer}}
\newcommand{\feed}{{\rm feed}}
\newcommand{\bandpass}{{\rm bandpass}}
\newcommand{\coldT}{{\rm coldT}}
\newcommand{\warmT}{{\rm warmT}}
\newcommand{\OMT}{{\rm OMT}}
\newcommand{\OMTfeed}{{\rm OMT+feed}}
\newcommand{\phase}{{\rm phase}}
\newcommand{\switch}{{\rm switch}}
\newcommand{\amp}{{\rm amp}}
\newcommand{\OFO}{{\rm OFO}}

\newcommand{\ximone}{x_{{\rm im},1}}
\newcommand{\ximtwo}{x_{{\rm im},2}}
 
\newcommand{\diagtwo}[2]
{\left(\begin{array}{cc}#1&0\\0&#2\end{array}\right)}
\newcommand{\blockdiagtwo}[2]
{\left(\begin{array}{cc}#1&\zero\\\zero&#2\end{array}\right)}

\newcommand{\diagfour}[4]
{\left(\begin{array}{cccc}#1&0&0&0\\0&#2&0&0\\0&0&#3&0\\0&0&0&#4
\end{array}\right)}

\newcommand{\pauliQ}{\pauli_3}
\newcommand{\pauliU}{\pauli_1}
\newcommand{\pauliV}{\pauli_2}

\newcommand{\colvec}[2]{\left(\begin{array}{c}{#1}\\{#2}\end{array}\right)}
\newcommand{\threebythree}[9]{\left(\begin{array}{ccc}{#1}&{#2}&{#3}\\
{#4}&{#5}&{#6}\\{#7}&{#8}&{#9}\end{array}\right)}
\newcommand{\xslm}[3]{{}_{#2}{#1}_{#3}}
\newcommand{\wignerthreej}[6]{\left(\begin{array}{ccc}{#1}&{#2}&{#3}\\
{#4}&{#5}&{#6}\end{array}\right)}
\newcommand{\yslm}[3]{{{}^{}_{#1}Y_{#2}^{#3}}}
\newcommand{\yslmcc}[3]{{{}^{}_{#1}Y_{#2}^{#3*}}}

\section{Radiometer Model}
\label{app:jones}

In this section we develop a simple model for the \WMAP\ instrument using
Jones matrices \citep{jones:1941,montgomery/dicke/purcell:POMC,
blum:1959,faris:1967,
sault/hamaker/bregman:1996,tinbergen:AP,hu/hedman/zaldarriaga:2003}.
In the following we assume that all circuit elements are matched
and ignore additive noise terms.
 
The Jones matrix $\Jones$ models the instrumental response to polarization,
\begin{equation}
\Eout = \Jones \Ein
\end{equation}
linearly relating the output electric field to the input.
WMAP is a differential instrument, so the input radiation vector $\Ein$
has four elements, $(E^A_x,E^A_y,E^B_x,E^B_y)$, corresponding to the electric
field seen by the A- and B-side feed pair.
The outputs $\Eout$ are the inputs to the detectors.
 
The first link in the chain is to model the optics, feeds, and
orthomode transducers (OMTs).
We consider them as a single unit, because ascribing effects
to the individual components is difficult and not well defined
in terms of observations. We include two effects,
loss imbalance and polarization leakage:
\begin{eqnarray}
\Jones^{A,B}_\OFO &=& \Jones^{A,B}_{\rm loss}\Jones^{A,B}_{\rm crosspol}
\label{eq:a2}
\\
\Jones^{A,B}_{\rm loss} &=& \diagtwo{L^{A,B}_x}{L^{A,B}_y}
\\
\Jones^{A,B}_{\rm crosspol} &=&
\twobytwo
{1}
{X^{A,B}_1 e^{iY^{A,B}_1}}
{-X^{A,B}_2 e^{-iY^{A,B}_2}}
{1}
\end{eqnarray}
Here $L^{A,B}_{x,y}$ is the loss for the particular polarization
and $X^{A,B}_{1,2}$ quantifies the level of cross-polarization 
(or polarization isolation) leakage,
which we model as a small rotation error. 
The matrix $\Jones^{A,B}_{\rm crosspol}$ is the first term in the expansion
of a general unitary matrix but is not unitary itself.
The subscripts ``1'' and ``2''
refer to the two orthogonally polarized radiometers which are differenced
to form $\Delta_P$. The matrix 
$\Jones^{A,B}_{\rm crosspol}$ is the first term in the expansion of a general
unitary matrix, though it is not unitary itself. 
The cross-polarization terms are allowed to have arbitrary phases
$Y^{A,B}_{1,2}$. It is possible for cross polarization to produce
circular polarization but \WMAP cannot detect it in $\Delta_I$ or
$\Delta_P$.
While in general there are four loss terms, two of them are calibrated out.
The two that remain are the radiometer loss imbalances,
$x_{{\rm im},1}$ and $x_{{\rm im},2}$.
\citet[Table 3]{jarosik/etal:2003b} measured the loss imbalances by fitting the
response to the common mode CMB dipole signal, and found them to be $\la1\%$.
The mean imbalance, $\bar{x}_{\rm im}=(x_{{\rm im},1}+x_{{\rm im},2})/2$,
is corrected for by the map-making algorithm,
while the ``imbalance in the imbalance'',
$\delta x_{\rm im}=(x_{{\rm im},1}-x_{{\rm im},2})/2$, is not
\citep[\S C.3]{hinshaw/etal:2003b}.
To connect the different notations,
$L^A_x = L_1(1+\ximone)$,
$L^A_y = L_2(1+\ximtwo)$,
$L^B_x = L_2(1-\ximtwo)$, and
$L^B_y = L_1(1-\ximone)$.
The $L_1$ and $L_2$ are calibrated out.

The next step is to model the radiometers.
They are described in detail in \citet{jarosik/etal:2003},
so we simply present the Jones representation of the radiometer and
refer the reader to the paper for more details.
\begin{equation}
\Jones_\radiometer = \Jones_\warmT\Jones_\switch\Jones_\amp\Jones_\coldT
\end{equation}
\begin{eqnarray}
\Jones_\coldT &=& \frac{1}{\sqrt{2}} \twobytwo{1}{1}{1}{-1}
\\
\Jones_\amp &=& \diagtwo{g_s}{g_d}
\\
\Jones_\switch &=& \diagtwo{1}{e^{i\phi}}
\\
\Jones_\warmT &=& \frac{1}{\sqrt{2}} \twobytwo{1}{1}{1}{-1}
\end{eqnarray}
Here, $g_s,g_d$ are the amplifier gains in the two legs of the radiometer,
and $\phi$ is the instantaneous phase of the (unjammed) phase switch.
We have lumped the warm and cold amplifiers together.

\begin{equation}
\Jones_\DA = \Jones_\bandpass\Jones_\radiometer\M_\connect\Jones_\OFO
\end{equation}

\begin{eqnarray}
\M_\connect &=&
\left(\begin{array}{cccc}
1&0&0&0\\
0&0&0&1\\
0&1&0&0\\
0&0&1&0
\end{array}\right)
\\
\Jones_\radiometer &=& \blockdiagtwo{\Jones^{(1)}_\radiometer}
{\Jones^{(2)}_\radiometer}
\\
\Jones_\bandpass &=&
\diagfour{f_{13}(\omega)}{f_{14}(\omega)}{f_{23}(\omega)}{f_{24}(\omega)}
\end{eqnarray}
 
The detector outputs in counts $(c_{13},c_{14},c_{23},c_{24})$ are the diagonal
elements of $\Polout=\langle \Eout\Eout^\dagger\rangle$,
multiplied by the responsivities $(s_{13},s_{14},s_{23},s_{24})$.
$\Jones_\OFO$ is a 4x4 matrix with $\Jones_\OFO^A$ (Equation~\ref{eq:a2})
filling the upper left
2x2 entries and $\Jones_\OFO^B$ filling the lower right 2x2 entries. 
\begin{equation}
\Polout = \Jones_\DA \Polin \Jones_\DA^{\dagger}
\end{equation}
\begin{equation}
\Polin = \left(\begin{array}{cc}
\Polin^A & 0\\
0 & \Polin^B
\end{array}\right) \\
\end{equation}
\begin{equation}
\Pol^X_{\rm in} = \left(\begin{array}{cc}
T^X+Q^X&U^X-iV^X\\
U^X+iV^X&T^X-Q^X
\end{array}\right).
\end{equation}
In this expression, Stokes $Q$, $U$, and $V$ refer to the quantities
measured in the radiometer reference frame; we drop the ``Rad'' notation
used in \S\ref{sec:syst} for notational convenience.
Before the outputs are recorded they are demodulated in phase with the
phase switch. We model this process as
\begin{equation}
c_{ij} \to \frac{1}{2}\left[c_{ij}(\phi_i)-c_{ij}(\phi_i+\pi-\delta_i)\right]
\end{equation}
where $\phi_i$ is the phase difference between the two radiometer legs,
and $\delta_i$ is the error between the two switch states.
 
Since the input radiation is incoherent,
\begin{eqnarray}
\Polout = \int{d\omega\,\frac{\partial\Polout}{\partial\omega}}.
\end{eqnarray}
Since $\Jones_\bandpass$ is the only frequency dependent 
component in the model, we make the substitution
$f^2_{ij}(\omega)\to \tilde{f}^2_{ij}$, where
\begin{equation}
\tilde{f}^2_{ij} = \int{d\omega\,f^2_{ij}(\omega)}.
\end{equation}

The calibrated detector outputs are $d_{ij} = c_{ij}/G_{ij}$,
where $G_{ij}$ is a gain for the temperature difference,
\begin{eqnarray}
G_{ij} &=& \frac{1}{2} L_i g_{is} g_{id} \tilde{f}^2_{ij} s_{ij}
        \cos(\delta_i/2) \cos(\phi_i-\delta_i/2)
        (1-\epsilon_{ij}).
\end{eqnarray}
Here $\epsilon_{ij}$ is the calibration uncertainty.
 
The radiometer signal channels are $\Delta T_i = (d_{i3}-d_{i4})/2$,
from which are formed the temperature and polarization signal channels
$\Delta T_I,\Delta T_P$.
Then to first order in the systematic uncertainties,
\begin{eqnarray}
\Delta T_I &=& 2\,\delta x_{\rm im}\,{Q^-} + {{\epsilon}^-}\,{Q^+} +
    \left( 1 + {{\epsilon}^+} \right) \,{T^-} +
    2\,\bar{x}_{\rm im}\,{T^+} + {Z^A_-}\,{U^A} - {Z^B_-}\,{U^B}
\\
\Delta T_P &=& 2\,\bar{x}_{\rm im}\,{Q^-} +
    \left( 1 + {{\epsilon}^+} \right) \,{Q^+} + {{\epsilon}^-}\,{T^-} +
    2\,\delta x_{\rm im}\,{T^+} + {Z^A_+}\,{U^A} + {Z^B_+}\,{U^B}
\end{eqnarray}
Here $T^\pm \equiv T^A\pm T^B$, $\{Q^\pm,U^\pm,L^\pm,\}$
are similarly defined,
$Z^{A,B}_{\pm}=X^{A,B}_1\cos(Y^{A,B}_1)\pm X^{A,B}_2\cos(Y^{A,B}_2)$
encodes the influence of the crosspol effects, and
$\epsilon^\pm\equiv((\epsilon_{13}+\epsilon_{14})
\pm(\epsilon_{23}+\epsilon_{24}))/4$.
The dominant $\Delta T_P$ component is $Q^{+}$, not $Q^{-}$,
because $Q^A\to -Q^B,Q^B\to -Q^A$ when the spacecraft rotates 180\arcdeg.
In the limit of no loss imbalance or calibration error, and similar
cross polarization for all components, $\Delta T_P=Q^+ + 2X\cos(Y)U^+$.

\section{Estimation of the Polarization Power Spectra}
\label{app:master}

The WMAP polarization power spectra at $l < 32$ incorporate an
extension of the MASTER quadratic estimator \citep{hivon/etal:2002},
which is used to account for mode coupling.  The original method
assumes that observations of every point on the sky give statistically
independent noise.  However, WMAP has a significant component of the noise
that is correlated between pointings due to its scan pattern and the $1/f$
noise, and thus the method needs to be modified as
described here to accommodate a full covariance matrix.  The most
conspicuous mathematical feature of the original method is Wigner
3-$j$ symbols, whereas in the extended method, these objects are not
used.  For more details of the original method, as well as the
application to polarization, see Appendix A of
\citet{kogut/etal:2003}, together with the references therein.

\subsection{Extended MASTER Algorithm for Temperature Power Spectrum}
\label{sec:masterT}

The original method is derived by modeling the sky brightness as a
continuous function of pointing.  For example, the observed cut-sky
spherical harmonic coefficients for Stokes $I$, denoted as $\tilde
T_{l m}$, are defined as follows:
\begin{equation}
  \tilde T_{l m} = \int \mathrm{d}\hat n ~w(\hat n) ~ 
T(\hat n) Y^*_{l m}(\hat n).
  \label{eq-3j}
\end{equation}
Here, $\hat{n}$ is the unit vector of the pointing, $w(\hat{n})$ is
the weighting function, $T(\hat n)$ is the sky brightness, and
$Y_{l m}(\hat{n})$ is a spherical harmonic basis function.
Expanding $T(\hat n)$ and $w(\hat n)$ in spherical harmonics gives a
series.  Each term of the series includes an integral of a product of
three spherical harmonic basis functions:
\begin{displaymath}
  \tilde T_{l m} = \sum_{l'm'}\sum_{l''m''} w_{l'' m''} T_{l' m'}
  \int \mathrm{d}\hat n ~ Y_{l'm'}(\hat n) 
   Y_{l'' m''}(\hat n)  Y^*_{l m}(\hat n).
\end{displaymath}
These distinctive integrals are what give rise to the 3-$j$ symbols.
The orthogonality relations of 3-$j$ symbols eliminate many terms in
the expression for the observed power spectrum.

When there is noise covariance, the weight is a function of two
pointings rather than just one, and the 3-$j$ symbols are not used.
This case is most easily treated by modeling the sky as a set of
discrete pixels.
The goal of the derivation is to form a mode-coupling matrix
$M^{XY,X'Y'}_{ll'}$, where $XY$ and
$X'Y'$ are each chosen from the nine correlations $TT$, $TE$,
$TB$, $ET$, $EE$, $EB$, $BT$, $BE$, and $BB$.  In order to introduce
the formalism, we first discuss the $TT$ correlation, which is the simplest.
Because there
is no coupling between $TT$ and the other eight correlations, only
$M^{TT,TT}_{ll'}$ needs to be considered.
We note here that we do not actually use this formalism for
$TT$ but only for the others, as the temperature power spectrum
at low-$l$ is dominated by the signal and the noise correlation
is not important. We use $TT$ here to illustrate the main point
of the method. The extension to the polarization power spectra
that follows $TT$ (\S~\ref{sec:masterP}) is what we 
use for the actual analysis.

The weighting is computed initially as the inverse
of the covariance matrix of the pixels.  The sky cut is
expressed by setting the appropriate rows and columns to a very
large number in the noise covariance matrix before inverting it
[Eq.~(\ref{eq:maskNinv})].
We call the resulting weight matrix $W$.  Further, let $Y_{lm,p}$ be
a matrix containing (appropriately normalized) values of a spherical
harmonic basis function evaluated at each pixel, $p$.  
index $lm$.  
The number of rows of $Y_{lm,p}$ is
$n_p$, which is the number of pixels in each sky map.
The observed Stokes $I$ sky map is $T_p$.

In this notation, the
observed spherical harmonic coefficients are expressed as
\begin{displaymath}
  \tilde T_{lm} = \sum_{pp'} Y_{lm,p}^{*} W_{pp^\prime} T_{p^\prime}.
\end{displaymath}
If the matrix $W$ is diagonal, this expression is simply the discrete
version of Eq. \ref{eq-3j} above.  Expanding $T_{p^\prime}$ in spherical
harmonics gives
\begin{displaymath}
  \tilde T_{lm} = \sum_{l'm'} 
  \left[\sum_{pp'}Y_{lm,p}^{*} W_{pp^\prime}Y_{l'm',p'}\right] T_{l'm'}.
\end{displaymath}
This expression suggests the utility of defining
\begin{equation}
  Z_{lm,l'm'} \equiv \sum_{pp'} Y_{lm,p}^{*} W_{pp^\prime}
  Y_{l'm',p'} \label{eq-zdef}
\end{equation}
so that
\begin{eqnarray}
  \tilde T_{lm} &=& \sum_{l'm'} Z_{lm,l'm'}T_{l'm'}. \nonumber
\end{eqnarray}
The value of the observed power spectrum
at $l$ is expressed as follows:
\begin{eqnarray}
(2l+1)\tilde{C}_l 
  &=& \sum_m\tilde{T}_{l m}^{*}\tilde{T}_{l m} \nonumber \\
  &=& \sum_{m}\sum_{l''m''}\sum_{l'm'} (Z_{l m, l''m''}T_{l''m''})^{*}
      Z_{l m, l'm'}T_{l'm'}. 
\label{eq-zz}
\end{eqnarray}
In order to get the true, underlying CMB power spectrum into the
equation, the next step is to take the expectation of Eq. \ref{eq-zz}:
\begin{eqnarray}
(2l+1)\langle\tilde{C}_l \rangle &=& \sum_{l''m''}\sum_{l'm'} 
  \sum_m Z^{*}_{lm,l''m''} Z_{l m, l'm'}
  \langle T_{l''m''}^{*}T_{l'm'}\rangle\nonumber\\
  &=& \sum_{l''m''}\sum_{l'm'} \sum_m Z^{*}_{lm,l''m''}Z_{l m, l'm'}
  \langle C_{l'}\rangle\delta_{l'l''}\delta_{m'm''}\nonumber\\
  &=& \sum_{l'} \left(\sum_{mm'} Z^{*}_{lm,l'm'}Z_{l m, l'm'}\right)
  \langle C_{l'}\rangle.
\end{eqnarray}
Therefore, we obtain the unbiased estimator of the underlying 
power spectrum as
\begin{equation}
  C_l = \sum_{l'}\left(M^{-1}\right)_{ll'}\tilde{C}_{l'},
  \label{eq:ttrelation}
\end{equation}
where
\begin{equation}
  M_{l l'}\equiv \frac{1}{2l+1}\sum_{mm'} \left|Z_{l m,l' m'}\right|^2.
  \label{eq-zsum}
\end{equation}
In order to apply this method to cross-correlations between DAs, 
one of the $Z$ matrices in Eq. \ref{eq-zsum} is computed from the noise
matrix of the first DA, and the other from that of the second DA.

\subsection{Extended MASTER Algorithm for Polarization Power Spectra}
\label{sec:masterP}

The same formalism accommodates polarization.  In what follows,
uppercase $X$ or $Y$ indicates one of the three harmonic transforms
$T$, $E$, or $B$, and lowercase $a$ or $b$ denotes the Stokes
parameter label $I$, $Q$, or $U$.
The following substitutions are made in the above derivation:
\begin{eqnarray}
  W_{pp'} &\to& W_{(ap)(a'p')} \\
  Y_{lm,p} &\to& \Upsilon_{(Xlm)(ap)}
\label{eq:b8}
\end{eqnarray}
where the non-zero elements of $\Upsilon$ are
\begin{eqnarray}
\Upsilon_{(Tlm)(Ip)} &=& Y_{lm,p}
\\
\Upsilon_{(Elm)(Qp)} &=&
-\frac{1}{2}\left( {}_{+2}Y_{lm,p} + {}_{-2}Y_{lm,p} \right)
\\
\Upsilon_{(Blm)(Qp)} &=&
-\frac{i}{2}\left( {}_{+2}Y_{lm,p} - {}_{-2}Y_{lm,p} \right)
\\
\Upsilon_{(Elm)(Up)} &=& -\Upsilon_{(Blm)(Qp)}
\\
\Upsilon_{(Blm)(Up)} &=& \Upsilon_{(Elm)(Qp)}
\end{eqnarray}
$_{\pm 2}Y_{lm,p}$ are spin-2 spherical harmonics in the
same matrix form as $Y_{lm,p}$.  
 
For each pair of DAs, a $Z$ matrix is computed by analogy with
Eq. \ref{eq-zdef}.  The derivation follows the general steps
above.  The analog of Eq. \ref{eq:ttrelation} is 
\begin{equation}
  C_l^{XY} = \sum_{X'Y'l'}\left(M^{-1}\right)^{XY,X'Y'}_{ll'}\tilde{c}^{X'Y'}_{l'},
\end{equation}
where
\begin{displaymath}
  M_{l l'}^{XY,X'Y'}
  = \frac{1}{2l+1}\sum_{mm'} Z^{XX'*}_{l m,l' m'} Z^{YY'}_{l m,l' m'},
\end{displaymath}
where

\begin{eqnarray}
Z^{XX'}_{lm,l'm'} \equiv
\sum_{ap,a'p'} \Upsilon^{*}_{(Xlm)(ap)} W_{(ap)(a'p')} \Upsilon_{(X'l'm')(a'p')}
\end{eqnarray}

For each DA pair, the 81 coupling submatrices 
$M_{ll'}^{XY,X'Y'}$ are combined in a grand coupling
matrix that takes into account all the coupling among the nine
correlation types.

\subsection{Analytical Approximation}

The expressions for the coupling matrices greatly simplify
when $W_{pp'}$ is diagonal in pixel space, $W_{pp'}=\delta_{pp'}N_{obs,p}$.
This limit is a good approximation to the WMAP data at high $l$,
where noise is approximately uncorrelated (diagonal in pixel space).
In this limit, one can evaluate the coupling matrices analytically.

It is convenient to write the $N_{obs}$ matrix as
\begin{equation}
  \left(\begin{array}{cc}
	N^{QQ}_{obs,p} & N^{QU}_{obs,p} \\
	N^{UQ}_{obs,p} & N^{UU}_{obs,p}
  \end{array}\right)
  = \left(\begin{array}{cc}
	N^{+}_{obs,p}+N^{-}_{obs,p} & N^{QU}_{obs,p} \\
	N^{UQ}_{obs,p} & N^{+}_{obs,p}-N^{-}_{obs,p}
  \end{array}\right),
\end{equation}
where
\begin{eqnarray}
  N^{+}_{obs,p} &\equiv& \frac{N^{QQ}_{obs,p}+N^{UU}_{obs,p}}2,\\
  N^{-}_{obs,p} &\equiv& \frac{N^{QQ}_{obs,p}-N^{UU}_{obs,p}}2.
\end{eqnarray}
One can show that under a rotation of basis by an angle $\theta$, 
these quantities transform as
\begin{eqnarray}
  N^{+}_{obs,p} &\rightarrow& N^{+}_{obs,p},\\
  N^{-}_{obs,p} \pm iN^{QU}_{obs,p} &\rightarrow& 
  e^{\mp 4i\theta}(N^{-}_{obs,p} \pm iN^{QU}_{obs,p}).
\end{eqnarray}
Therefore, we expand them into spin harmonics as follows:
\begin{eqnarray}
  N^{+}_{obs,p} &=& \sum_{lm}n^+_{lm}Y_{lm,p},\\
  N^{-}_{obs,p} \pm iN^{QU}_{obs,p} &=&
  \sum_{lm}{}_{\mp 4}n_{lm}{}_{\mp 4}Y_{lm,p}.
\end{eqnarray}
We obtain
\begin{eqnarray}
\nonumber
  Z_{lm,l'm'}^{EE}
  &=& \frac12\sum_{LM}I^{LM}_{lm,l'm'}\left\{
    n^+_{LM}\left[1+(-)^{L+l+l'}\right]
  \wignerthreej{L}{l}{l'}{0}{2}{-2}\right.\\
  & &	\left.+	\left[
	  \xslm{n}{+4}{LM}
	+ (-)^{L+l+l'}\xslm{n}{-4}{LM}
	\right]\wignerthreej{L}{l}{l'}{-4}{2}{2}\right\},\\
\nonumber
  Z_{lm,l'm'}^{BB}
  &=& \frac12\sum_{LM}I^{LM}_{lm,l'm'}\left\{
    n^+_{LM}\left[1+(-)^{L+l+l'}\right]
  \wignerthreej{L}{l}{l'}{0}{2}{-2}\right.\\
  & &	\left.-	\left[
	  \xslm{n}{+4}{LM}
	+ (-)^{L+l+l'}\xslm{n}{-4}{LM}
	\right]\wignerthreej{L}{l}{l'}{-4}{2}{2}\right\},\\
\nonumber
  Z_{lm,l'm'}^{EB}
  &=& \frac{i}2\sum_{LM}I^{LM}_{lm,l'm'}\left\{
    n^+_{LM}\left[1-(-)^{L+l+l'}\right]
  \wignerthreej{L}{l}{l'}{0}{2}{-2}\right.\\
  & &	\left.-	\left[
	  \xslm{n}{+4}{LM}
	- (-)^{L+l+l'}\xslm{n}{-4}{LM}
	\right]\wignerthreej{L}{l}{l'}{-4}{2}{2}\right\},
\end{eqnarray}
where
\begin{equation}
  I^{LM}_{lm,l'm'} \equiv
  (-)^m \sqrt{\frac{(2L+1)(2l+1)(2l'+1)}{4\pi}}
  \wignerthreej{L}{l}{l'}{M}{-m}{m'}.
\end{equation}
Using the identity
\begin{equation}
  \sum_{mm'}I^{LM}_{lm,l'm'} I^{L'M'}_{lm,l'm'} 
  =\frac{(2l+1)(2l'+1)}{4\pi}\delta_{LL'}\delta_{MM'},
\end{equation}
it is straightforward to evaluate all the relevant coupling matrices
analytically:
\begin{eqnarray}
\nonumber
  M_{l l'}^{EE,EE}
  &=& \frac1{2l+1}\sum_{mm'}|Z_{lm,l'm'}^{EE}|^2\\
  \nonumber
  &=& \frac{2l'+1}{16\pi}\sum_{LM}
  \left|n_{LM}^+\left[1+(-)^{L+l+l'}\right]
  \wignerthreej{L}{l}{l'}{0}{2}{-2}\right.\\
  & & + \left.
  \left[{}_4n_{LM}+(-)^{L+l+l'}{}_{-4}n_{LM}\right]
  \wignerthreej{L}{l}{l'}{-4}{2}{2}\right|^2,\\
\nonumber
  M_{l l'}^{BB,BB}
  &=& \frac1{2l+1}\sum_{mm'}|Z_{lm,l'm'}^{BB}|^2\\
  \nonumber
  &=& \frac{2l'+1}{16\pi}\sum_{LM}
  \left|n_{LM}^+\left[1+(-)^{L+l+l'}\right]
  \wignerthreej{L}{l}{l'}{0}{2}{-2}\right.\\
  & & -\left.
  \left[{}_4n_{LM}+(-)^{L+l+l'}{}_{-4}n_{LM}\right]
  \wignerthreej{L}{l}{l'}{-4}{2}{2}\right|^2,\\
\nonumber
  M_{l l'}^{EE,BB}
  &=& \frac1{2l+1}\sum_{mm'}|Z_{lm,l'm'}^{EB}|^2\\
  \nonumber
  &=& \frac{2l'+1}{16\pi}\sum_{LM}
  \left|n_{LM}^+\left[1-(-)^{L+l+l'}\right]
  \wignerthreej{L}{l}{l'}{0}{2}{-2}\right.\\
  & & -\left.
  \left[{}_4n_{LM}-(-)^{L+l+l'}{}_{-4}n_{LM}\right]
  \wignerthreej{L}{l}{l'}{-4}{2}{2}\right|^2,\\
\nonumber
  M_{l l'}^{EB,EB}
  &=& \frac1{2l+1}\sum_{mm'}Z_{lm,l'm'}^{EE*}Z_{lm,l'm'}^{BB}\\
  \nonumber
  &=& \frac{2l'+1}{16\pi}\sum_{LM}
  \left\{\left|n_{LM}^+\left[1+(-)^{L+l+l'}\right]
  \wignerthreej{L}{l}{l'}{0}{2}{-2}\right|^2\right.\\
  & & -\left.
  \left|\left[{}_4n_{LM}+(-)^{L+l+l'}{}_{-4}n_{LM}\right]
  \wignerthreej{L}{l}{l'}{-4}{2}{2}\right|^2\right\}.
\end{eqnarray}

\section{Polarization Fisher and Covariance Matrix}
\label{app:fisher}

In this Appendix, we derive expressions for the 
Fisher and covariance matrices of the temperature and polarization power spectra.
Our derivation extends the derivation of the TT matrices
given in \citet{hinshaw/etal:2003} to all combinations of polarization 
power spectra.

Note that we do {\it not} use these results for evaluating the likelihood
that is used in the cosmological analysis.
At low multipoles, $l\le 23$, we evaluate 
the likelihood of polarization
data directly from the maps using the exact method described in
Appendix~\ref{app:ektau}. 
Why do we not use the Fisher or covariance matrix for the cosmological
analysis, except for TT and TE spectra at $\ell>23$? 
The reason is because the form of the likelihood
function for the power spectra is not a Gaussian at low multipoles,
and therefore the Fisher or covariance matrix, which only characterize 
the second-order moment of the power spectrum, is not sufficient to 
fully specify the likelihood function. This was pointed out after the 
first year release by
\citet{efstathiou:2004a} and \citet{slosar/seljak/makarov:2004}
and is discussed in \citet{hinshaw/etal:prep}.
As we do not know the precise
form of the likelihood for the power spectra, we evaluate the likelihood
of the temperature and polarization maps directly, which is a Gaussian,
at low multipoles, $l\le 23$.
For high multipoles, $l>23$, the likelihood function may be approximated
as a Gaussian and therefore we use a Gaussian likelihood with 
the Fisher or covariance matrices.
While we do not use the EE or BB power spectra at $l>23$,
as they contain very little signal compared to noise,
we do use the covariance matrix of the TE power spectrum
at $l>23$ in the likelihood code, for which we adopt the analytical
{\it ansatz} given in Equation~\ref{eq:TEansatz}, which was also used
in the first-year analysis of the TE power spectrum \citep{kogut/etal:2003}.
For the evaluation of the TT likelihood, see \citet{hinshaw/etal:prep}.

\subsection{Fisher Matrix: Exact formula}

The Fisher matrix, $F_{ll'}$, is given by
\begin{equation}
 F_{ll'}^{XY,X'Y'}
  = \frac12 \sum_q \left[\sum_{q_i}(C^{-1})_{qq_1}
\frac{\partial C_{q_1q_2}}{\partial S_l^{XY}}(C^{-1})_{q_2q_3}
\frac{\partial C_{q_3q}}{\partial S_{l'}^{X'Y'}}\right],
\label{eq:fisher}
\end{equation}
where the covariance matrix $C_{qq'}$ consists of the covariance matrices of 
all the bilinear combinations of $T$, $Q$, and $U$:
\begin{equation}
 C_{qq'} = \left(\begin{array}{ccc}
	C^{TT}_{pp'}&C^{TQ}_{pp'}&C^{TU}_{pp'}\\
	C^{QT}_{pp'}&C^{QQ}_{pp'}&C^{QU}_{pp'}\\
	C^{UT}_{pp'}&C^{UQ}_{pp'}&C^{UU}_{pp'}
  \end{array}\right),
  \label{eq:fullcovariance}
\end{equation}
and the covariance includes the signal and noise,
$C_{qq'}=S_{qq'}+N_{qq'}$.
Here $S_l^{XY}$ is the angular (cross) power spectrum of the signal 
where $X$ and $Y$ denote $T$, $E$, or $B$.
The inverse covariance matrix in harmonic space 
is then given by the harmonic transform of $(C^{-1})_{qq'}$:
\begin{equation}
  (C^{-1})^{XY}_{lm,l'm'} = 
\sum_{ap,a'p'}\Upsilon_{(Xlm)(ap)}(C^{-1})_{ap,a'p'}\Upsilon_{(Y^*l'm')(a'p')},
\end{equation}
where $\Upsilon$ is given by the equations following~(\ref{eq:b8}).

Using these quantities, 
each term of the Fisher matrix (Eq.~[\ref{eq:fisher}]) evaluates to 
\begin{eqnarray}
 \label{eq:Fxx}
 F_{ll'}^{XX,XX}
  &=& \frac12 \sum_{mm'}\left[(C^{-1})^{XX}_{lm,l'm'}\right]^2,\\
 F_{ll'}^{XX,XY}
  &=& \sum_{mm'}\left[(C^{-1})^{XX}_{lm,l'm'}(C^{-1})^{XY}_{lm,l'm'}\right],\\
 F_{ll'}^{XX,YY}
  &=& \frac12 \sum_{mm'}\left[(C^{-1})^{XY}_{lm,l'm'}\right]^2,\\
 \label{eq:Fxy}
 F_{ll'}^{XY,XY}
  &=& \sum_{mm'}\left[(C^{-1})^{XY}_{lm,l'm'}\right]^2
+ \sum_{mm'}\left[(C^{-1})^{XX}_{lm,l'm'}(C^{-1})^{YY}_{lm,l'm'}\right],
\end{eqnarray}
where $X\neq Y$. 
In general cases where $S_{qq'}$ or $N_{qq'}$ (or both) are 
non-diagonal, one must calculate $(C^{-1})_{qq'}$  by directly inverting
the covariance matrix given by equation~(\ref{eq:fullcovariance}).
In reality, however, the matrix inversion requires $n_p^3$ operations
and thus it become computationally too expensive to evaluate for the full
WMAP resolution. On the other hand, 
if one considers only large scale anisotropies at low $l$, then the 
matrix inversion can be done in a reasonable computational time.
We use Eq.~(\ref{eq:Fxx})--(\ref{eq:Fxy}) for 
computing the Fisher matrices for $C_l^{TT}$, 
$C_l^{TE}$, $C_l^{TB}$, $C_l^{EE}$, $C_l^{EB}$, and $C_l^{BB}$,
at low multipoles, $l\le 32$.

\subsection{Fisher Matrix: Analytical Approximation}

The expressions for the Fisher matrices can be evaluated
analytically when $C_{qq'}$ is diagonal in pixel space.
This limit is a good approximation to the \WMAP\ data at high $l$,
where $C_{qq'}$ is dominated by noise and noise is approximately
uncorrelated (diagonal in pixel space). In this limit, one obtains
the following analytical formulae:
\begin{eqnarray}
\nonumber
  F_{l l'}^{EE,EE}
  &=& \frac1{2}\sum_{mm'}|(N^{-1})_{lm,l'm'}^{EE}|^2\\
  \nonumber
  &=& \frac{(2l+1)(2l'+1)}{32\pi}\sum_{LM}
  \left|n_{LM}^+\left[1+(-)^{L+l+l'}\right]
  \wignerthreej{L}{l}{l'}{0}{2}{-2}\right.\\
  & & +\left.
  \left[{}_4n_{LM}+(-)^{L+l+l'}{}_{-4}n_{LM}\right]
  \wignerthreej{L}{l}{l'}{-4}{2}{2}\right|^2,\\
\nonumber
  F_{l l'}^{BB,BB}
  &=& \frac1{2}\sum_{mm'}|(N^{-1})_{lm,l'm'}^{BB}|^2\\
  \nonumber
  &=& \frac{(2l+1)(2l'+1)}{32\pi}\sum_{LM}
  \left|n_{LM}^+\left[1+(-)^{L+l+l'}\right]
  \wignerthreej{L}{l}{l'}{0}{2}{-2}\right.\\
& & -\left.
  \left[{}_4n_{LM}+(-)^{L+l+l'}{}_{-4}n_{LM}\right]
  \wignerthreej{L}{l}{l'}{-4}{2}{2}\right|^2,\\
\nonumber
  F_{l l'}^{EE,BB}
  &=& \frac1{2}\sum_{mm'}|(N^{-1})_{lm,l'm'}^{EB}|^2\\
  \nonumber
  &=& \frac{(2l+1)(2l'+1)}{32\pi}\sum_{LM}
  \left|n_{LM}^+\left[1-(-)^{L+l+l'}\right]
  \wignerthreej{L}{l}{l'}{0}{2}{-2}\right.\\
  & & -\left.
  \left[{}_4n_{LM}-(-)^{L+l+l'}{}_{-4}n_{LM}\right]
  \wignerthreej{L}{l}{l'}{-4}{2}{2}\right|^2,\\
\nonumber
  M_{l l'}^{EB,EB}
  &=&\sum_{mm'}|(N^{-1})_{lm,l'm'}^{EB}|^2
  +\sum_{mm'}(N^{-1})_{lm,l'm'}^{EE*}(N^{-1})_{lm,l'm'}^{BB}\\
  \nonumber
  &=& \frac{(2l+1)(2l'+1)}{32\pi}\sum_{LM}
  \left\{
  \left|n_{LM}^+\left[1-(-)^{L+l+l'}\right]
  \wignerthreej{L}{l}{l'}{0}{2}{-2}\right.\right.\\
\nonumber
  & & -\left.
  \left[{}_4n_{LM}-(-)^{L+l+l'}{}_{-4}n_{LM}\right]
  \wignerthreej{L}{l}{l'}{-4}{2}{2}\right|^2\\
\nonumber
  &+& \left|n_{LM}^+\left[1+(-)^{L+l+l'}\right]
  \wignerthreej{L}{l}{l'}{0}{2}{-2}\right|^2\\
  & & -\left.
  \left|\left[{}_4n_{LM}+(-)^{L+l+l'}{}_{-4}n_{LM}\right]
  \wignerthreej{L}{l}{l'}{-4}{2}{2}\right|^2\right\}.
\end{eqnarray}

\subsection{Covariance Matrix: Ansatz}
\label{app:covar}

The inverse of the Fisher matrix gives the covariance matrix,
$\Sigma$. While we use the map-based exact likelihood
described in Appendix~\ref{app:ektau} for the cosmological analysis,
it is still useful to have an approximate method to evaluate
the likelihood of the data given theory and noise model 
from the power spectra.
For this purpose, we use the following {\it ansatz}:
    
\begin{equation}    
\Sigma_\ell^{TE\,TE}=\frac{\left(S_{\ell}^{TT} +     
n_{eff\;\ell}^{TT}\right)\left(S_{\ell}^{EE}+    
n_{eff\;\ell}^{EE}\right)+\left(S_{\ell}^{TE}\right)^2}    
{(2\ell+1)\left[f_{{\rm sky}\;eff}^{TE}(\ell)\right]^2}     
\label{eq:TEansatz}
\end{equation}    
    
\begin{equation}    
\Sigma_\ell^{TB\,TB}=\frac{\left(S_{\ell}^{TT} +     
n_{eff\;\ell}^{TT}\right)\left(S_{\ell}^{BB}+    
n_{eff\;\ell}^{BB}\right)}    
{(2\ell+1)\left[f_{{\rm sky}\;eff}^{TB}(\ell)\right]^2}     
\end{equation}    
    
\begin{equation}    
\Sigma_\ell^{EE\,EE}=\frac{2\left(S_{\ell}^{EE} +     
n_{eff\;\ell}^{EE}\right)^2}{(2\ell+1)\left[f_{{\rm sky}\;eff}^{EE}
(\ell)\right]^2}
\end{equation}    
    
\begin{equation}    
\Sigma_\ell^{BB\,BB}=\frac{2\left(S_{\ell}^{BB} +     
n_{eff\;\ell}^{BB}\right)^2}{(2\ell+1)\left[f_{{\rm sky}\;eff}^{BB}
(\ell)\right]^2}
\end{equation}    

\begin{equation}    
\Sigma_\ell^{EB\,EB}=\frac{\left(S_{\ell}^{EE} +     
n_{eff\;\ell}^{EE}\right)\left(S_{\ell}^{BB} +     
n_{eff\;\ell}^{BB}\right)}{(2\ell+1)\left[f_{{\rm sky}\;eff}^{EB}
(\ell)\right]^2}
\end{equation}    
In these expressions $n_{eff\;\ell}$ denotes the effective noise as a     
function of $\ell$ and $f_{{\rm sky}\; eff}$ denotes the effective fraction     
of the sky observed.  These are obtained from comparing the {\it ansatz}
to the inverse of the Fisher matrices derived in the previous sections.
We have found that $f_{{\rm sky}}^{XY}\simeq 
\sqrt{f_{\rm sky }^{XX}f_{\rm sky}^{YY}}$
to a very good approximation.
See also \citet{kogut/etal:2003} for the evaluation of
$\Sigma_\ell^{TE\,TE}$
and \citet{hinshaw/etal:prep} for the evaluation of
$\Sigma_\ell^{TT\,TT}$.

\section{Exact Likelihood Evaluation at Low Multipoles}
\label{app:ektau}

At low multipoles, $l\le 23$, we evaluate the likelihood
of the data for a given theoretical model exactly from the 
temperature and polarization maps.
The standard likelihood is given by
\begin{equation}
 L(\vec{m}|S)d\vec{m} = 
\frac{\exp\left[-\frac12\vec{m}^t(S+N)^{-1}\vec{m}\right]}{|S+N|^{1/2}}
\frac{d\vec{m}}{(2\pi)^{3n_p/2}},
\label{eq:like}
\end{equation}
where $\vec{m}$ is the data
vector containing the temperature map, $\vec{T}$, as well as the 
polarization maps, $\vec{Q}$, and $\vec{U}$, 
$n_p$ is the number of pixels of each map, and $S$ and $N$ are
the signal and noise covariance matrix ($3n_p\times 3n_p$), 
respectively.
As the temperature data are completely dominated
by the signal at such low multipoles, noise in temperature 
may be ignored. This simplifies the form of likelihood as
\begin{equation}
 L(\vec{m}|S)d\vec{m} =
\frac{\exp\left[-\frac12\vec{\tilde{m}}^t(\tilde{S}_P+N_P)^{-1}
\vec{\tilde{m}}\right]}{|\tilde{S}_P+N_P|^{1/2}}
\frac{d\vec{\tilde{m}}}{(2\pi)^{n_p}}~ 
\frac{\exp\left(-\frac12\vec{T}^tS_{T}^{-1}\vec{T}\right)}{|S_{T}|^{1/2}}
\frac{d\vec{T}}{(2\pi)^{n_p/2}},
\label{eq:dnslike}
\end{equation}
where $S_T$ is the temperature signal matrix ($n_p\times n_p$),
the new polarization data vector, 
$\vec{\tilde{m}}=(\tilde{Q}_p,~\tilde{U}_p)$, is given by
\begin{eqnarray}
 \tilde{Q}_p &\equiv& Q_p - \frac12\sum_{l=2}^{23}
  \frac{S_l^{TE}}{S_l^{TT}}\sum_{m=-l}^lT_{lm}({}_{+2}Y_{lm,p}+{}_{-2}Y^*_{lm,p}),
\\
 \tilde{U}_p &\equiv& U_p - \frac{i}2\sum_{l=2}^{23}
  \frac{S_l^{TE}}{S_l^{TT}}\sum_{m=-l}^lT_{lm}({}_{+2}Y_{lm,p}-{}_{-2}Y^*_{lm,p}),
\end{eqnarray}
and $\tilde{S}_P$ is the signal matrix for the new polarization vector
with the size of $2n_p\times 2n_p$. 
As $T_{lm}$ is totally signal dominated,
the noise matrix for $(\vec{\tilde{Q}},~\vec{\tilde{U}})$
 equals that for $(\vec{Q},~\vec{U})$, $n_p$.
To estimate $T_{lm}$, we used the full-sky internal linear 
combination (ILC) temperature map \citep{hinshaw/etal:prep}. 

One can show that equation~(\ref{eq:like}) and (\ref{eq:dnslike})
are mathematically equivalent when the temperature noise is ignored.
The new form, equation~(\ref{eq:dnslike}), allows us to factorize
the likelihood of temperature and polarization, with
the information in their cross-correlation, $S_l^{TE}$, fully retained.
We further rewrite the polarization part of the likelihood as
\begin{equation}
 L(\vec{\tilde{m}}|\tilde{S}) 
= \frac{\exp\left[-\frac12(N_P^{-1}\vec{\tilde{m}})^t
(N_P^{-1}\tilde{S}_PN_P^{-1}+N_P^{-1})^{-1}
(N_P^{-1}\vec{\tilde{m}})\right]}{|N_P^{-1}
\tilde{S}_PN_P^{-1}+N_P^{-1}|^{1/2}}
\frac{|N_P^{-1}|d\vec{\tilde{m}}}{(2\pi)^{n_p}}.
\label{eq:eklike}
\end{equation}
This form is operationally more useful, as it contains only $N_P^{-1}$.
\citet{hinshaw/etal:prep} describes the method to evaluate the temperature
part of the likelihood.

The effect of {\it P06} mask is included in $N_P^{-1}$. 
Suppose that the structure of $N_P^{-1}$ is given by
\begin{equation}
  N_P^{-1}=\left(\begin{array}{cc}A&B\\B&D\end{array}\right),
\end{equation}
where $A$ is the noise matrix for unmasked pixels, $D$ is for
masked pixels, and $B$ is for their correlations.
We assign infinite noise to the masked pixels such that 
$N_P\rightarrow N_P+\lambda(I-M)$, where $M$ is the diagonal 
matrix whose elements
are zero for masked pixels and unity otherwise.
In the limit of $\lambda\rightarrow \infty$, the inverse of $N_P$ is
given by 
\begin{equation}
 N_P^{-1} \rightarrow \left(\begin{array}
{cc}A-B^tDB&0\\0&0\end{array}\right).
\label{eq:maskNinv}
\end{equation}
We have checked that this form of $N_P^{-1}$ yields the unbiased
estimates of the signal matrix from simulated realizations of the 
{\sl WMAP} data. When the masked pixels were simply 
ignored (i.e., $B^tDB=0$), 
on the other hand, the estimated signal matrix was found
to be biased high.
As the likelihood form is sensitive to the precise form
of $N_P^{-1}$, it is important to treat the mask in this way 
so that the estimated signal matrix from the data is unbiased.

We mask the polarization maps as follows. We first mask the maps
at the full resolution, $n_{side}=512$, and 
then degrade the masked maps using the weight that is diagonal in
pixel space, $N^{-1}_{P,pp}$, to a lower resolution, $n_{side}=16$.
(Note that while the weight is diagonal in pixel space, it contains
noise covariance between $Q_p$ and $U_p$. The spurious polarization
term, $S$, is ignored in this process.)
The degraded mask is redefined such that it takes on 1 when
the lower resolution pixel contains more than half of the original
full resolution pixels, and 0  otherwise. 
We degrade these maps further to the resolution of 
$n_{side}=8$ using the full noise matrix, and also degrade the mask
and the noise matrix. (The noise matrix has been masked using 
Eq~[\ref{eq:maskNinv}].) We use the resulting maps and noise matrix
in the likelihood function given in equation~(\ref{eq:eklike}).

\section{An estimate of $\Omega_{GW}$} 
\label{app:GW}

Tensor perturbations generated by inflation are stochastic in nature,
so the gravity wave perturbation can be expanded in plane waves
\begin{equation}
h_{ij} (\eta,\mathrm{\bf x}) = \int \frac{d^3 k}{(2\pi)^3} \left[h_+
(\eta,\mathrm{\bf k})\epsilon_{ij}^{+}e^{-i\mathrm{\bf k.x}} +
h_\times (\eta,\mathrm{\bf k})\epsilon_{ij}^{\times}e^{-i\mathrm{\bf
k.x}}\right], 
\end{equation}
where $\epsilon_{ij}^{a}$ is the polarization tensor, and $a=+,\times$
are the two polarizations in the transverse traceless (tt) gauge (in which
$h_{ij,j}=h^i_i=0$; we also set $h_{00}=h_{0j}=0$). 
The stress-energy tensor for gravity waves is
defined as 
\begin{equation}
T_{\mu\nu} = \frac{1}{32\pi G} \langle h_{\alpha \beta,\mu}h^{\alpha
\beta}{}_{,\nu}\rangle,
\end{equation}
and in the tt gauge, we have
\begin{equation}
T_{00} = \frac{1}{32\pi G} \langle \dot{h}_{ij}\dot{h}^{ij}\rangle.
\end{equation}
Thus, 
\begin{eqnarray}
\langle \dot{h}_{ij}\dot{h}^{ij} \rangle &=& \int \frac{d^3
k}{(2\pi)^3} \int \frac{d^3 k'}{(2\pi)^3} e^{i\mathrm{({\bf k-k'})\bf{.x}}} 
\nonumber \\
 & & \left[ \langle \dot{h}_+
(\eta,\mathrm{\bf k}) \dot{h}_+ (\eta,\mathrm{\bf k'})\rangle
\epsilon_{ij}^{+}\epsilon^{+ij} + 
 \langle \dot{h}_\times (\eta,\mathrm{\bf k}) \dot{h}_\times (\eta,\mathrm{\bf k'})\rangle \epsilon_{ij}^{\times}\epsilon^{\times ij} \right]    
\end{eqnarray}
The variance of the perturbations in the $h$ fields can be written as
\begin{equation}
\langle \dot{h}_a (\eta,\mathrm{\bf k}) \dot{h}_a (\eta,\mathrm{\bf
k'})\rangle = \langle|\dot{h}_a (\eta,\mathrm{\bf k})|^2\rangle (2\pi)^3
\delta^3({\bf k-k'}),
\end{equation}
and since $\epsilon_{ij}^{a}\epsilon^{aij} = 2$, we obtain
\begin{equation}
\langle \dot{h}_{ij}\dot{h}^{ij} \rangle = \int \frac{d^3
k}{(2\pi)^3} 2 \left[\langle|\dot{h}_+ (\eta,\mathrm{\bf k})|^2\rangle +
\langle|\dot{h}_\times (\eta,\mathrm{\bf k})|^2\rangle \right].  
\end{equation}
Writing
\begin{equation}
h_a (\eta,\mathrm{\bf k}) = h_a (0,\mathrm{\bf k}) T(\eta,k),
\end{equation}
where $T$ is the transfer function, we have
\begin{eqnarray}
\langle \dot{h}_{ij}\dot{h}^{ij} \rangle &=& \int \frac{4\pi k^2
dk}{(2\pi)^3} 2 \left[\langle|h_+ (0,\mathrm{\bf k})|^2\rangle +
\langle|h_\times (0,\mathrm{\bf k})|^2\rangle \right]
\dot{T}^2(\eta, k) \nonumber \\
 & = & \int \frac{dk}{k} \frac{2k^3}{2\pi^2} \left[\langle|h_+
(0,\mathrm{\bf k})|^2\rangle +\langle |h_\times (0,\mathrm{\bf k})|^2\rangle \right]
\dot{T}^2(\eta, k).
\end{eqnarray}
From the definition of the primordial tensor power spectrum,
\begin{equation}
\Delta_h^2(k) = \frac{2k^3}{2\pi^2} \left[\langle|h_+
(0,\mathrm{\bf k})|^2\rangle + \langle|h_\times (0,\mathrm{\bf
k})|^2\rangle
 \right],
\end{equation}
we obtain
\begin{equation}
\langle \dot{h}_{ij}\dot{h}^{ij} \rangle = \int d\ln k
\Delta_h^2(k) \dot{T}^2(\eta,k).
\end{equation}
Now
\begin{equation}
T_{00} = \rho_{GW} \equiv \int d \ln k \frac{d \rho_{GW}}{d \ln k},
\end{equation}
thus we have
\begin{equation}
\frac{d\rho_{GW}}{d \ln k} = \frac{\Delta_h^2(k) \dot{T}^2(\eta,k)}{32\pi G}. 
\end{equation}

Remembering that $\Omega = \rho \times (8\pi G/3 H_0^2)$, we obtain
\begin{equation}
\frac{d\Omega_{GW}}{d \ln k} = \frac{\Delta_h^2(k)
\dot{T}^2(\eta,k)}{12 H_0^2}. 
\end{equation}
Therefore, 
\begin{equation}
\Omega_{GW} = \int d \ln k \frac{\Delta_h^2(k)
\dot{T}^2(\eta,k)}{12 H_0^2}. 
\end{equation}

The transfer function $T$ and its time derivative $\dot{T}$ can be
calculated easily by numerically integrating the evolution equation
for the polarization states, which, neglecting the neutrino anisotropic
stress, is given by
\begin{equation}
\label{eq:KGeqn4h}
h''_a + 2 \left(\frac{a'}{a}\right)h'_a + k^2 h_a = 0,
\end{equation}
where prime denotes derivatives with respect to conformal time $\eta$, 
related to the time derivative by $d\eta = dt/a(\eta)$. This expression 
may be numerically integrated. In the following, however,
we derive an analytic estimate relating a given limit
on the tensor-to-scalar ratio, $r$, and the measured amplitude of the
primordial scalar power spectrum, $A$, to a limit on the current
energy density in primordial gravitational radiation.

There are several approaches taken in the literature to derive
analytic expressions for the tensor transfer function, though these results
are obtained in almost all cases for a universe containing only matter
and radiation. These include using (1) an instantaneous transition
from radiation to matter domination
\citep[e.g., ]{abbott/harari:1986,ng/speliotopoulos:1995, grishchuk:2001,
pritchard/kamionkowski:2005} (2) a ``transfer function'' to account
for the smooth transition from radiation domination to matter
domination \citep[e.g., ]{turner/white/lidsey:1993, wang:1996,
turner:1997}, and (3) WKB methods
\citep[e.g., ]{ng/speliotopoulos:1995,pritchard/kamionkowski:2005}. In
the following derivation, we will apply the sudden transition
approximation to a $\Lambda$CDM universe
\citep[see also]{zhang/etal:2005}, which is a good approximation for
gravitational waves with wavelengths much longer than the time taken
for the transition to happen.

In a universe which undergoes a set of piecewise instantaneous
transitions in the scale-factor, given by $a(\eta) \propto
\eta^{-\nu}$, the solution to eq.~\ref{eq:KGeqn4h} is given by
\begin{equation}
\label{eq:hpiecewise}
h\left(\eta,k\right) = \left(k \eta\right)^{\nu+1} \left[C\ j_\nu(k\eta) 
+ D\ y_\nu(k\eta) \right],
\end{equation}
where $j_\nu$ and $y_\nu$ are spherical Bessel functions of order
$\nu$ of the first and second kinds, respectively. Here, $\nu=-1$ for
radiation-domination (RD, $\eta < \eta_{eq1}$), $\nu=-2$ for
matter-domination (MD, $\eta_{eq1} < \eta < \eta_{eq2}$), and $\nu=+1$
for $\Lambda$-domination (LD, $\eta >
\eta_{eq2}$). $\eta_{eq1}$ is the conformal time at radiation-matter
equality, with a scale-factor corresponding to $a_{eq1} =
(\Omega_r/\Omega_m)$, and $\eta_{eq2}$ is the conformal time at
matter-$\Lambda$ equality, with a scale-factor corresponding to
$a_{eq2} = (\Omega_m/\Omega_\Lambda)^{1/3}$. For a concordance
cosmology with $\left\{\Omega_r, \Omega_m, \Omega_\Lambda, h\right\} =
\left\{4.18\times10^{-5}/h^2,\ 0.3,\ 0.7,\ 0.72\right\}$, $\eta_{eq1} = 103$
Mpc$^{-1}$ and $\eta_{eq2} = 12270$ Mpc$^{-1}$ (115 and 12030
Mpc$^{-1}$ respectively in the instantaneous approximation).

To obtain the coefficients $C$ and $D$, we require $h$ and $h'$
to be continuous at each of the transitions, $\eta_{eq1}$ and
$\eta_{eq2}$. Thus, denoting $x\equiv k\eta_0$ and making use of
special properties of spherical Bessel functions, we obtain the
transfer function and its derivative at present:
\begin{eqnarray}
T(x) & = & x^2 \left[C\ j_1(x) + D\ y_1(x) \right], \\ \label{eq:tffinal}
\dot{T}(k, x) &=& k x^2 \left[C\ j_0(x) 
+ D\ y_0(x) \right]. \label{eq:tdotfinal}
\end{eqnarray}
The coefficients are given by
\begin{equation}
C = \frac{1}{2 x_2^6} \left[2 A x_2^3 + 3 B (1+x_2^2) 
+ 3\cos (2 x_2)\left(B+2 A x_2-B x_2^2\right) 
+ 3\sin (2 x_2)\left(2 B x_2 + A(x_2^2-1)\right)\right]
\end{equation}
\begin{equation}
D = \frac{1}{2 x_2^6} \left[2 B x_2^3 - 3 A (1+x_2^2) 
+ 3\cos (2 x_2)\left(A-2B x_2 - A x_2^2\right) 
+ 3\sin (2x_2)\left(B+2 A x_2 - B x_2^2\right) \right]
\end{equation}
\begin{eqnarray}
A &=& \frac{3 x_1 - x_1\cos (2 x_1) + 2 \sin (2 x_1)}{2 x_1}\\
B &=& \frac{2 - 2 x_1^2 - 2\cos (2 x_1) - x_1 \sin (2x_1)}{2 x_1},
\end{eqnarray}
where $x_1\equiv k\eta_{eq1}$ and $x_2 \equiv k\eta_{eq2}$.

Further, we have the following definitions:
\begin{eqnarray}
 \label{eq:P_h} \Delta^2_h(k) &=& \Delta^2_h(k_0)
 \left(\frac{k}{k_0}\right)^{n_t(k_0)} \\
 \label{eq:rdef} r &\equiv& \frac{\Delta^2_h(k_0)}{\Delta^2_{\cal
 R}(k_0)},   
\end{eqnarray}
where 
\begin{equation}
 \label{eq:Adef} \Delta^2_{\cal R}(k_0) \simeq 2.95\times10^{-9}  A(k_0).
\end{equation}
To eliminate $n_t$, we use the inflationary single-field consistency relation,
$n_t=-r/8$. 

Combining these equations, and evaluating them at the present conformal 
time $\eta_0$ (with $a=1$) for modes within our current horizon, we are 
left with 
\begin{equation}
\Omega_{GW} = \frac{1}{12 H_0^2}\int_{2\pi/\eta_0}^\infty 
\frac{dk}{k}\ r \Delta^2_{\cal
R}(k_0) \left(\frac{k}{k_0}\right)^{-r/8} k^2 (k\eta_0)^4
\left[C(k, \eta_{eq1}, \eta_{eq2}) j_0(k\eta_0) 
+ D(k, \eta_{eq1}, \eta_{eq2}) y_0(k\eta_0)\right]^2,
\end{equation}
where $k$ and $\eta_0$ are to be evaluated in units of 1/Mpc and
$k_0=0.002$ Mpc$^{-1}$. We can now change to the dimensionless
variable $x\equiv k\eta_0$ and obtain
\begin{equation}
\Omega_{GW} \simeq \frac{2.95\times10^{-9}\ r
A(k_0)\ x_0^{r/8}}{12 H_0^2\eta_0^2} \int_{2\pi}^\infty dx\ x^{5-r/8}
\left[C j_0(x) + D y_0(x)\right]^2,
\end{equation}
where $x_0=k_0\eta_0$. We also have the result
\begin{eqnarray}
\frac{d\Omega_{GW}}{d\ln k}\left(k,\eta_0\right) 
&=& 2.21 \times 10^{-3} (rA) \left( \frac{k}{k_0} \right)^{-r/8} 
\left[\dot{T}\left(k,\eta_0\right)\right]^2 \label{eq:domdlnk} \\
&\simeq&  2.21 \times 10^{-3} (rA) \left( \frac{k}{k_0} 
\right)^{-r/8} \left\{k x^2 \left[C\ j_0(x) + D\ y_0(x)\right]\right\}^2.
\end{eqnarray}
Now
\begin{equation}
H_0\eta_0 = \int_0^1 \frac{1}{\sqrt{\Omega_r + \Omega_m a 
+\Omega_\Lambda a^4}},
\end{equation}
and $H_0\eta_0=3.25$ for the concordance $\Lambda$CDM model. 
Taking the concordance model and $k_0=0.002$ Mpc$^{-1}$, for given 
upper limits on
$r$ and $A$, the upper limit on $\Omega_{GW}$ is given by \citep{peiris:phd},
\begin{equation}
\Omega_{GW} \leq 2.33 \times 10^{-11} \left(r\ A\right) 
(27.05)^{r/8} \left[ 0.1278 - 0.0835\ (\log r) - 0.0671\ 
(\log r)^2 -0.0248\ (\log r)^3 \right],
\label{eq:ogw}
\end{equation}
where the logarithm is taken in base ten.

\begin{deluxetable}{ccccccc}
\tabletypesize{\footnotesize}
\tablecaption{Polarization of \TauA\label{tab:crab}}
\tablehead{
\colhead{Measurement} 
& \colhead{$\nu$ [GHz]}
& \colhead{$I$ [Jy]} 
& \colhead{$Q$ [Jy]} 
& \colhead{$U$ [Jy]} 
& \colhead{$P/I$ [\%]} 
& \colhead{$\gamma_{PA}$ [deg]}}
\startdata
\WMAP &22.5 (K) & $352\pm11$ & $-24.7\pm0.8$  & $1.3\pm0.9$ & $7.0\pm 0.3$ 
      & $-88\fdg$ ($150^\circ$) \\
\WMAP &32.8 (Ka) & $322\pm6$ & $-22.2\pm2.0$ & $1.9\pm1.1$ & $6.9\pm 0.3$ 
      & $-87\fdg$ ($151^\circ$) \\
\WMAP &40.4 (Q) & $299\pm6$ & $-19.6\pm2.6$  & $0.5\pm2.4$ & $6.6\pm 0.9$ 
      & $-89\fdg$ ($149^\circ$)\\
\WMAP &60.2 (V) & $265\pm7$ & $-18.5\pm2.7$   & $-1.9\pm6.2$ & $7.0\pm 1.1$ 
      & $-93\fdg$ ($145^\circ$)\\
\WMAP &92.9 (W) & $229\pm11$ & $-17.5\pm4.4$   & $-1.3\pm7.2$ & $7.6\pm 2.0$ 
      & $-92\fdg$ ($146^\circ$)\\
\citet{mayer/hollinger:1968} & 19& & & & 6.6 [15.5]&($140^{\circ}\pm10$) \\
\citet{wright/forster:1980} & 23  & & & & 9 & ($152^{\circ}$) \\
\citet{johnston/hobbs:1969} & 31 & &&  & 8.1 [17] & ($158^{\circ}$) \\
\citet{flett/henderson:1979} & 33 &  && &  [16] & ([$154\fdg8\pm2$]) \\
\citet{matveenko/conklin:1973} & 86  & && & ([$23\pm3$]) &  \\
\citet{montgomery/etal:1971} & 88 &  & && 13 & ($152^{\circ}$)  \\
\citet{hobbs/maran/brown:1978} &99& &&&[$11.9\pm0.9$] & ([$123^{\circ}$])\\
\citet{flett/murray:1991} &273 & && & [$27\pm1$] & ([$146^{\circ}\pm2$])\\ 
\citet{greaves/etal:2003} & 363 & && & $25\pm5$ & ($150^{\circ}\pm6$) \\
\enddata
\tablecomments{
The fluxes are integrated over pixels within a radius that includes 
99\% of the beam solid angle, $r99=[2\fdg525, 1\fdg645, 1\fdg517, 
1\fdg141, 0\fdg946]$ degrees in K through W bands. 
The errors are $1\sigma$ estimates calculated as a 
quadrature sum of statistical error, error due to background uncertainty, 
confusion error, 0.5\% calibration error, and an additional 1\% error since 
the aperture radius does not include all of the beam solid angle.  
Confusion error was calculated as the maximum difference in derived flux 
when the aperture radius and annulus radius are both decreased by 20\% 
or increased by 20\%.  Confusion error is usually the largest contribution 
to the total error.  The frequencies are band center
frequencies for Tau A's antenna temperature spectral index, $\beta=-2.3$.  
The two numbers for $\gamma_{PA}$ correspond to Galactic and 
equatorial (in parentheses) coordinates. 
Non-\WMAP\ measurements are generally done with 
arcminute resolution and therefore have different average and peak 
(in square brackets) fractional polarization. Their polarization directions
are all in equatorial coordinates.}
\end{deluxetable}
\clearpage

\begin{deluxetable}{cccc}
\tablecaption{Temperatures in the Galactic Center Region\label{tab:gcent}}
\tablehead{
\colhead{Band} & 
\colhead{$I$ [mK]} &
\colhead{$Q$ [mK]} & 
\colhead{$U$ [mK]}}
\startdata
K  &  33  & 0.69 & -0.25  \\
Ka &  14  & 0.21   &   -0.086  \\ 
Q  &  8.7  & 0.10   &  -0.041  \\ 
V  &  4.0  & 0.037   & $-0.01<U<0.01$  \\
W  &  3.6  &  0.043  & $-0.01<U<0.01$ \\
\enddata
\tablecomments{The table gives the average values for the temperature
and $Q$ and $U$ Stokes parameters in a 
$\delta b=2^\circ$~by~$\delta l  = 10 ^\circ$ 
region centered on $(l,b)=(0,0)$. The values are
in thermodynamic units relative to the CMB. To convert to antenna temperature,
divide by 1.014, 1.029, 1.044, 1.100, 1.251 in K through W bands 
respectively.} 
\end{deluxetable}
\clearpage

\begin{deluxetable}{ccccc}
\tablecaption{Fit Coefficients to Foreground Templates\label{tab:fgfit}}
\tablehead{
\colhead{Band} & 
\colhead{$\alpha_{s,\nu}$} & 
\colhead{$\beta_s(\nu_K,\nu)$} & 
\colhead{$\alpha_{d,\nu}$} &
\colhead{$\beta_d(\nu,\nu_W)$}}
\startdata
Ka &  0.3103  & $-3.22$  & 0.0148  & 1.54 \\ 
Q  &  0.1691  & $-3.12$  & 0.0154  & 1.89 \\ 
V  &  0.0610  & $-2.94$  & 0.0343  & 1.92 \\
W  &  0.0358  & $-2.51$  & 0.0891  & \nodata \\
\noalign{\smallskip\hrule\smallskip}
Ka &  0.2973  & $-3.33$  & 0.0148  & 1.54 \\ 
Q  &  0.1492  & $-3.33$  & 0.0154  & 1.89 \\ 
V  &  0.0414  & $-3.33$  & 0.0343  & 1.92 \\
W  &  0.0112  & $-3.33$  & 0.0891  & \nodata \\
\enddata
\tablecomments{The top of the Table gives the coefficients for a direct 
fit to the 
polarization maps. The $\alpha$ are dimensionless and produce model maps in
thermodynamic units.  The spectral indices $\beta$ refer to antenna 
temperature. The bottom half of the Table gives the same numbers for 
when the synchrotron fit is constrained to follow a power law. 
The fits were evaluated outside the processing mask.}
\end{deluxetable}
\clearpage

\begin{deluxetable}{ccccc}
\tablecaption{Comparison of $\chi^2$ Between Pre-cleaned and 
cleaned Maps\label{tab:fgchi}}
\tablehead{
\colhead{Band} & 
\colhead{$\chi^2/\nu$ Pre-cleaned} & 
\colhead{$\chi^2/\nu$ Cleaned} & 
\colhead{$\nu$} & 
\colhead{$\Delta\chi^2$}}
\startdata
Ka &  10.65  & 1.20  & 6144  & 58061  \\ 
Q  &  3.91  & 1.09  & 6144  & 17326  \\ 
V  &  1.36  & 1.19  & 6144  &  1045 \\
W  &  1.38  & 1.58  & 6144  & -1229 \\
\noalign{\smallskip\hrule\smallskip}
Ka &  2.142  & 1.096  & 4534  & 4743 \\ 
Q  &  1.289  & 1.018  & 4534  & 1229 \\ 
V  &  1.048  & 1.016  & 4534  & 145 \\
W  &  1.061  & 1.050  & 4534  & 50 \\
\enddata
\tablecomments{The top half of the table compares $\chi^2/\nu$ for the full-sky
pre-cleaned map to $\chi^2/\nu$ for full-sky cleaned map.  
The bottom half makes a similar comparison for the region outside the 
P06 mask.}
\end{deluxetable}
\clearpage

\begin{deluxetable}{ccccccccc}
\tabletypesize{\small}
\tablecaption{{\WMAP} EE$_{\ell=2}$ and BB$_{\ell=5}$ Values for 
$\ell(\ell+1)C_\ell/2\pi$\label{tab:eebb}}
\tablehead{
\colhead{Cross} & 
\colhead{Bin} & 
\colhead{$f_{eff}$} & 
\colhead{EE$_{\ell=2}$}  &  
\colhead{y1-y2 EE$_{\ell=2}$} & 
\colhead{BB$_{\ell=5}$} & 
\colhead{y1-y2 BB$_{\ell=5}$} & 
\colhead{EE$_{\ell=2}^{Cleaned}$} & 
\colhead{BB$_{\ell=5}^{Cleaned}$} \\
\colhead{} &
\colhead{} &
\colhead{GHz} & 
\colhead{${\rm (\mu K)^2}$} &  \colhead{${\rm (\mu K)^2}$} &
\colhead{${\rm (\mu K)^2}$} &  \colhead{${\rm (\mu K)^2}$} &
\colhead{${\rm (\mu K)^2}$} &  \colhead{${\rm (\mu K)^2}$}
}
\startdata
KK  &1 & 22.8&$306.6\pm0.12$ &\nodata &$ 38.1\pm0.18$ & \nodata &
\nodata &\nodata \\
KKa &2 & 27.4&$93.3\pm0.07$ &$  0.0\pm0.22$ &$ 14.9\pm0.10$ &$  0.6\pm0.30$ &
$ 0.8\pm0.10$ &$  0.6\pm0.14$\\
KQ  &2 & 30.5&$ 53.6\pm0.09$ &$ -1.6\pm0.27$ &$  8.5\pm0.11$ &$ -1.3\pm0.32$ &
$ 3.0\pm0.10$ &$  0.2\pm0.11$\\
KV  &3 & 37.2&$ 21.8\pm0.10$ &$ -0.7\pm0.29$ &$  2.0\pm0.13$ &$ -0.6\pm0.38$ &
$ 1.6\pm0.10$ &$  -0.7\pm0.13$\\
KW  &  & 46.2&$ 10.4\pm0.13$ &$ -3.8\pm0.4$ &$  0.1\pm0.17$ &$  -0.3\pm0.52$ &
$  -7.4\pm0.14$ &$  -1.9\pm0.18$\\
KaKa& 2 & 33.0&$ 30.5\pm0.13$ &\nodata &$  4.8\pm0.17$ & \nodata &
$  0.7\pm0.26$ &$  -0.1\pm0.35$\\
KaQ & 3 & 36.6&$ 17.2\pm0.09$ &$ -0.0\pm0.27$ &$  2.7\pm0.11$ &$ -0.7\pm0.32$ &
$  0.6\pm0.15$ &$  -0.1\pm0.18$\\
KaV &4 & 44.8&$ 8.2\pm0.10$ &$  0.2\pm0.30$ &$  0.7\pm0.12$ &$  0.2\pm0.37$ &
$  0.1\pm0.15$ &$  -0.2\pm0.19$\\
KaW &4 & 55.5&$  5.9\pm0.14$ &$  0.6\pm0.41$ &$  0.6\pm0.17$ &$  0.0\pm0.51$ &
$  0.4\pm0.20$ &$  -0.1\pm0.25$\\
QQ  &4 & 40.7&$  9.6\pm0.17$ &$  -0.1\pm0.67$ &$  1.8\pm0.17$ &$ 0.3\pm0.68$ &
$  0.3\pm0.23$ &$ 0.0\pm0.24$\\
QV  &4 & 49.7&$  4.5\pm0.12$ &$  -0.1\pm0.37$ &$  0.6\pm0.13$ &$  0.9\pm0.40$ &
$  -0.1\pm0.15$ &$  0.0\pm0.16$\\
QW  &4 & 61.7&$  3.3\pm0.17$ &$  0.2\pm0.5$ &$  0.7\pm0.18$ &$  -0.1\pm0.55$ &
$  0.1\pm0.20$ &$  0.2\pm0.21$\\
VV  &4 & 60.8&$  2.4\pm0.21$ &$  -0.5\pm0.81$ &$  0.2\pm0.21$ &$ -0.2\pm0.65$ &
$  0.5\pm0.19$ &$ 0.2\pm0.23$\\
VW  &4 & 75.4&$  2.3\pm0.18$ &$  1.0\pm0.55$ &$  0.2\pm0.21$ &$ -0.2\pm0.65$ &
$0.5\pm0.19$ &$  0.2\pm0.23$\\
WW  &4 & 93.5&$  2.2\pm0.37$ &$  1.5\pm1.27$ &$  -0.4\pm0.44$ &$ -0.3\pm1.48$ &
$  0.3\pm0.38$ & $  -0.7\pm0.45$ \\
\enddata
\tablecomments{For $\nu>40~$GHz, the largest foreground signals are at
$\ell=2$ of EE and $\ell=5$ of BB. This table shows the ``raw'' 
and ``cleaned'' values. The column labeled ``bin''
indicates which cross spectra are coadded into frequency bins.
Because K band is used as a foreground template, there
are no foreground corrected values. Also, as there are only single
K and Ka band polarization channels, it is not possible to form
cross spectra of year one minus year two. The y1-y2 notation refers to year one
minus year two. KW is not used in 
any of the averages over frequency.}
\end{deluxetable}
\clearpage

\begin{deluxetable}{ccccc}
\tablecaption{$\chi^2/\nu$ for TE, TE (2003), TB, EE, BB, and EB\label{tab:chisq}}
\tablehead{
\colhead{} &
\colhead{r4\tablenotemark{a}} & 
\colhead{r9\tablenotemark{{\rm b,f}}} & 
\colhead{r9} & 
\colhead{r9} \\
\colhead{} & 
\colhead{$l=2-16$} &
\colhead{$l=17-100$}& 
\colhead{$l=17-500$}&  
\colhead{$l=17-800$} \\
\colhead{} & 
\colhead{($\nu=15$ dof)} &
\colhead{($\nu=84$ dof)}& 
colhead{($\nu=484$ dof)}&  
\colhead{($\nu=784$ dof)}}
\startdata
TE\tablenotemark{{\rm c,e,g,h}}   & 0.31 (0.99)\tablenotemark{{\rm f}} & 1.01 (0.46) & 1.20($0.01$) & 1.08(0.06)\\
TE (2003)\tablenotemark{{\rm d}} & 1.88 (0.03) & 1.18 (0.25) & 2.06 (0) & \nodata \\
TB\tablenotemark{{\rm c,g}}  & 0.57 (0.90) & 0.72 (0.97) & 0.97 (0.70) & 0.97 (0.74) \\
EE\tablenotemark{{\rm c}}
        &  1.34 (0.17) &  1.06 (0.33) & 0.98 (0.59) & 0.96 (0.76)\\
BB\tablenotemark{{\rm c}}
        & 0.72  (0.77) & 1.28 (0.04) & 0.96 (0.73) &  0.95 (0.81) \\
EB\tablenotemark{{\rm c}}
        & 0.41 (0.98) & 1.21 (0.09) & 1.03 (0.34) & 0.96 (0.76) \\
\enddata
\tablecomments{$\chi^2/\nu$ is computed for the null model
($C_l^{XX}=0$).}
\tablenotetext{a}{r4 HEALPix maps are used for $\ell<32$. We limit
this to $\ell<17$ to avoid pixel window effects.}
\tablenotetext{b}{r9 HEALPix maps are used for  $16<\ell<800$.}
\tablenotetext{c}{For all results a model of the foreground emission has been removed.}
\tablenotetext{d}{TE (2003) corresponds to \citet{kogut/etal:2003}.}
\tablenotetext{e}{The numbers in parentheses are the PTEs.}
\tablenotetext{f}{For $\ell>16$, 
we use the binned diagonal elements of the covariance matrices
in Appendix~\ref{app:covar}.}
\tablenotetext{g}{For TE and TB, the E and B are comprised of a combination
of Q and V bands and the T is from V and W bands.}
\tablenotetext{h}{The TE signal is in this $\ell$ range and so the PTE
should be low.}
\end{deluxetable}
\clearpage

\begin{deluxetable}{cccc}
\tablecaption{$\chi^2/\nu$ for r4 Yearly Difference 
Null Maps\label{tab:chisq2}}
\tablehead{
\colhead{} &
\colhead{y1 - y2} &
\colhead{y2 - y3} &
\colhead{y1 - y3} \\
\colhead{} &
\colhead{$l=2-16$ ($\nu=15$ dof)} & 
\colhead{$l=2-16$ ($\nu=15$ dof)} & 
\colhead{$l=2-16$ ($\nu=15$ dof)}}
\startdata
TE   & 1.70 (0.04) & 1.05 (0.40) & 1.87 (0.02) \\
TB   & 1.95 (0.02) & 1.20 (0.26) & 1.08 (0.37) \\
EE   & 1.55 (0.08) & 0.89 (0.58) & 0.55 (0.91) \\
BB   & 0.56 (0.90) & 1.50 (0.09) & 0.76 (0.72) \\
EB   & 0.62 (0.86) & 1.04 (0.41) & 0.84 (0.63) \\
\enddata
\tablecomments{$\chi^2/\nu$ is computed for the null model,
$C_l^{XX}=0$.}
\end{deluxetable}
\clearpage

\begin{deluxetable}{cccccccc}
\tablewidth{6.5in}
\tabletypesize{\tiny}
\tablecaption{Binned Data for ${\cal B}^{EE}/\ell$ 
for $\ell>20$\label{tab:highee}}
\tablehead{
\colhead{} &
\colhead{$30\leq\ell\leq50$} & 
\colhead{$51\leq\ell\leq150$} &
\colhead{$151\leq\ell\leq250$} &
\colhead{$251\leq\ell\leq350$} &
\colhead{$351\leq\ell\leq450$} &
\colhead{$451\leq\ell\leq650$} &
\colhead{$651\leq\ell\leq1023$}}
\startdata
QV    &  $0.010\pm 0.007$  
      &  $0.011\pm 0.005$ 
      & $-0.001\pm 0.012$  & $-0.003\pm 0.026$
      & $-0.014\pm 0.058$  & $0.16\pm 0.12$      
      & $-0.73\pm 0.66$  \\
VW    &  $0.013\pm 0.011$  
      &  $0.004\pm 0.004$ 
      & $0.017\pm 0.009$  & $0.027\pm 0.018$
      & $0.031\pm 0.037$  & $0.095\pm 0.065$      
      & $0.13\pm 0.22$  \\
QVW   & $0.013\pm 0.006$  
      & $0.004\pm 0.004$ 
      & $0.017\pm 0.009$  & $0.027\pm 0.018$
      & $0.031\pm 0.037$  & $0.095\pm 0.065$      
      & $0.13\pm 0.22$  \\
KaQVW & $0.016\pm 0.004$  
      & $0.011\pm 0.003$ 
      & $0.012\pm 0.007$  & $0.020\pm 0.016$
      & $0.065\pm 0.035$  & $0.097\pm 0.064$      
      & $0.12\pm 0.22$  \\
\noalign{\smallskip\hrule\smallskip}
QV$^a$    &  $0.005\pm 0.009$ 
          &  $0.018\pm 0.007$ & \nodata & \nodata 
      & \nodata & \nodata  & \nodata \\
VW$^a$    &  $0.013\pm 0.011$     
          &  $0.001\pm 0.008$ & \nodata & \nodata 
      & \nodata & \nodata  & \nodata\\
QVW$^a$    & $0.012\pm 0.007$   
           & $0.006\pm 0.005$ & \nodata & \nodata 
      & \nodata & \nodata  & \nodata\\
KaQVW$^a$  &  $0.005\pm 0.005$ 
           &  $0.020\pm 0.004$ & \nodata & \nodata 
      & \nodata & \nodata  & \nodata \\
\enddata
\tablecomments{All entries have units of ($\mu$K)$^2$. The
top set is for combinations of the pre-cleaned data.
Sample variance is not included. The bottom set is for data 
cleaned with the KD3Pol model.
Note that the cleaning has little affect on the 
$51\leq\ell\leq150$ bin other than to increase the
uncertainty.}
\end{deluxetable}
\clearpage

\begin{deluxetable}{ccccc}
\tablecaption{Optical Depth vs. Data Selection\label{tab:taucomp}}
\tablehead{
\colhead{Combination} &
\colhead{Exact EE Only} &
\colhead{Exact EE \& TE} &
\colhead{Simple tau EE} &
\colhead{Simple tau, no $\ell=5,7$}}
\startdata
KaQV   &  $0.111\pm0.022$ & $0.111\pm0.022$  &  \nodata  &\nodata\\
Q      &  $0.100\pm0.044$ & $0.080\pm0.043$ & $0.08\pm0.03$ 
       & $0.085\pm0.03$ \\
QV     &  $0.100\pm0.029$ & $0.092\pm0.029$  
&  $0.110\pm0.027$  &   $0.085^{+0.045}_{-0.015}$\\
QV+VV     &  \nodata & \nodata & $0.145\pm0.03$ & $0.14^{+0.02}_{-0.06}$\\
V   & $0.092\pm0.048$ & $0.095\pm0.043$ & $0.09^{+0.03}_{-0.07}$ 
                                        & $0.10^{+0.03}_{-0.07}$    \\
QVW    &  $0.109\pm0.022$ & $0.099\pm0.023$  
&  $0.090\pm0.012$  &   $0.090\pm0.015$ \\
KaQVW  &  $0.107\pm0.019$ & $0.105\pm0.019$  
&  $0.095\pm0.015$ &   $0.095\pm0.015$  \\
\enddata
\tablecomments{The values of simple tau are computed for $2\leq\ell\leq 11$.
The models are computed in steps of $\Delta\tau=0.005$ and linearly
interpolated. The last column is computed with the errors on $\ell=5,7$ 
multiplied by ten. The QV+VV is the QV combination without the QQ
component. Since the exact likelihood is based on the Ka, Q, V, and W maps, 
there is no corresponding entry for QV+VV. Note that the maximum likelihood
values are independent of frequency combination indicating that
foreground emission is not biasing the determination of $\tau$. The 
calculations for the first two columns include the effects of marginalization
over synchrotron foreground emission and projecting out the small loss 
imbalance signal\citep{jarosik/etal:prep}. }
\end{deluxetable}
\clearpage

\begin{figure}
\epsscale{0.85}
\plotone{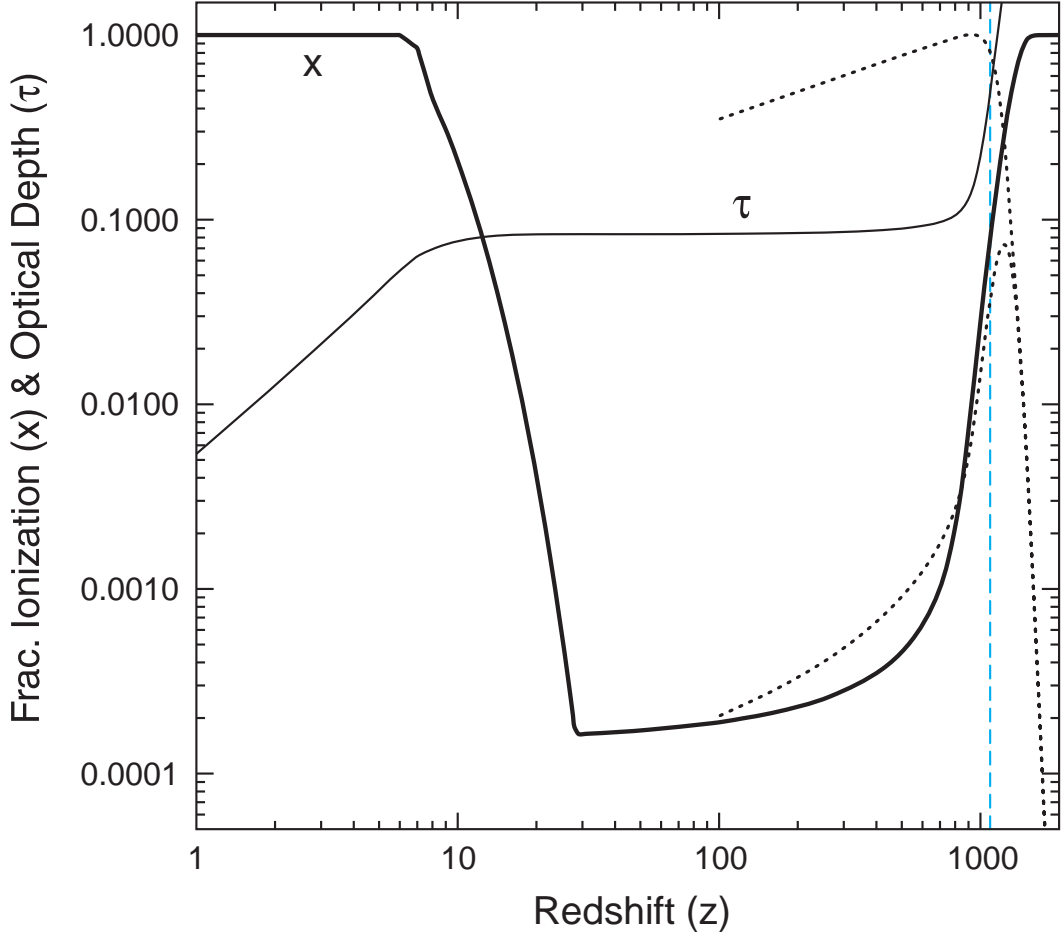}
\caption{\small A model of the ionization history of the universe. 
The line marked ``x'' is 
the ionization fraction, $x=n_e/n$ where $n_e$ is the number of electrons
and $n=11.2\omega_b(1+z)^3~{\rm m^{-3}}$ is the number of protons
with $\omega_b$ the baryon density. 
From quasar absorption systems we know 
the universe has been fully ionized since at least 
$z\approx6$. Between $6\la z\la30$ the first 
generation of stars ionized the universe.  We show a possible model 
inspired by \citet{holder/etal:2003}.
The history for this period 
is uncertain though the reionization produces a characteristic signature
in the CMB polarization. For $30<z<2000$, we
show decoupling as described in \citet{peebles:POPC}. The line marked
$\tau$ is the net optical depth, $\tau(z)$. The dashed curves are the 
integrands in the numerator (bottom)
and denominator (top) of equation~\ref{eq:ih} (divided by 200) for 
the $100<z<2000$ region. By eye, one can see that
the ratio of the integrals at the maximum, and thus the fractional 
polarization, is $\approx5$\%. The vertical line marks the 
redshift of decoupling, 
$z_{dec}=1088$, at the maximum of the visibility function (not shown).}
\end{figure}
\clearpage

\begin{figure*}[tb]
\plotone{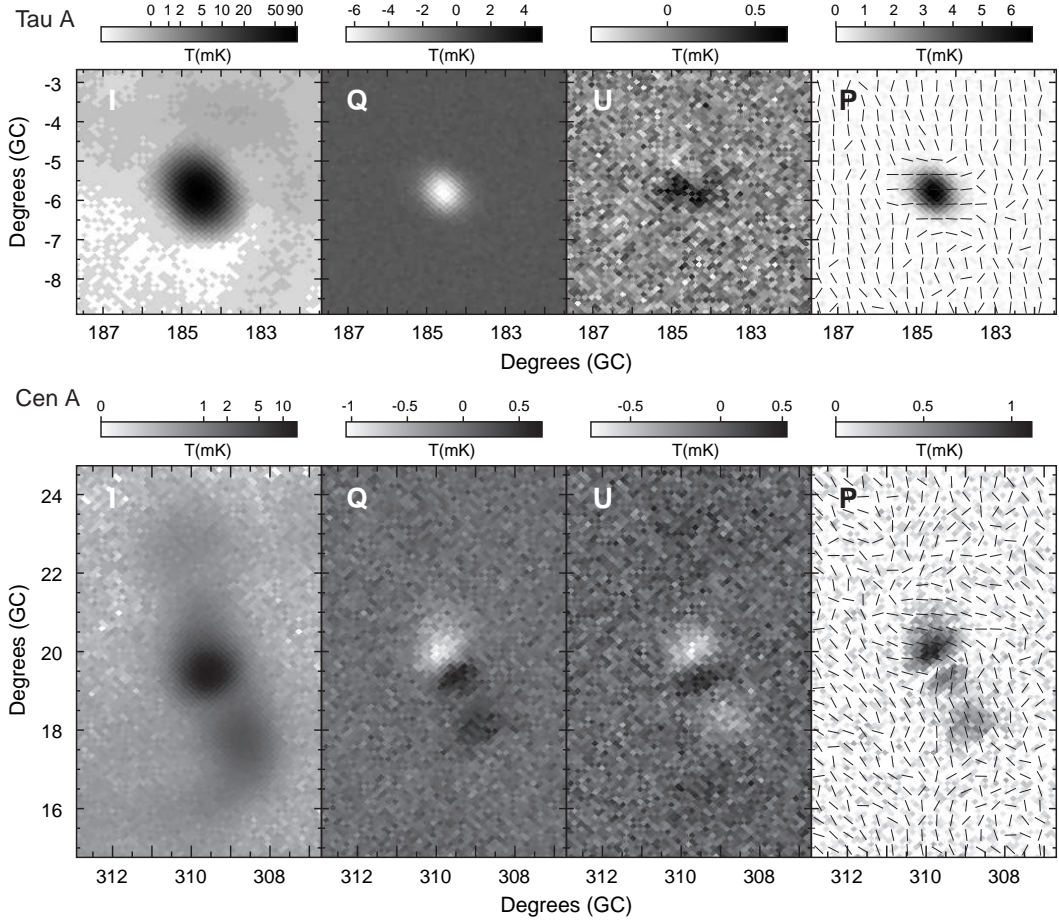}
\caption{ {\it Top:} Map of {\TauA} in Galactic coordinates
at 41 GHz in Stokes $I$, $Q$, $U$, $P$, smoothed to $1^{\circ}$. 
Since {\TauA} is polarized parallel to the Galactic plane
it is negative in $Q$ and small in $U$. {\it Bottom:} Map of 
Centaurus A in Stokes $I$, $Q$, $U$, and $P$. For both sets of plots, 
Stokes $I$ is scaled logarithmically and all the others are 
scaled linearly.  The scaling in mK is
indicated above the grayscale wedge for each panel.
A map of the noise bias has been subtracted from the P images.}
\label{fig:crab} 
\end{figure*}
\clearpage 

\begin{figure*}[tb]
\epsscale{0.6}
\plotone{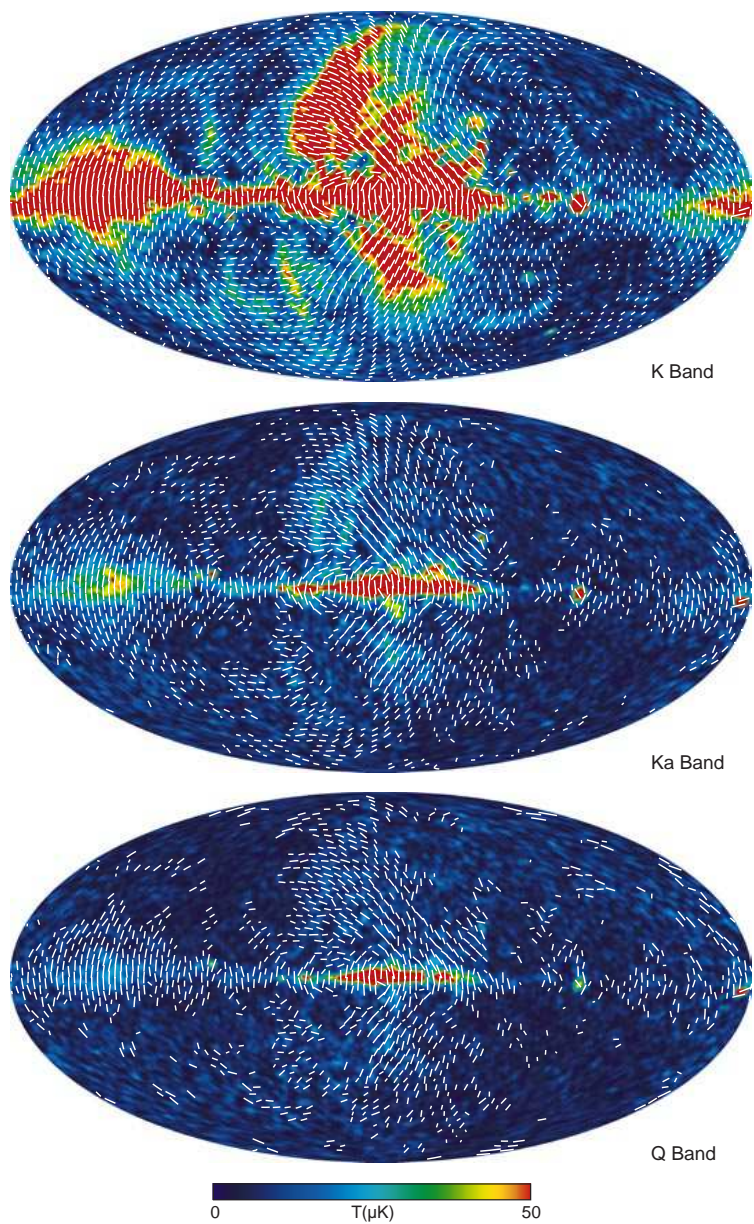}
\caption{P and $\gamma$ maps for K, Ka, and Q bands in Galactic coordinates.
See \citet[Figure 4]{bennett/etal:2003} for features and coordinates.
There is only one polarization map for K and Ka bands. 
For Q band, there are two maps which have been coadded.
The maps are smoothed to $2^{\circ}$. The polarization vectors are plotted
whenever a r4 HEALPix pixel 
(see \S\ref{sec:mask}, roughly $4\deg\times4\deg$) and three of its
neighbors has a signal to noise (P/N) greater than unity.
The length of the arrow is logarithmically dependent on the magnitude 
of $P$. Note that $P$ is positive.
Maps of the noise bias have been subtracted in these images.   
\label{fig:pgmaps1} }
\end{figure*}
\clearpage 

\begin{figure*}[tb]
\epsscale{0.9}
\plotone{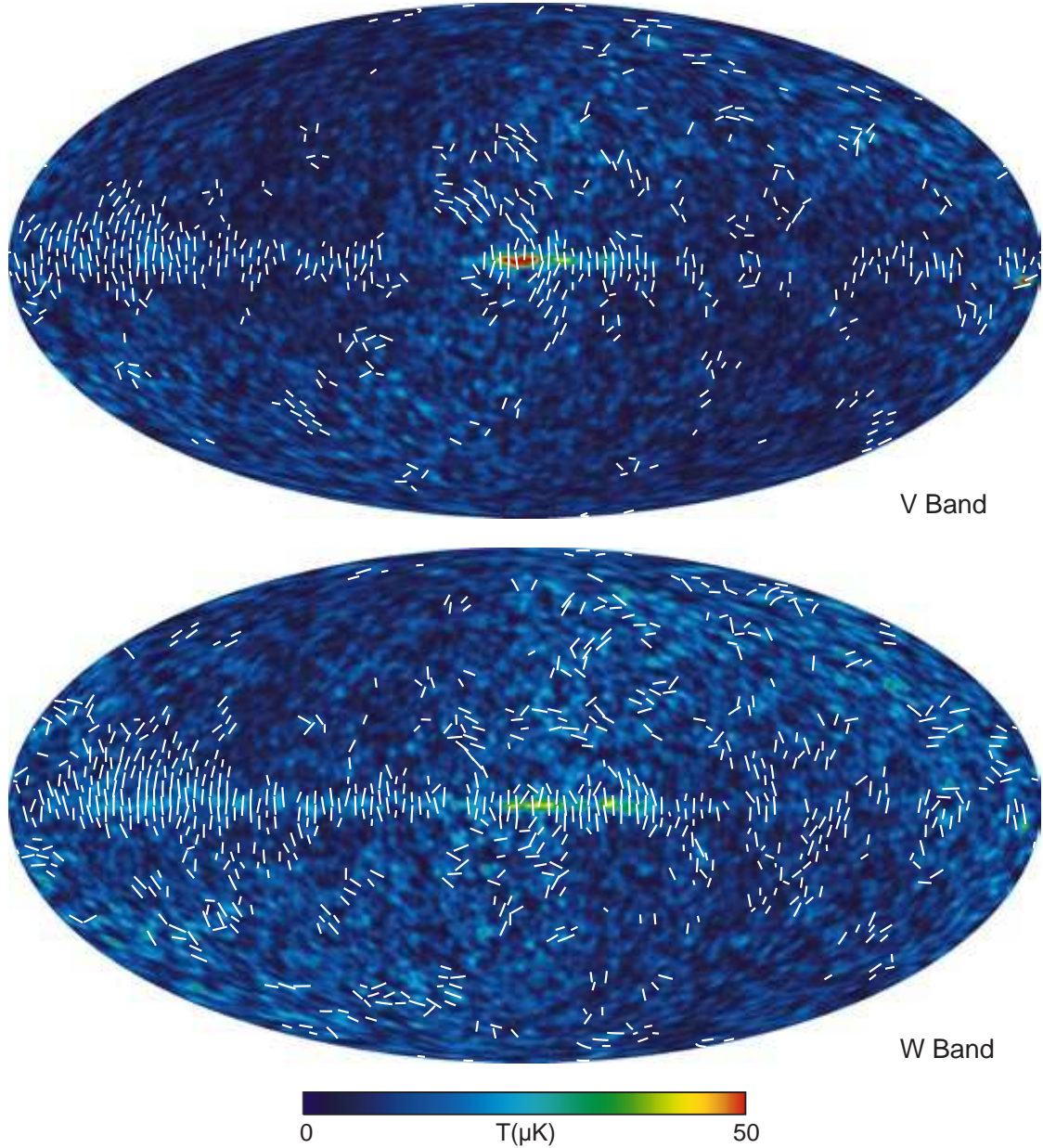}
\caption{Similar to Figure~\ref{fig:pgmaps1} but for V and W bands.
The two V-band maps have been coadded as have the four W-band maps. 
The relatively higher noise in the ecliptic plane is evident.
Maps of the noise bias have been subtracted in these images.  
\label{fig:pgmaps2} }
\end{figure*}
\clearpage 

\begin{figure*}[tb]
\epsscale{1.0}
\plotone{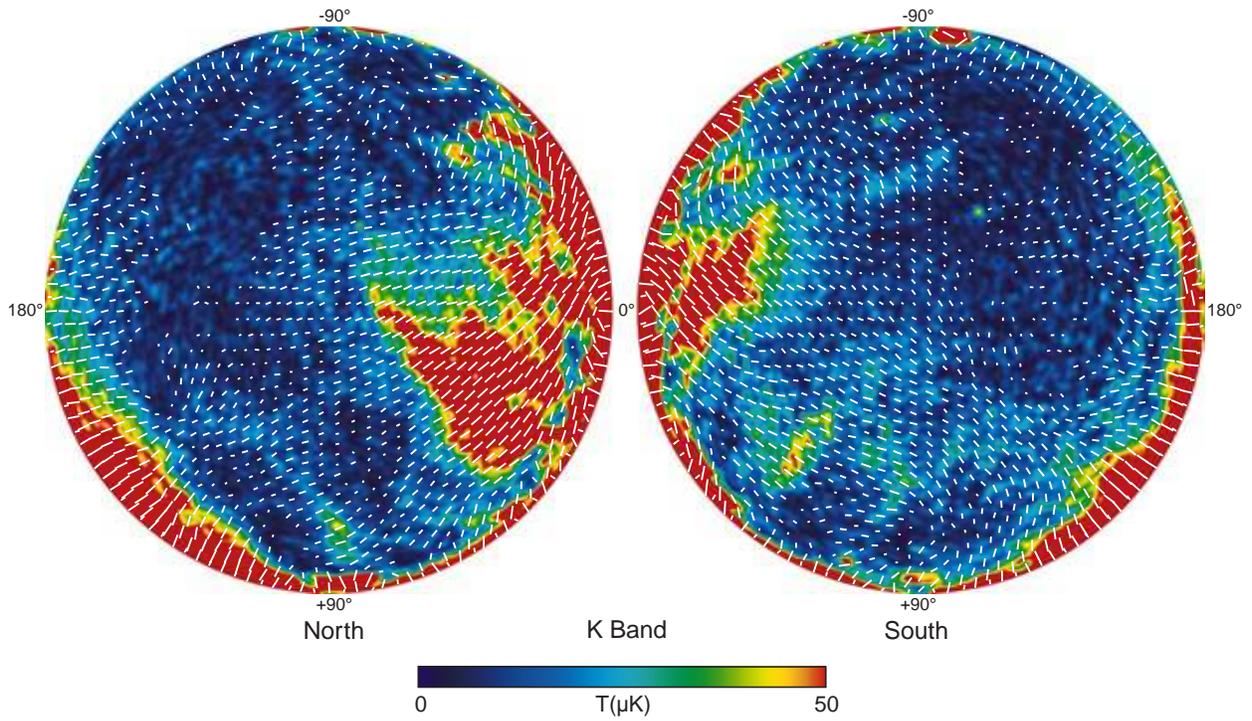}
\caption{A Lambert azimuthal equal area projection of the Galactic poles
({\it left:} north) showing the K-band polarization. The circumference
of each map is at zero Galactic latitude.
The convention in this plot is to use bars to indicate
the polarization direction. It is clear that the polarization extends 
to high Galactic latitudes.  
A map of the noise bias is subtracted from this image.
\label{fig:kstereo} }
\end{figure*}
\clearpage 

\begin{figure*}[tb]
\epsscale{1.0}
\plotone{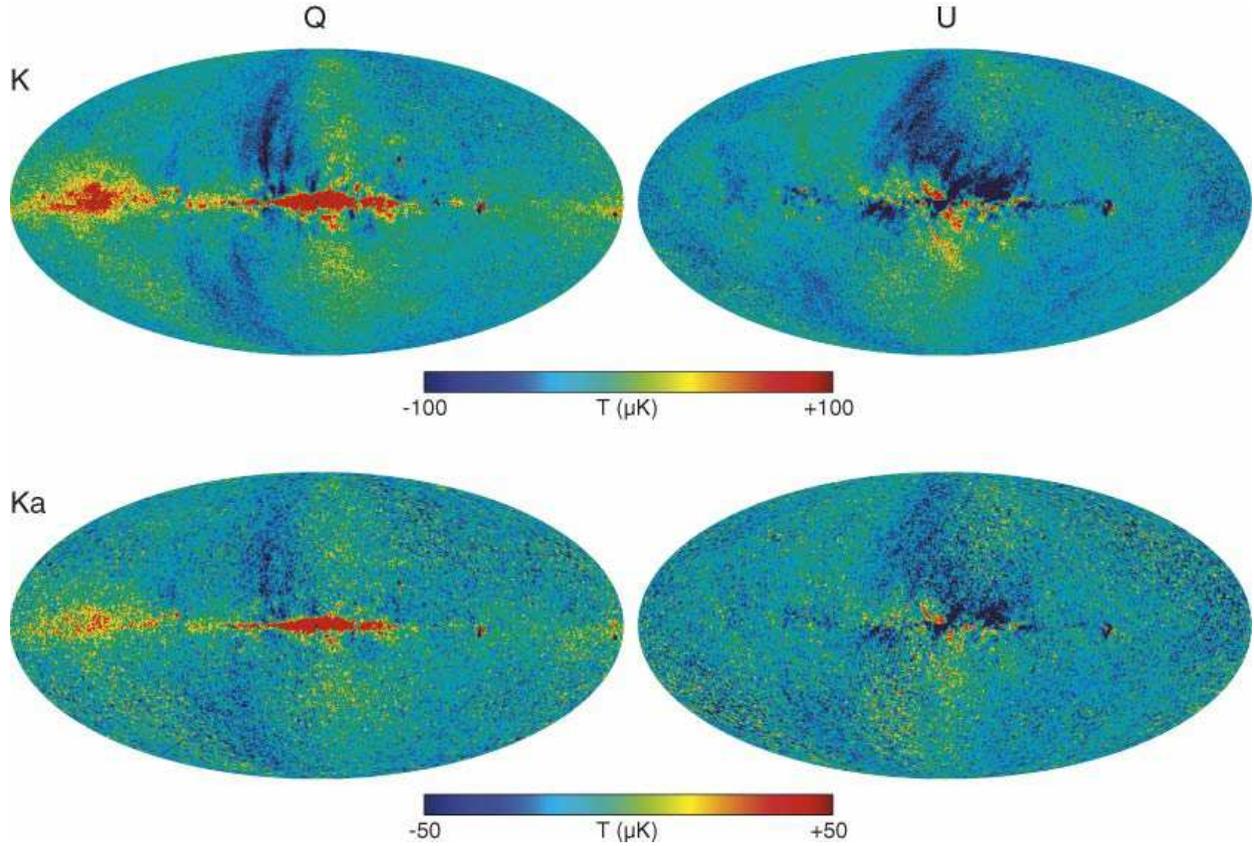}
\caption{Stokes $Q$ and $U$ maps in K and Ka bands. The Galactic 
plane is dominated
by positive Stokes $Q$ because the foreground polarization direction is 
perpendicular to the plane. As discussed in \S\ref{sec:foregrounds},
this is expected because the Galactic magnetic field is predominantly
parallel to the plane. For comparison, the Stokes $Q$ and $U$
maps of a noiseless CMB simulation have peak-to-peak 
values of less than $6~\mu$K. These maps have been smoothed to $1^{\circ}$.
\label{fig:qumaps} }
\end{figure*}
\clearpage 

\begin{figure*}[tb]
\epsscale{1}
\plotone{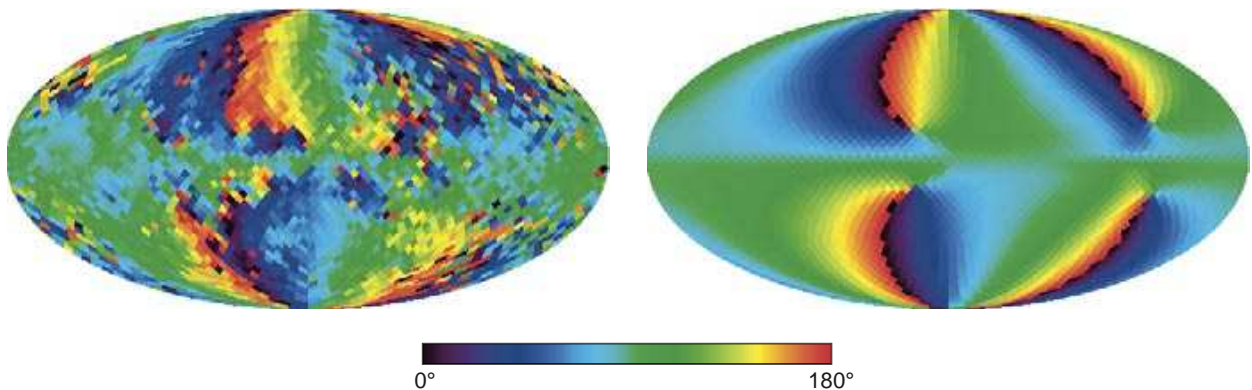}
\caption{{\it Left:} The angle of the magnetic field, 
$\gamma_M=\gamma_{PA} + 90\fdg$, derived from the synchrotron 
radiation in the K-band map (smoothed with a $4^\circ$ beam) shown 
in Figure 3. (We do not distinguish between $\pm 180^{\circ}$ in the
field direction.) The predominant low Galactic latitude magnetic field 
direction is parallel to the Galactic plane ($\gamma_M=90^{\circ}$)
and thus the synchrotron
(and dust) polarization directions have $\gamma\approx 0^\circ$.
In the North Polar Spur region, the magnetic field is perpendicular
to the Galactic plane corresponding to $\gamma_M\approx0^{\circ}$ or 
$180^{\circ}$. Note the large scale coherency of the field.
{\it Right:} The predicted magnetic field direction
given by a simple model of the electron distribution 
and the logarithmic spiral arm model (Equation~\ref{eq:bfield}) for 
the magnetic field. 
\label{fig:coherence} }
\end{figure*} 
\clearpage 

\begin{figure*}[tb]
\epsscale{1}
\plotone{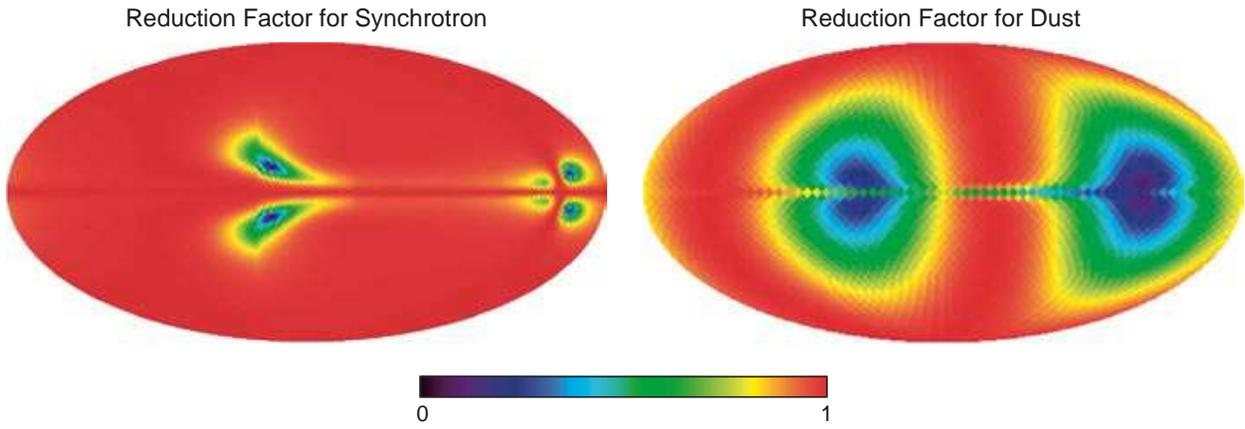}
\caption{The left panel shows 
the geometric suppression factor, $g_{sync}(\hat{n})$, 
in the polarization
due to the magnetic field geometry.  The right panel
shows a similar geometric suppression factor for polarized dust
emission, $g_{dust}(\hat{n})$, see \S\ref{sec:thde}.
\label{fig:supgeom} }
\end{figure*}
\clearpage 

\begin{figure*}[tb]
\epsscale{1}
\plotone{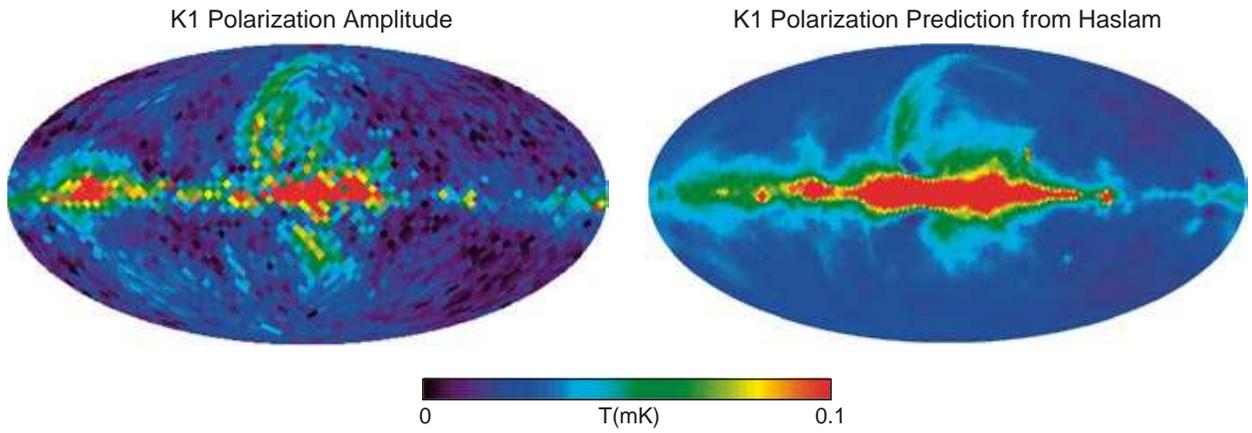}
\caption{ {\it Left:} The observed K-band polarization, $P$. The color scale 
ranges from 0 to 0.1 mK.
{\it Right:} The model prediction of the K-band polarization based on 
the Haslam {\it intensity} map. The model has one effective free parameter,
the ratio of the homogeneous field strength to the total field
strength as shown in Equation~\ref{eq:qumod}. 
This plot shows the results for $\beta_s = -2.7$ \& $q = 0.7$.
\label{fig:haslam} }
\end{figure*}
\clearpage 

\begin{figure*}[tb]
\epsscale{0.75}
\plotone{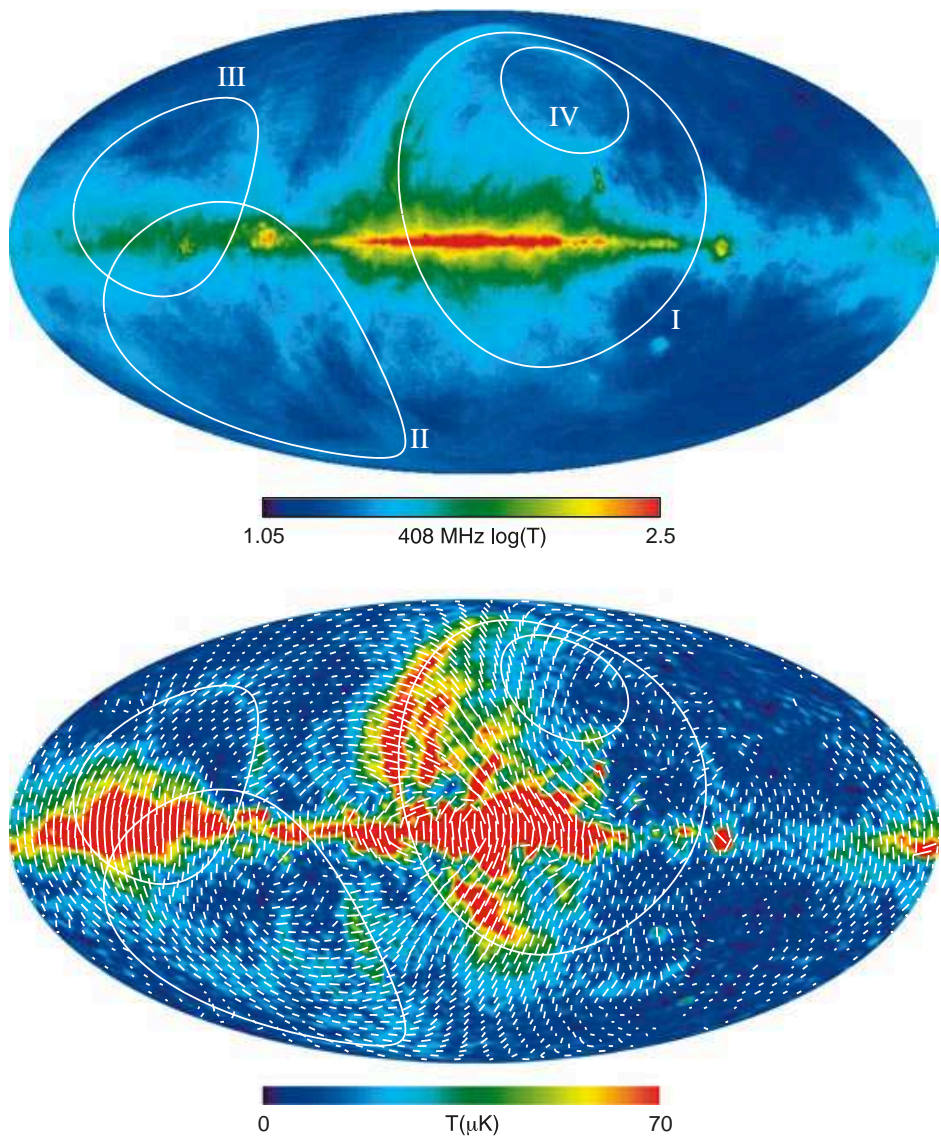}
\caption{{\it Top:} The Haslam 408~MHz map is shown with circles indicating loops
from \citet{berkhuijsen/haslam/salter:1971}. 
These ridges of enhanced Galactic radio 
emission are seen across the sky at low radio frequencies. 
The North Polar Spur (``Loop I'') and the Cetus arc (``Loop II'') are 
examples of these features, which have been described as the remnants of 
individual supernovae, or of correlated supernovae outbursts that 
produce blowouts, or as helical patterns that follow the local magnetic
fields projecting out of the plane.  Four such loops can be seen in 
the Haslam 408 MHz radio map and the \WMAP\ map. Note that the color stretch
is logarithmic in temperature.
{\it Bottom:} The \WMAP\ K-band
polarization map with the same loops superimposed. Note that the
highly polarized southern feature is close to the North Polar
Spur circle and may be related to the same physical structure. 
Note also that the polarization direction is perpendicular 
to the main ridge arc of the North Polar Spur, indicating a tangential 
magnetic field.  This is also seen in the southern feature.  
Whether or not they are physically related remains unclear. 
\label{fig:loops} }
\end{figure*}
\clearpage 

\begin{figure*}[tb]
\epsscale{0.5}
\plotone{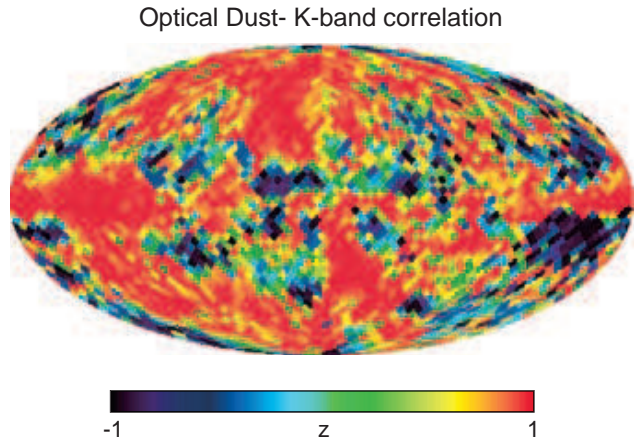}
\caption{A map of the correlation, $Z$, between the polarization angle
derived from the polarization of starlight, and the polarization 
angle in K-band. In the regions of high K-band polarization,
the correlation is strong. The polarization directions are 
anti-correlated in the Orion-Eridanus region near $l=-165^\circ$,
suggesting spatially distinguished regions of dust and synchrotron
emission.
\label{fig:zzz} }
\end{figure*}
\clearpage 

\begin{figure*}[tb]
\epsscale{1.0}
\begin{center}
\plotone{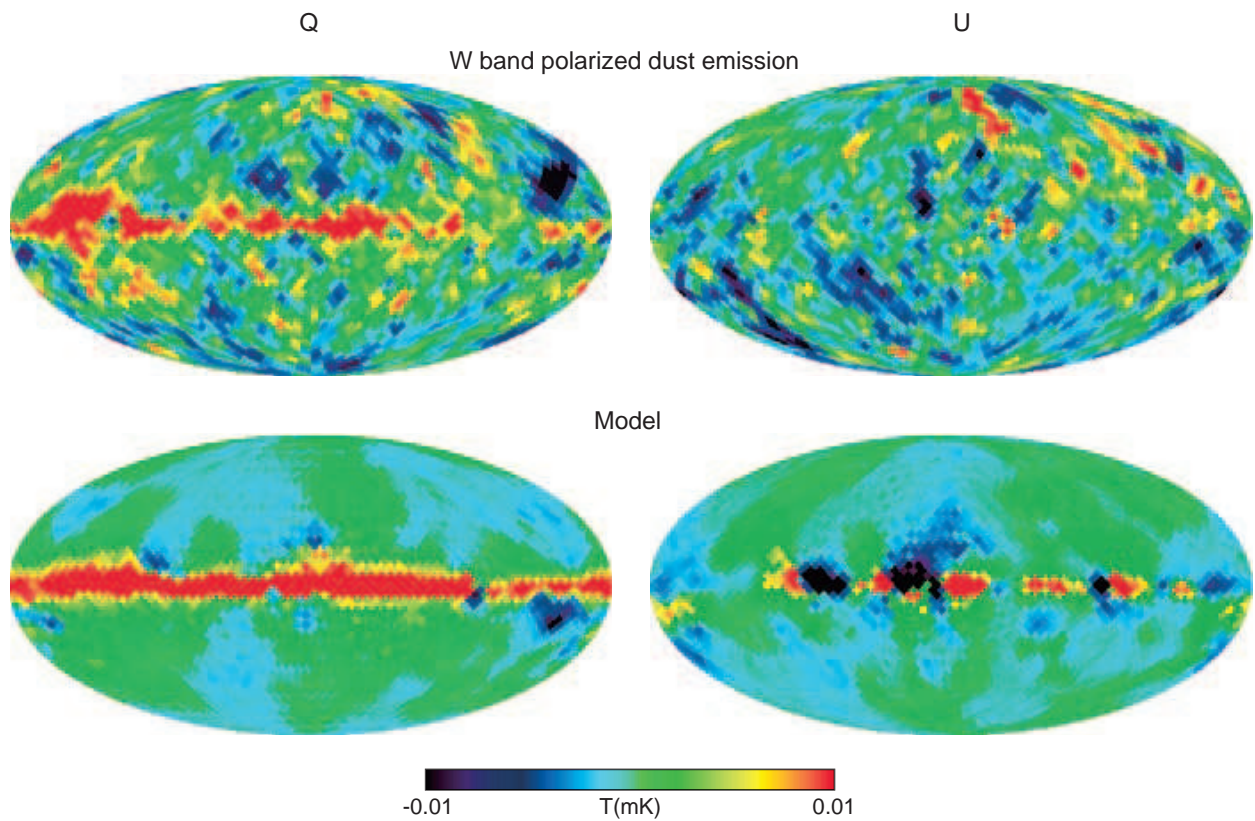}
\end{center}
\caption{The upper panels show the polarization signal at
W band with the CMB and synchrotron signal removed
(smoothed with a $10^\circ$ Gaussian beam).  
The left and right panels show Stokes $Q$ and $U$ polarization components 
respectively. There is a clear preponderance of Stokes $Q$ emission
in the plane. The lower panels show the 
predicted dust polarization based on Equation~\ref{eq:dust_template}. 
For $|b|<10^{\circ}$, the stars do not sample the dust column well
and the model is not accurate, especially for Stokes $U$.
For $|b|>10^{\circ}$, there are regions where the data and model agree
to the eye. However, a fit (\S\ref{sec:fgr}) is used to 
assess the level of polarized dust emission in the maps. 
\label{fig:dust_template} }
\end{figure*}
\clearpage 

\begin{figure*}[tb]
\epsscale{1.0}
\plotone{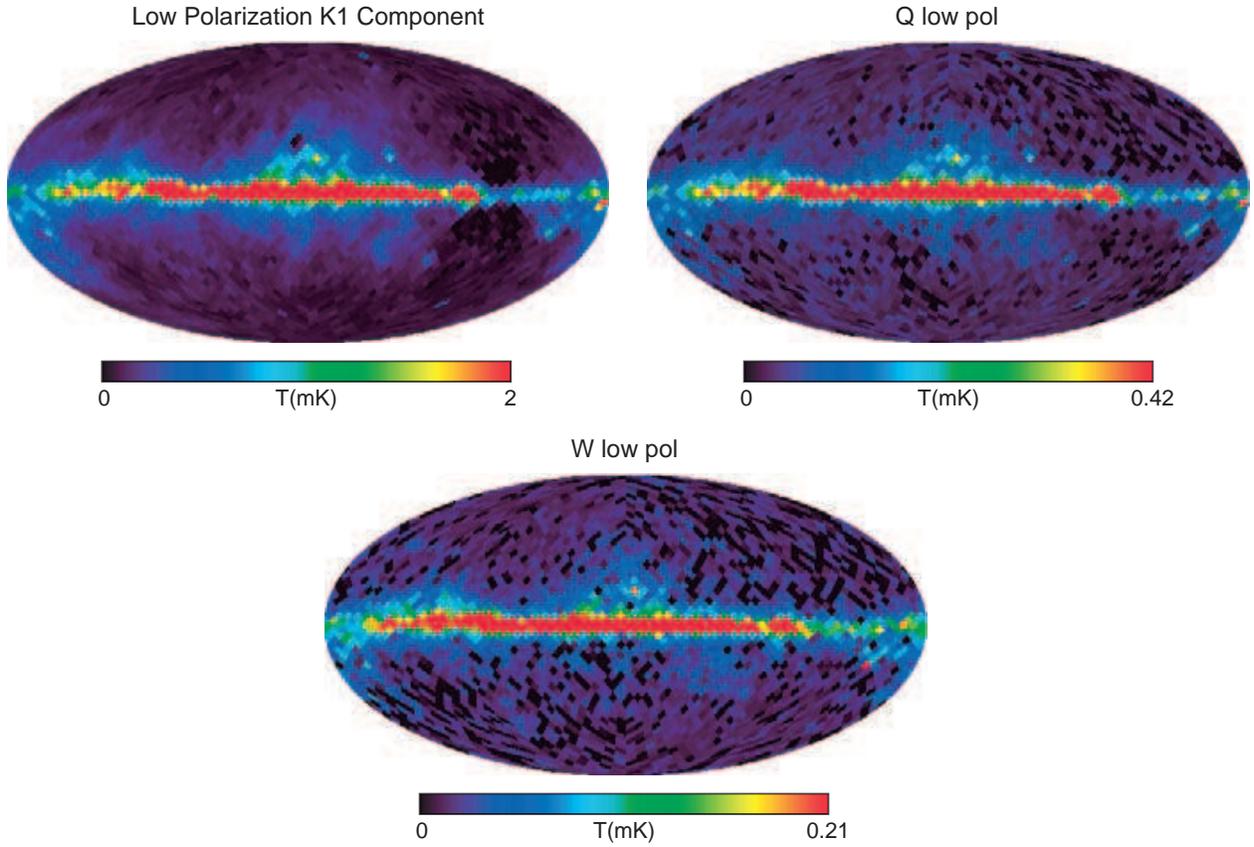}
\caption{Temperature maps of the low polarization components for
K, Q, and W bands.  The maps are computed  using
equation (\ref{eq:low_high}). The color scale is in mK.
Near the Galactic center, the low polarization component is
approximately 6\%, 3\%, and 6\% of the unpolarized emission
in K, Q, and W bands respectively.
\label{fig:lowpol} }
\end{figure*}
\clearpage 

\begin{figure*}[tb]
\epsscale{1}
\plotone{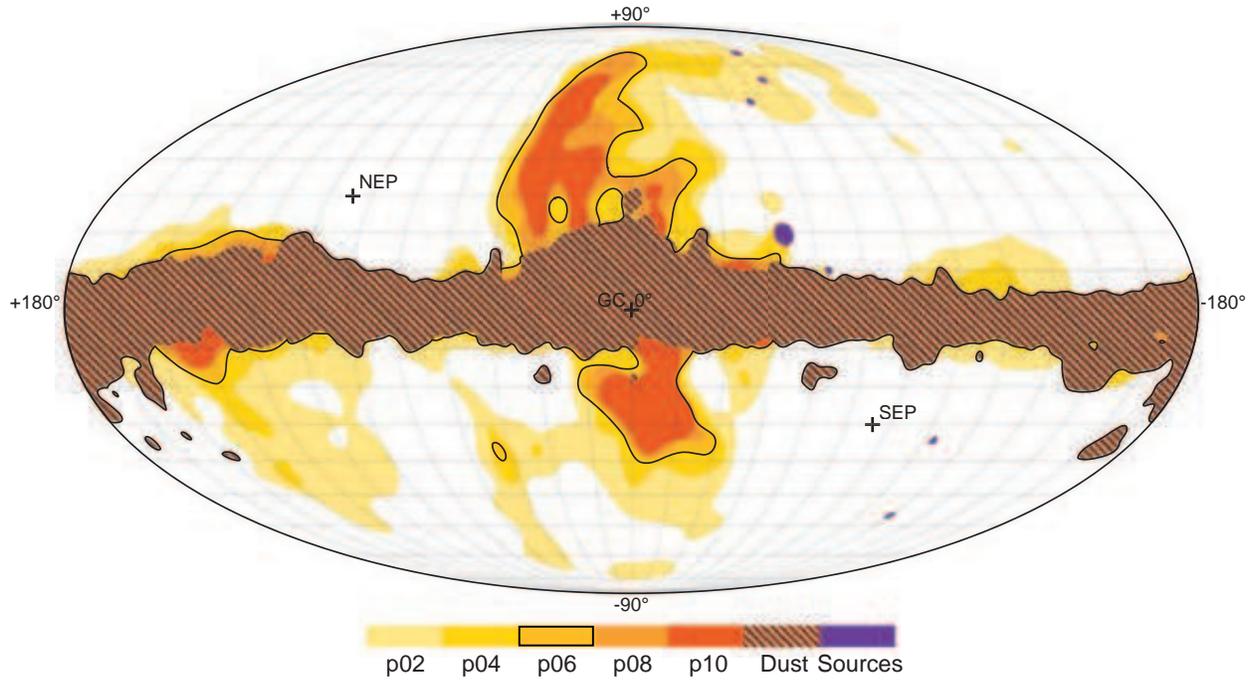}
\caption{The polarization masks, in Galactic coordinates, are shown for
the P02, P04, P06, and P10 cut levels. The cross hatched region along the 
Galactic plane, common to all polarization masks, shows the dust 
intensity cut. The P06 cut is outlined by the black curve. The masked
sources are in violet. The North 
Ecliptic Pole (NEP), and South Ecliptic Pole (SEP), and Galactic Center (GC)
are indicated.
\label{fig:polmasks} }
\end{figure*} 
\clearpage 

\begin{figure*}[tb]
\epsscale{1}
\plotone{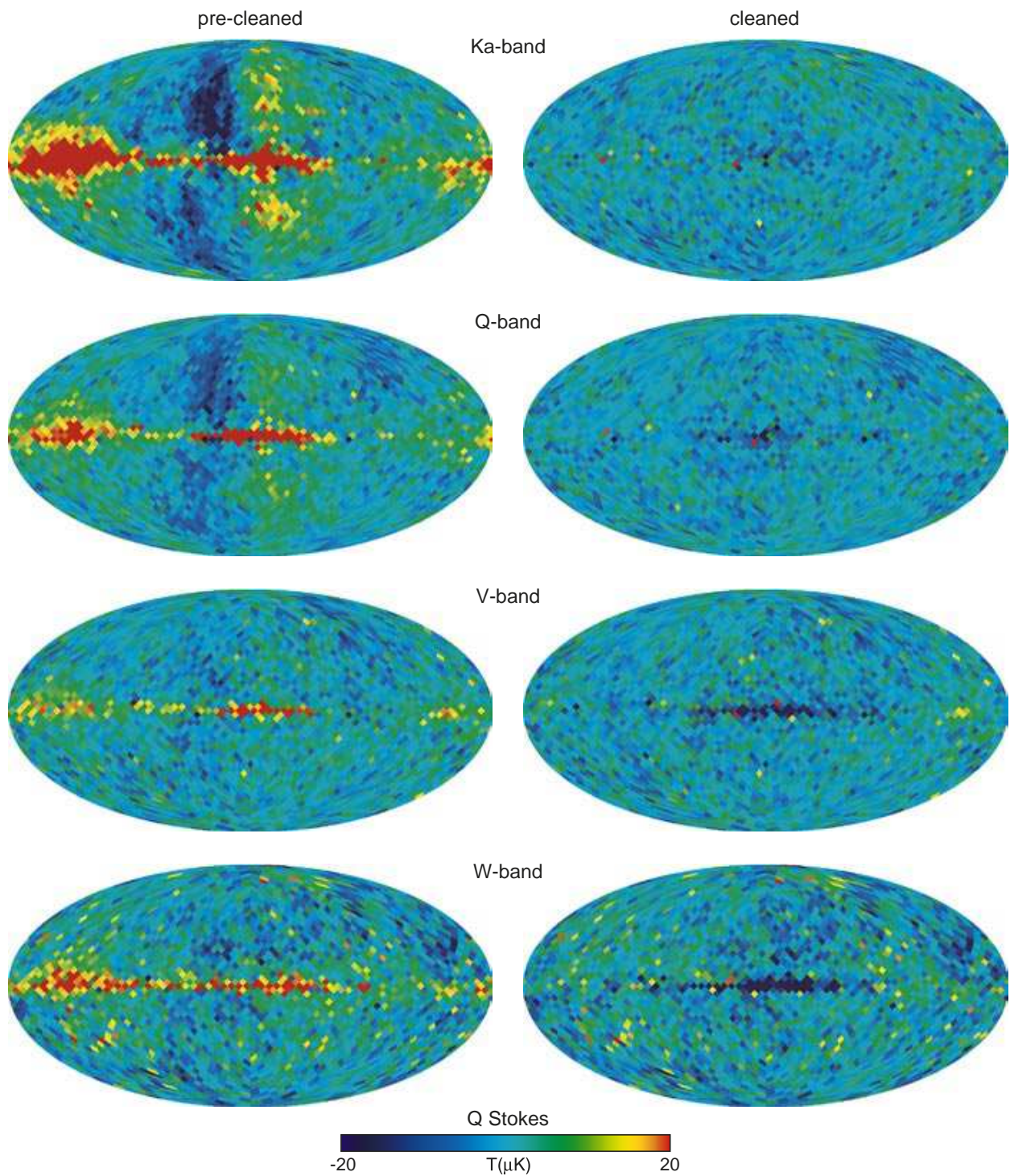}
\caption{The Ka, Q, V, and W band $Q$ Stokes Parameter 
maps before and after foreground subtraction using the method
outlined in \S\ref{sec:fgr}. There is a possible residual signal 
in W band though the noise is not yet sufficiently low to be certain.
The $U$ maps look similar. The cleaning for the cosmological analysis 
was done outside the processing cut \citep{jarosik/etal:prep}
and was based on the K-band maps and the starlight-based dust template.
The over-subtracted dark regions on the galactic plane are inside the
processing cut. 
\label{fig:inout} }
\end{figure*} 
\clearpage 

\begin{figure*}[tb]
\epsscale{0.8}
\plotone{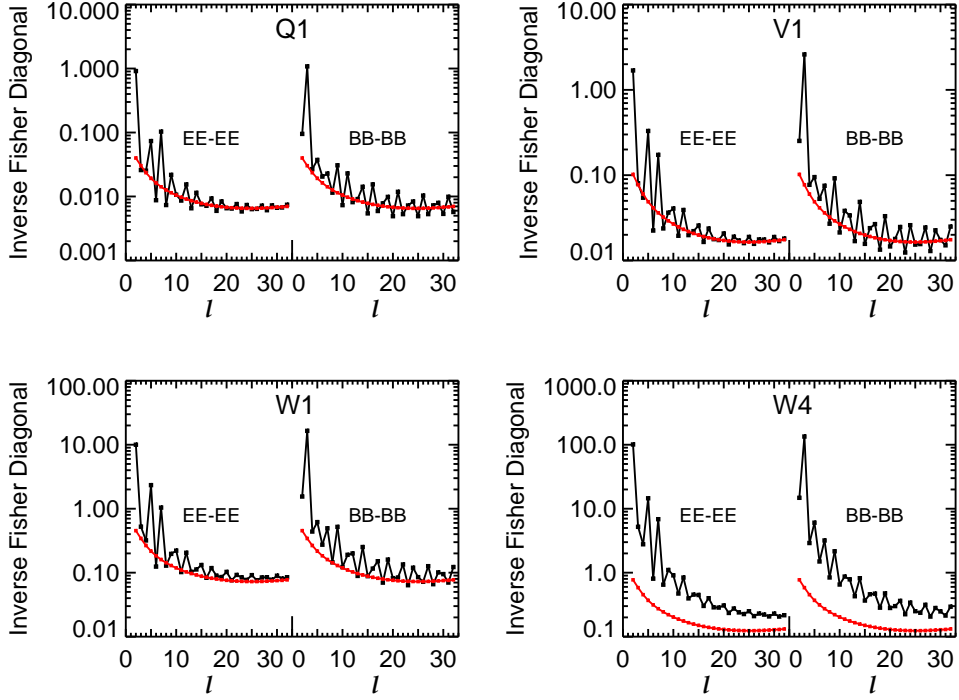}
\caption{A comparison of the predicted $C_\ell$ errors with (black)
and without (red) assuming correlated noise in the polarization sky maps. 
On the y-axis is plotted 
the diagonal element of the inverse of the Fisher matrix for one year
of data.  The units 
are ${\rm (\mu K)^4}$. Note that the y-axis scale for each plot is different.
In each panel EE and BB are shown. The variations
in the ${\bf N^{-1}}$ weighting are due to the scan pattern combined with 
the sky cut. There is less variation for B-modes than there is for E-modes.
W4 has the largest $1/f$ noise of all radiometers. One can see that the 
combination of $1/f$ noise coupled with \WMAP's scan strategy
leads to a larger uncertainty than one would get from considering
just the effects of $1/f$ noise alone. 
\label{fig:nobsninv} }
\end{figure*}
\clearpage 

\begin{figure*}[tb]
\epsscale{0.75}
\plotone{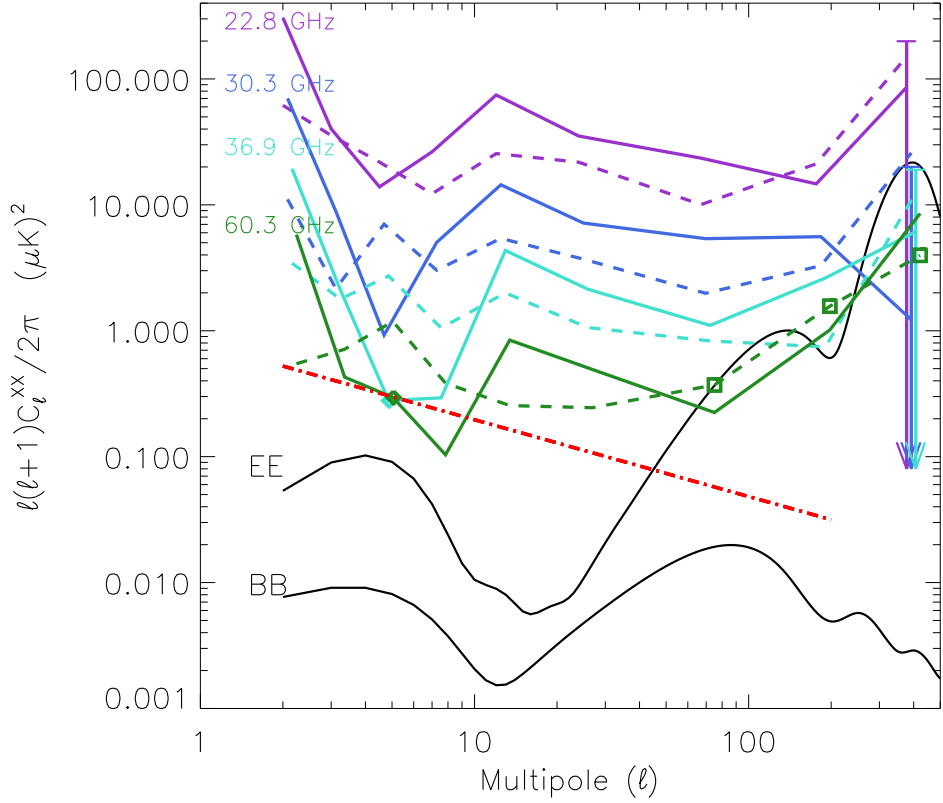}
\caption{The absolute value of the EE (solid, violet through green) 
and BB (dashed, violet through green) 
polarization spectra for the region outside the P06 mask. The best fit
$\Lambda$CDM model to TT, TE, and EE data
with $\tau=0.09$ and an additional tensor contribution with $r=0.3$
is shown in black. 
The cross spectra have been combined
into frequency bins according to Table~\ref{tab:eebb}
and into the following $\ell$ bins: [2, 3, 4-5, 6-8, 9-15, 16-32,
33-101, 102-251, 252-502]. In the presence of a dominant synchrotron 
spectrum, the averages over frequency are dominated by contributions 
from the lowest frequencies as can be seen by comparing the above
at $\ell=2$ to Figure~\ref{fig:rawps_fspace}.
Diamonds (EE) and boxes (BB) denote 
the data points that are negative.
The points are plotted at their 
absolute value to limit clutter. They should be interpreted as
indicating the approximate noise level of the measurement. The
$1\sigma$ upper bounds and downward arrows
mark points that are positive but consistent with zero.
The general rise in the data for $\ell>100$ is due to the
large noise term.
The red line corresponds to Equation~\ref{eq:fgmod} evaluated
for $\nu=60~$GHz for the BB foreground emission.
\label{fig:rawps_lspace} }
\end{figure*}
\clearpage 

\begin{figure*}[tb]
\epsscale{0.75}
\plotone{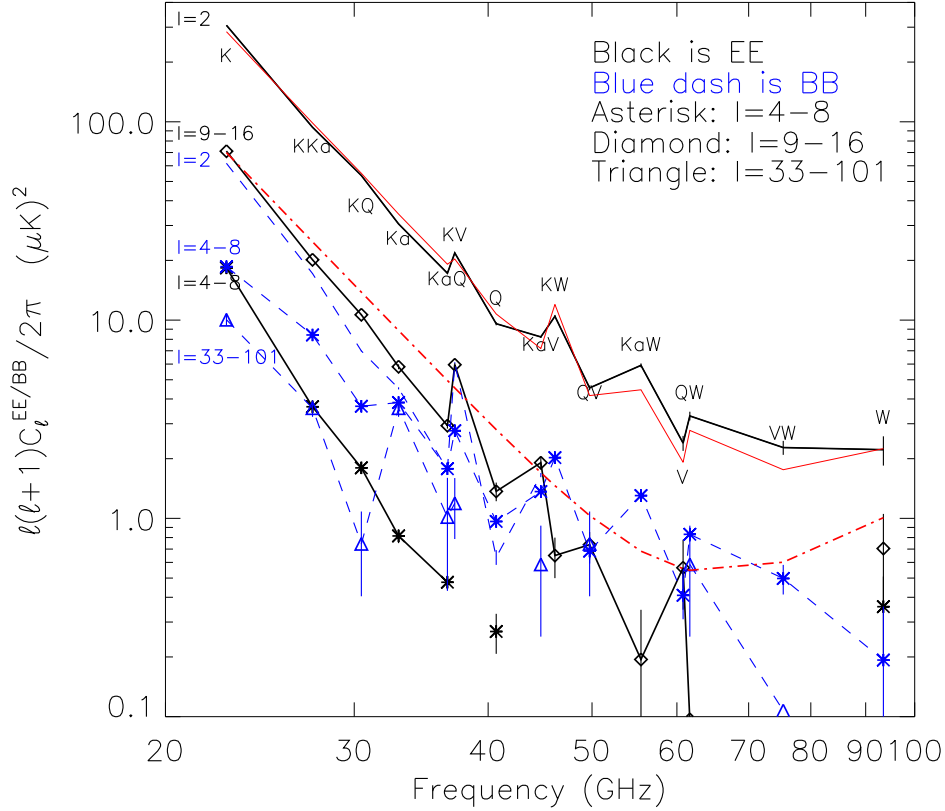}
\caption{The frequency spectrum of the EE and BB  
power spectra for the region outside the P06 mask. To increase the 
signal to noise
ratio, multiple values of $\ell$ are averaged as indicated.
Only statistical errors
are shown. Negative values are not plotted. The frequency
band combinations are given in Table~\ref{tab:eebb}.
The thin red line running close to the $\ell=2$ EE spectrum
is the model in  Equation~\ref{eq:dscorr}.
The dot-dash red line corresponds to Equation~\ref{eq:fgmod} evaluated
for BB at $\ell=2$.
\label{fig:rawps_fspace} }
\end{figure*}
\clearpage

\begin{figure*}[tb]
\epsscale{0.45}
\plotone{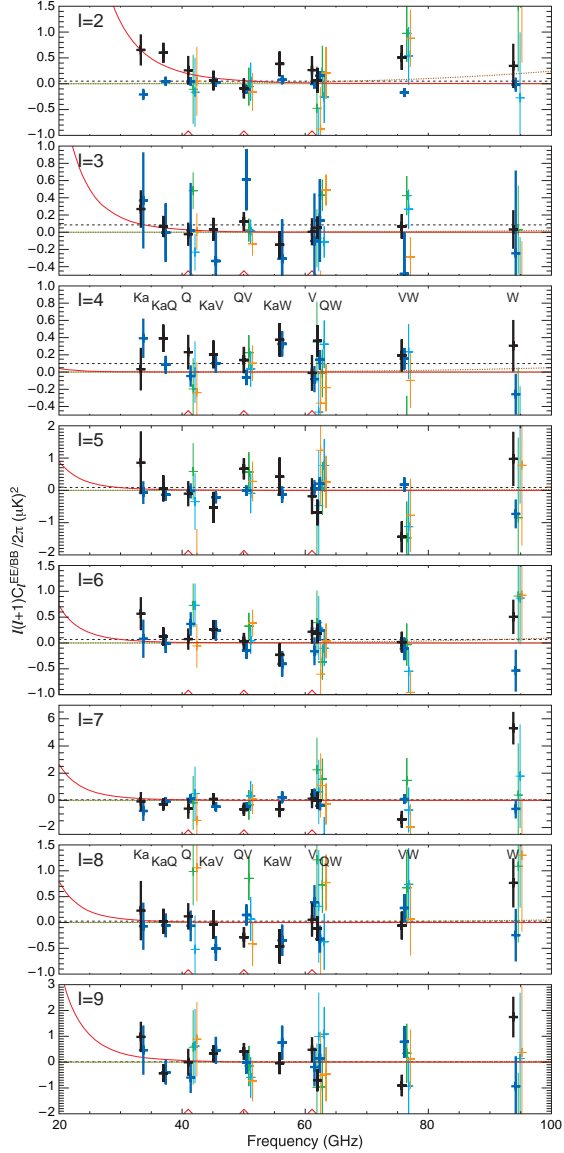}
\caption{The frequency spectrum of the foreground cleaned EE and BB  
power spectra outside the P06 mask for $\ell=2-9$. Black
shows EE, blue shows BB, and green, cyan, and orange show
the EE $yi-yj$ spectra (the BB ones are similar). For cosmological 
analysis, only the QQ,
QV, and VV frequency channels are used (the ``QV combination,''
indicated by red triangles 
on the bottom of each panel). 
The dotted black line shows the EE signal for $\tau=0.09$.
The dashed brown line shows the MEM dust temperature spectrum scaled 
by $0.0025$
to indicate the level of 5\% polarization (most evident near 90~GHz 
at $\ell=2$). Averaged over the region outside the P06 mask, this
is most likely an overestimate. The red curve shows the synchrotron 
spectrum scaled to 0.15 the pre-cleaned K-band temperature value. 
Based on the foreground model and discussion in text, it is unlikely 
that there is a significant residual foreground contamination in Q and V bands.
Note that for all frequency combinations above 40~GHz (excluding KW), 
BB is clearly consistent with zero, also 
indicating the efficacy of the foreground cleaning.
\label{fig:lbylEEBB} }
\end{figure*}
\clearpage 

\begin{figure*}[tb]
\epsscale{1.0}
\plotone{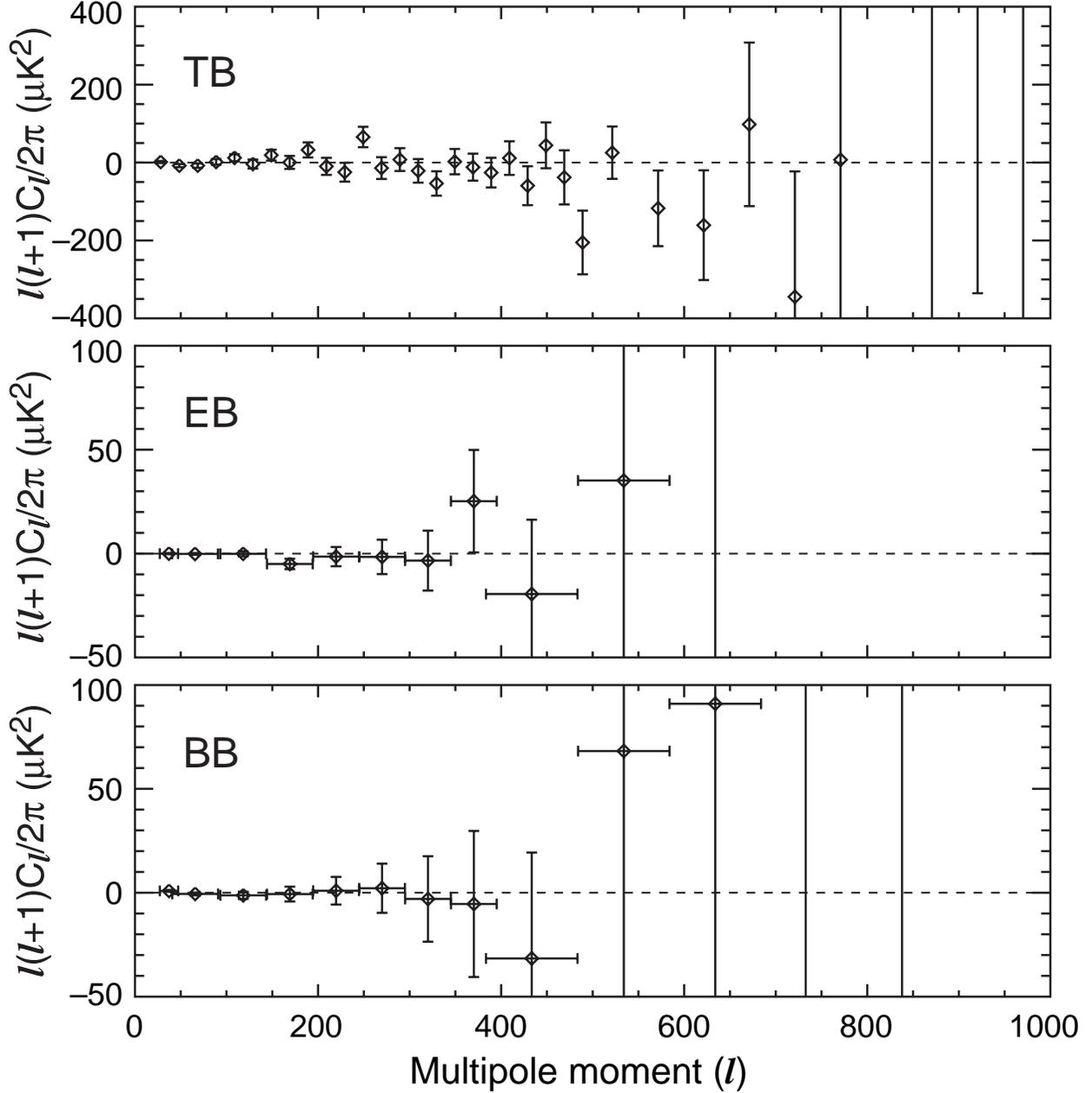}
\caption{Plots of all the noise for the expected null combinations
of TB, EB, BB for the region outside the P06 mask. For 
T the foreground-cleaned 
V and W bands have been combined bands. For E and B, the foreground-cleaned
Q and V bands have been combined.
Cosmic variance is included for all plots. For each plot there are
999 $\ell$ values that have been averaged into 33 bins
for TB and 12 bins for EB and BB. For TB, $\chi^2/\nu=41.6/33$ and
$\chi^2/\nu=931/999$ with corresponding $PTE = 0.15$ and $0.94$
for the two binnings.  For EB, $\chi^2/\nu=7.5/12$ and
$\chi^2/\nu=956/999$ with corresponding $PTE = 0.82$ and $0.84$.
For BB, $\chi^2/\nu=6.2/12$ and
$\chi^2/\nu=1000/999$ with corresponding $PTE = 0.91$ and $0.49$.
The polarization maps have been cleaned as described in 
Section~\S\ref{sec:fgr}.
See also Table~\ref{tab:chisq}. 
\label{fig:chisq} }
\end{figure*}
\clearpage 

\begin{figure*}[tb]
\epsscale{0.9}
\plotone{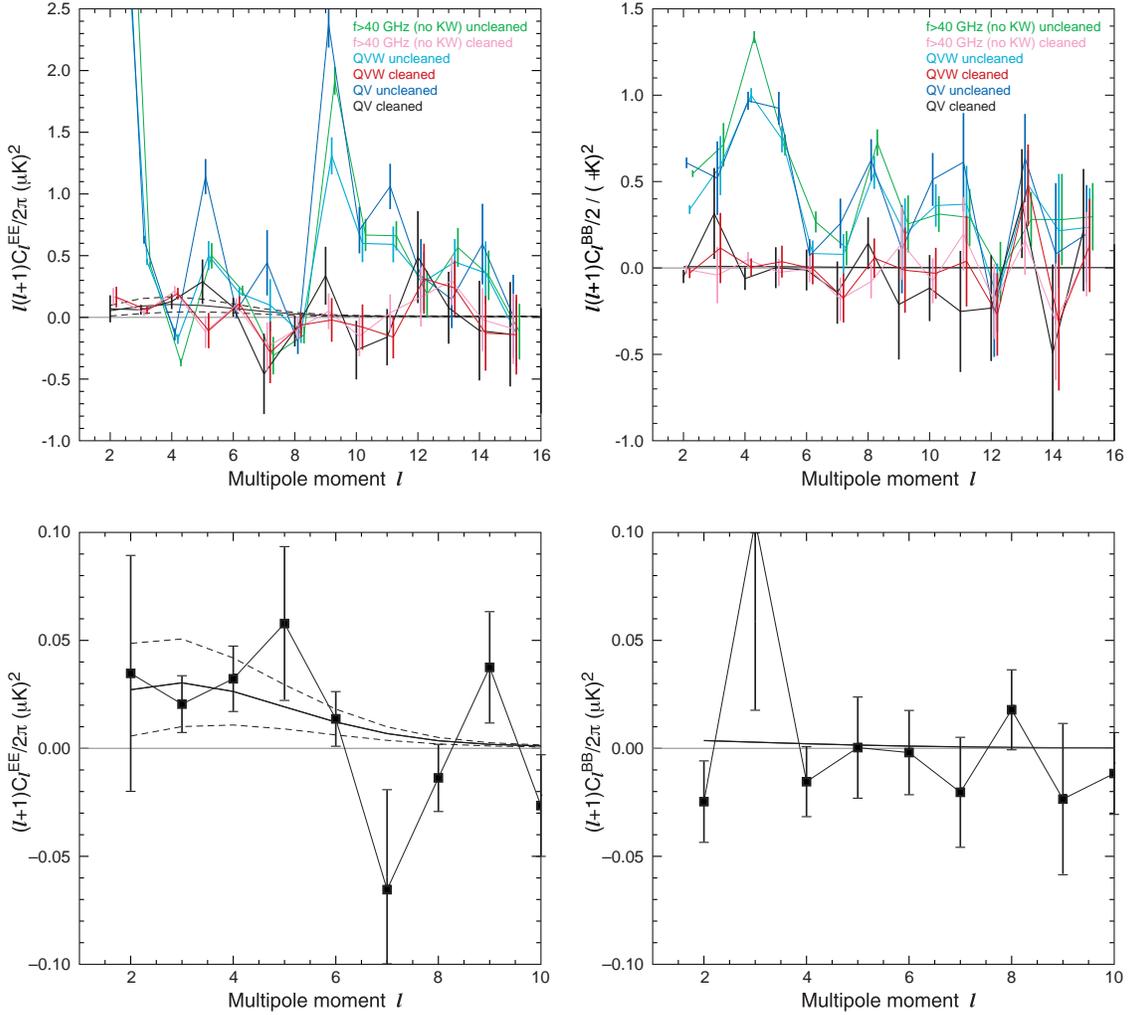}
\caption{{\it Top:} The EE and BB power spectra outside the P06 mask
before and after applying the KD3Pol foreground model. Different
colors show different frequency combinations. Negative values are possible
due to anticorrelations between foreground components, and to a lesser degree,
from the coupling between different values of $\ell$. Only statistical
uncertainties are shown. For EE,
the smooth black lines are the best fit model 
to the TT, TE, and EE data. The cosmic variance uncertainty is indicated
by the dashed lines. The EE values at $\ell=2$ are
5.8, 4.5, \& 5.5 ~$\mu$K$^2$ for f>40 (no KW), QVW, and QV combinations
respectively. To clean these to a level of 0.1~$\mu$K$^2$ requires
cleaning the Stokes $Q$ and $U$ maps to one part in eight.
The BB foreground emission is generally less than half the EE emission.
{\it Bottom:} Expanded plots of the QV data
for the P06 cut. The models are for $\tau=0.09$ and $r=0.3$.
The absence of any signal in BB is another indication that foreground
emission is not a significant contaminant. 
Note that the y-axis of the bottom plot has been divided by
one power of $\ell$ relative to the top plot.  
\label{fig:correebb} }
\end{figure*}
\clearpage 

\begin{figure*}[tb]
\epsscale{0.9}
\plotone{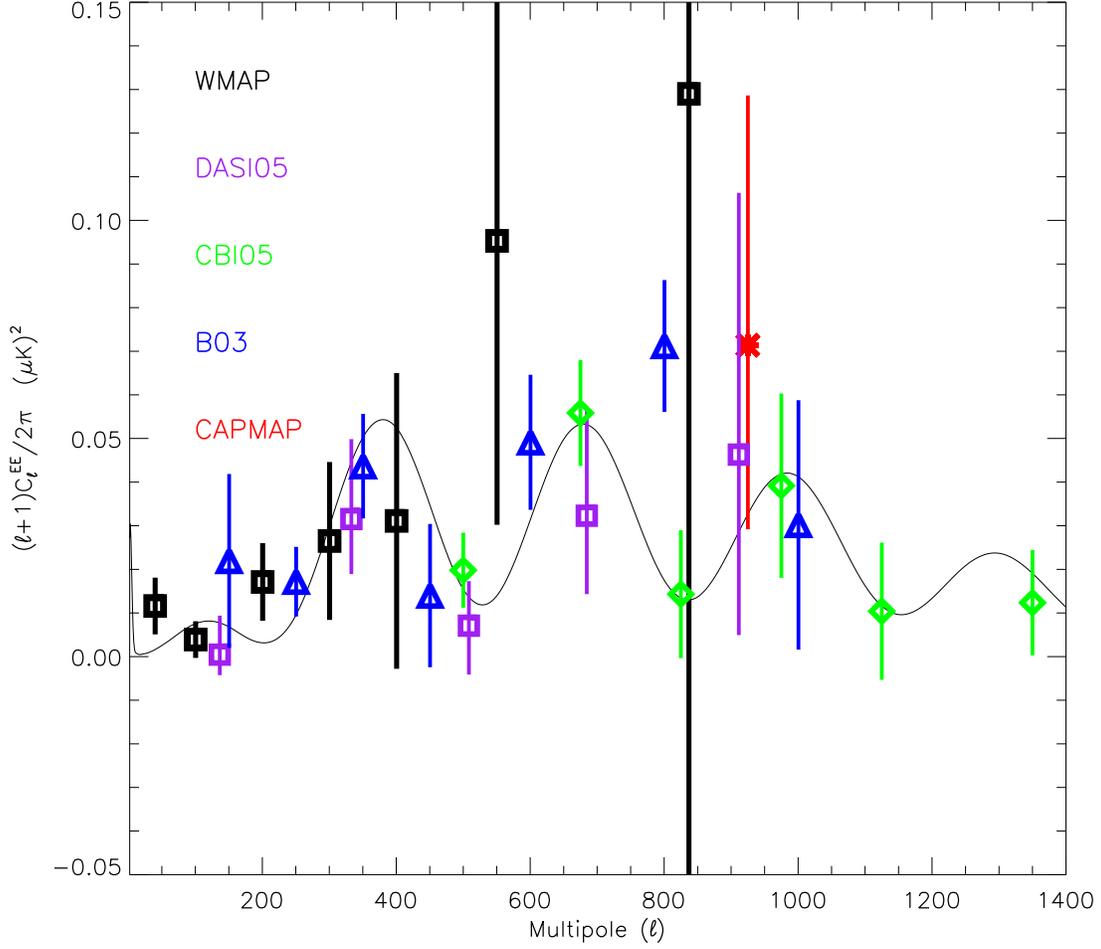}
\caption{ The EE spectrum at $\ell>40$ for all measurements of the CMB 
polarization. The curve is the best fit EE spectrum.
Note that the y axis has only one power of $\ell$.
The black boxes are the {\sl WMAP} data; the triangles are the BOOMERanG data;
the squares are the DASI data; the diamonds are the CBI data; and 
the asterisk is the CAPMAP data. The {\sl WMAP} data are the QVW combination.
For the first point, the cleaned value is used. For other
values, the raw values are used. The data are given in 
Table~\ref{tab:highee} 
\label{fig:highee} }
\end{figure*}
\clearpage 

\begin{figure*}[tb]
\plottwo{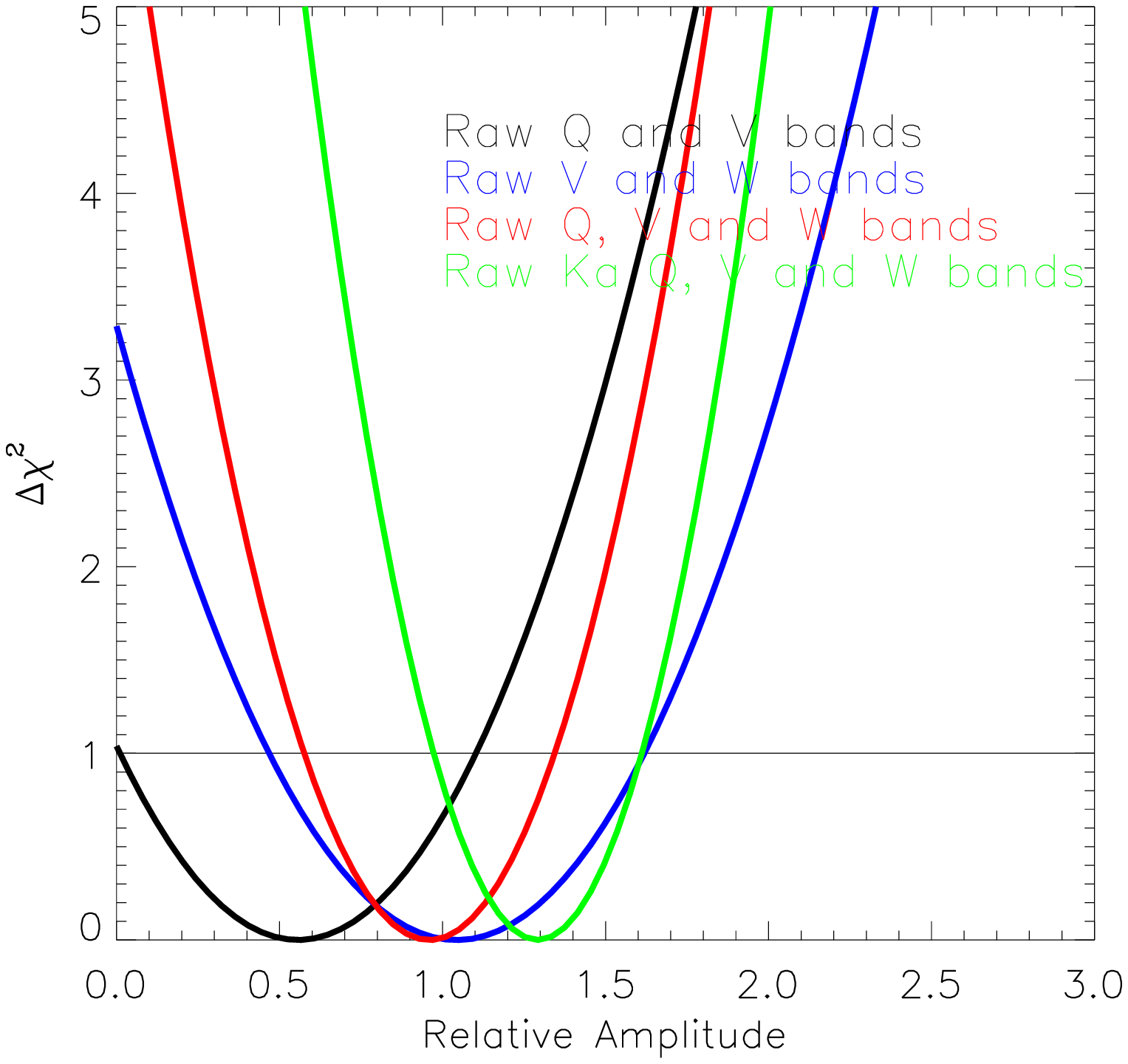}{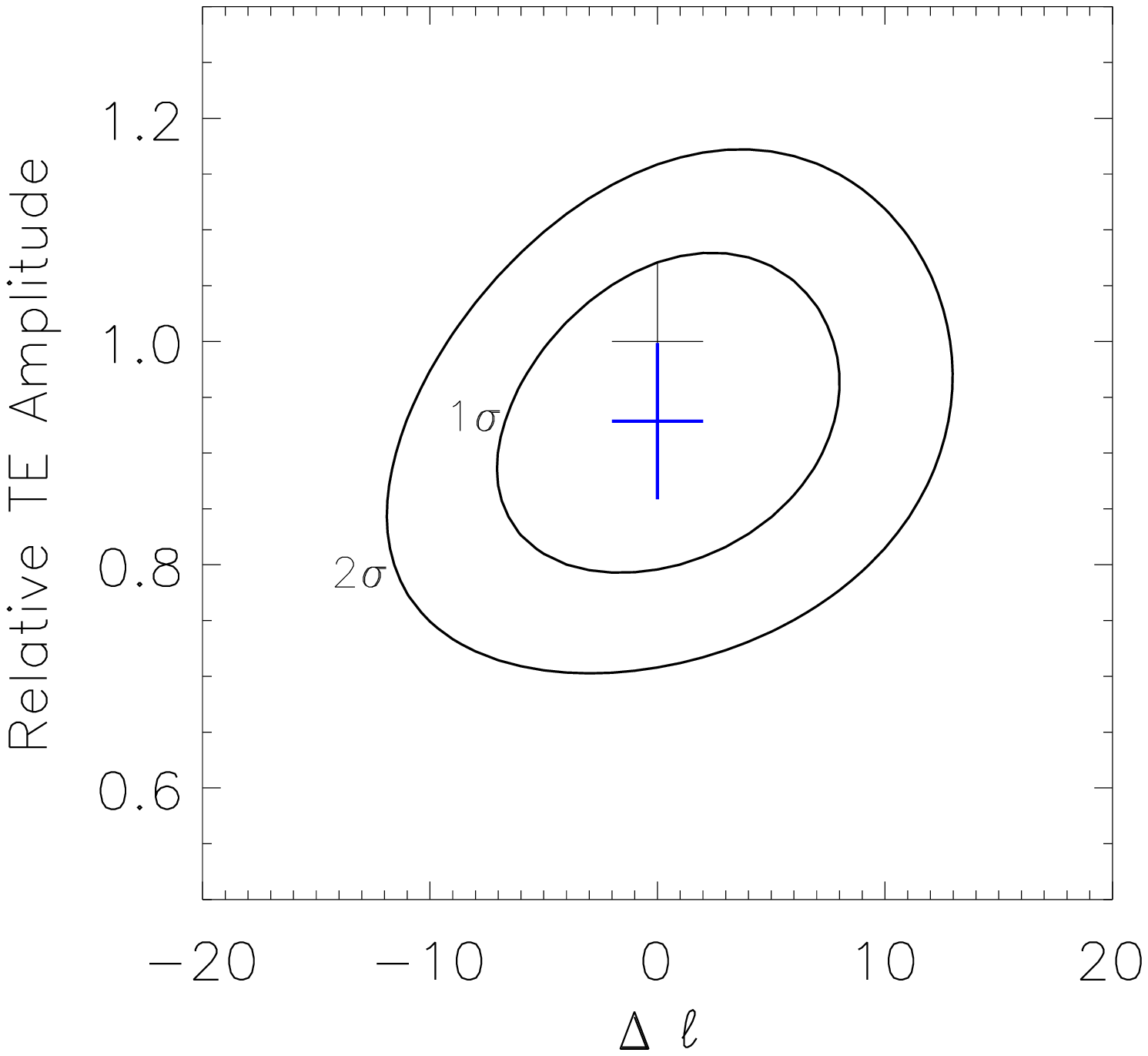}
\caption{ {\it Left:} Lines of $\Delta\chi^2$ for the 
fit given in Equation~\ref{eq:aee} vs. $A^{EE}$.
The different colors correspond to different frequency 
combinations. If the EE prediction from the TT measurements
describes the EE measurements, then the minimum would be
at $A^{EE}=1$. The line at $\Delta\chi^2=1$ corresponds
to the $1\sigma$ error. One can see that the $\ell>50$ EE
data are consistent with the model. {\it Right:} 
The amplitude and phase of the TE measurement with respect to the 
model predicted by the TT data. If the TE were completely 
predicted by the model based on TT, the contours would be
consistent with $A^{TE},\Delta\ell^{TE} = (1,0)$. It is clear that
the TT model describes the TE data as well.
The reduced $\chi^2$ for the best fit model is 0.67.
To convert $\Delta\ell$ to a phase angle in degrees, multiply 
by 1.18. 
\label{fig:eeteampphase} }
\end{figure*}
\clearpage 

\begin{figure*}[tb]
\epsscale{1.1}
\plottwo{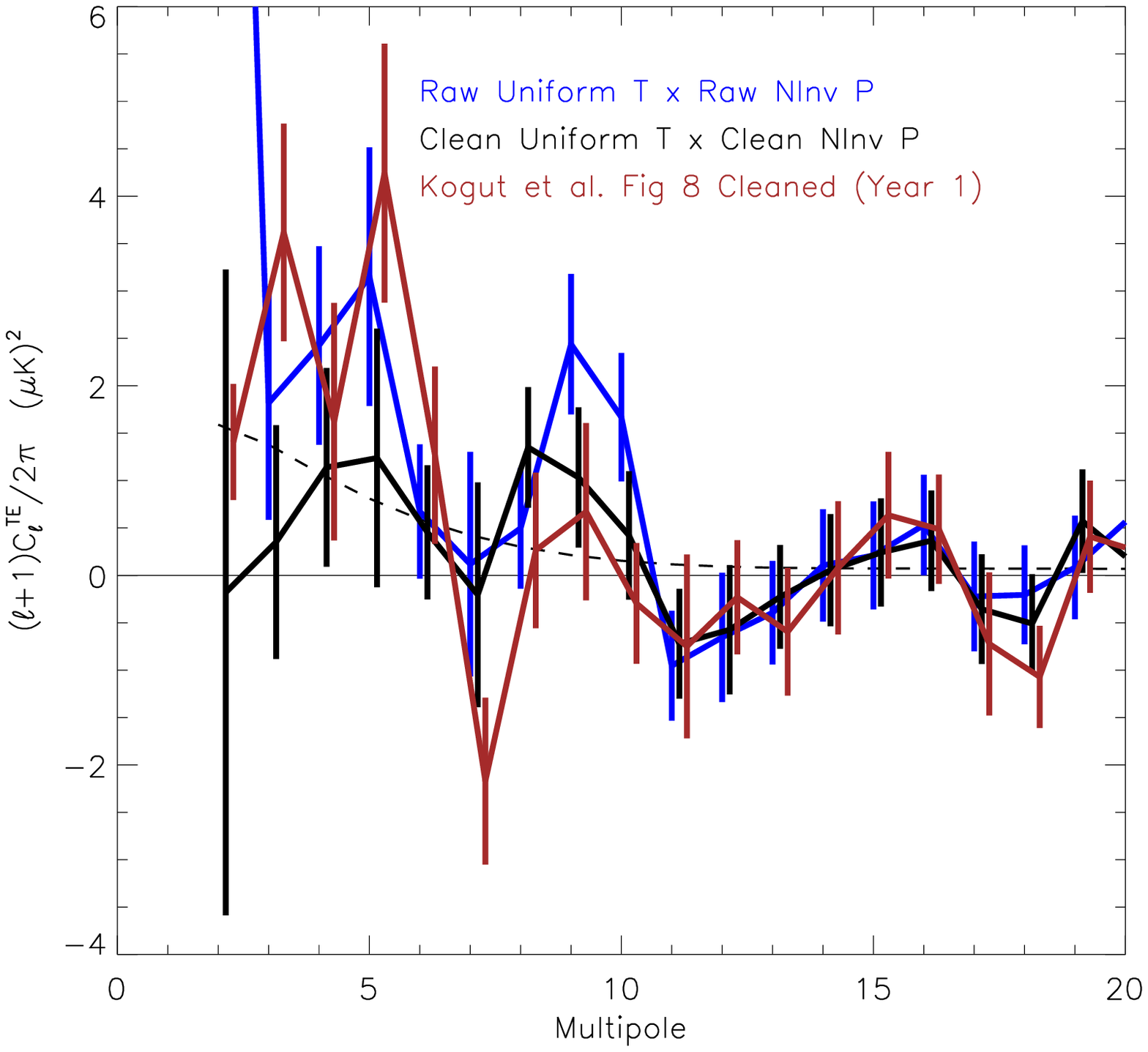}{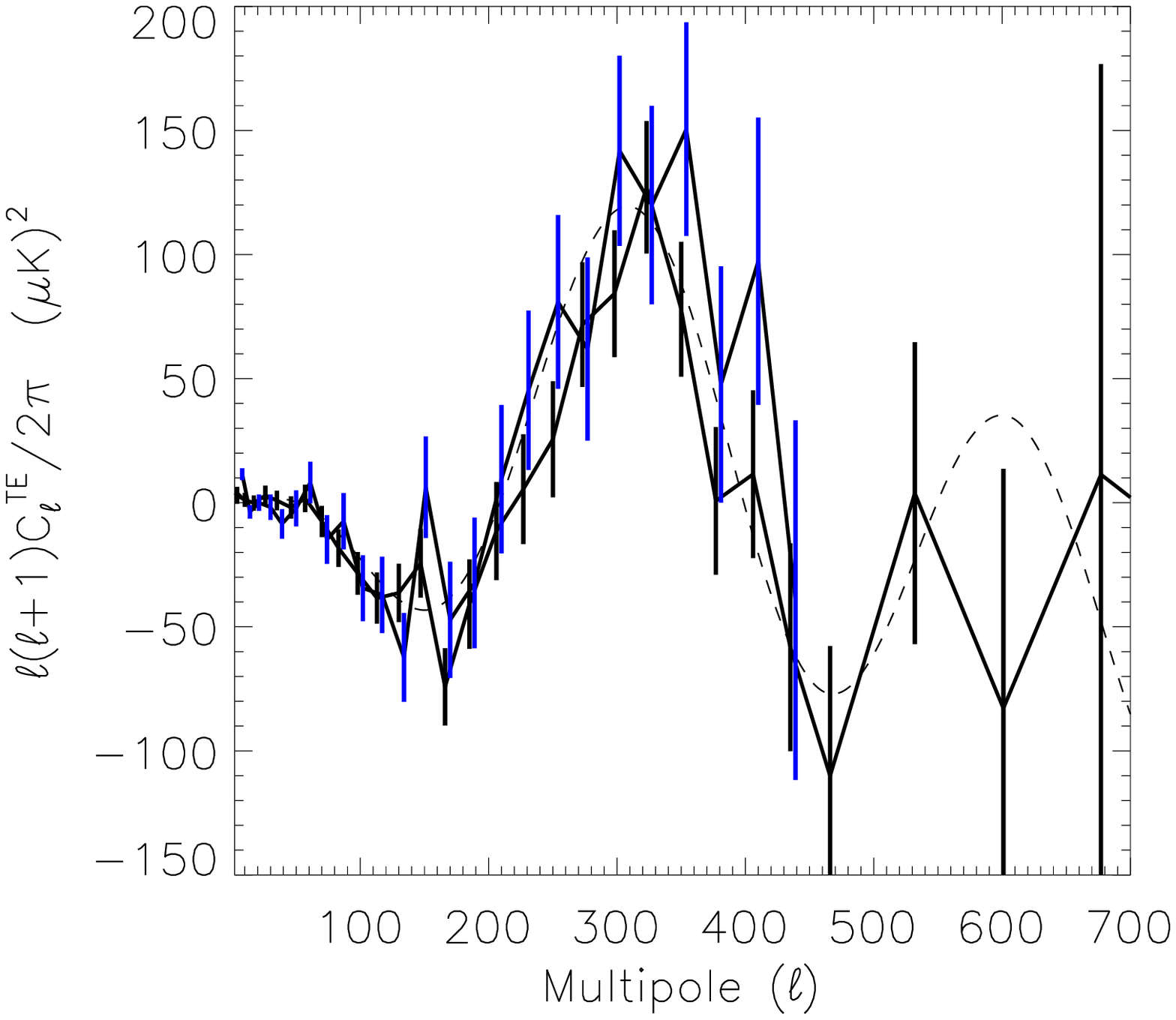}
\caption{The TE power spectra for high and low $\ell$ ranges
for the region outside the P06 mask in EE and TT.
{\it Left:} At low $\ell$ we use the QV combination
for polarization with full ${\bf N^{-1}}$ weighting, and for temperature
we use V band with uniform weighting. The black data points 
correspond to spectrum made with the 
KD3Pol cleaned polarization maps; the blue correspond to 
the same spectrum but without cleaning (the $\ell=2$ point is
at $17.8\pm3.4~\mu$K), and the brown are  
from \citet{kogut/etal:2003}. The black dashed line 
is the best fit model to all the \WMAP\ data. For the first-year data,
$\chi^2=35.3 $ for $\ell=2-10$ with a corresponding $PTE\approx0$.
For the three-year data, $\chi^2=9.4$ for $\ell=2-10$ evaluated 
relative to a null signal. The corresponding 
$PTE$ is $0.4$. When the three-year data are evaluated with respect to
the best-fit model, $\chi^2=5.4$ with a corresponding $PTE=0.79$. 
We find that the data sets are consistent with each other and that 
the three-year data prefer the $\tau=0.09$ model over the null signal
at the $2\sigma$ ($\Delta \chi^2=4$) level. However, the three-year data 
are also consistent with a null signal.
{\it Right:} The black data points show the three-year
TE spectrum. This was made using V band for temperature and the QV combination
for the E-mode of polarization. The blue data points are 
from \citet{kogut/etal:2003}.
The smooth dashed curve is the best fit model to the \WMAP\  data.
An additional zero crossing near $\ell=400$ is now present.
\label{fig:tespec} }
\end{figure*}
\clearpage 

\begin{figure*}[tb]
\epsscale{0.90}
\plotone{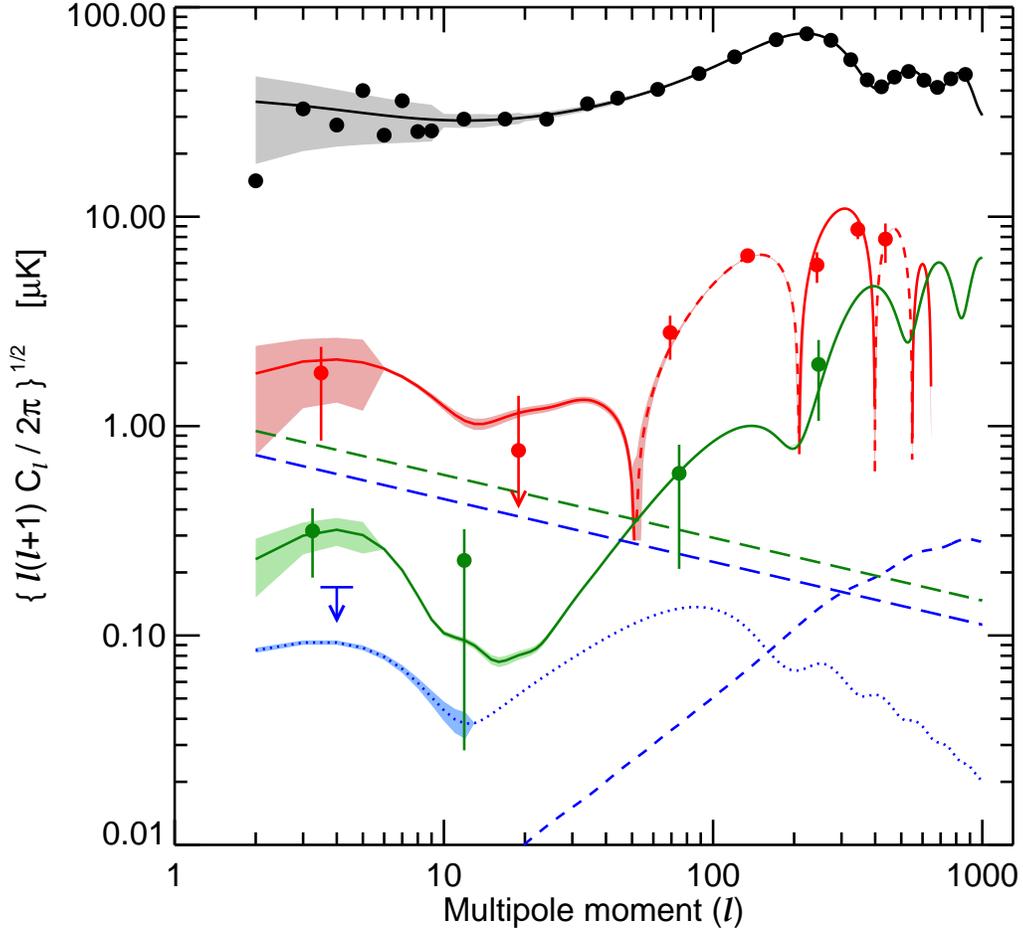}
\caption{Plots of signal for TT (black), TE (red), EE (green) for the 
best fit model.
The dashed line for TE indicates areas of anticorrelation. 
The cosmic variance is shown as a light swath around each model. It
is binned in $\ell$ in the same way as the data. Thus, its variations
reflect transitions between $\ell$ bin sizes. All error bars include 
the signal times noise term. The $\ell$ at which each point is plotted
is found from the weighted mean of the data comprising the bin.
This is most conspicuous for EE where the data are divided into
bins of $2\leq\ell\leq 5$, $6\leq\ell\leq 49$, $50\leq\ell\leq 199$,
and $200\leq\ell\leq 799$. The lowest $\ell$ point shows the cleaned QV data,
the next shows the cleaned QVW data, and the last two show the pre-cleaned
QVW data. There is possibly residual foreground contamination in the 
second point because our model is not so effective in this range
as discussed in the text. The level of foreground contamination in 
rightmost two EE points could be roughly $\sigma/2$.  
For BB (blue dots), we show a model with $r=0.3$.
It is dotted to indicate that at this time \WMAP\ only limits the signal.
We show the $1\sigma$ limit of $0.17~\mu$K for the weighted average 
of $\ell=2-10$. The BB lensing signal is shown as a blue dashed line.
The foreground model (Equation~\ref{eq:fgmod}) for synchrotron plus dust
emission is shown as straight dashed lines with green for EE and blue 
for BB. Both are evaluated at $\nu=65~$GHz. Recall that this is an 
average level and 
does not emphasize the $\ell$s where the emission is low.
\label{fig:grandspec} }
\end{figure*}
\clearpage 

\begin{figure*}[tb]
\epsscale{0.8}
\plotone{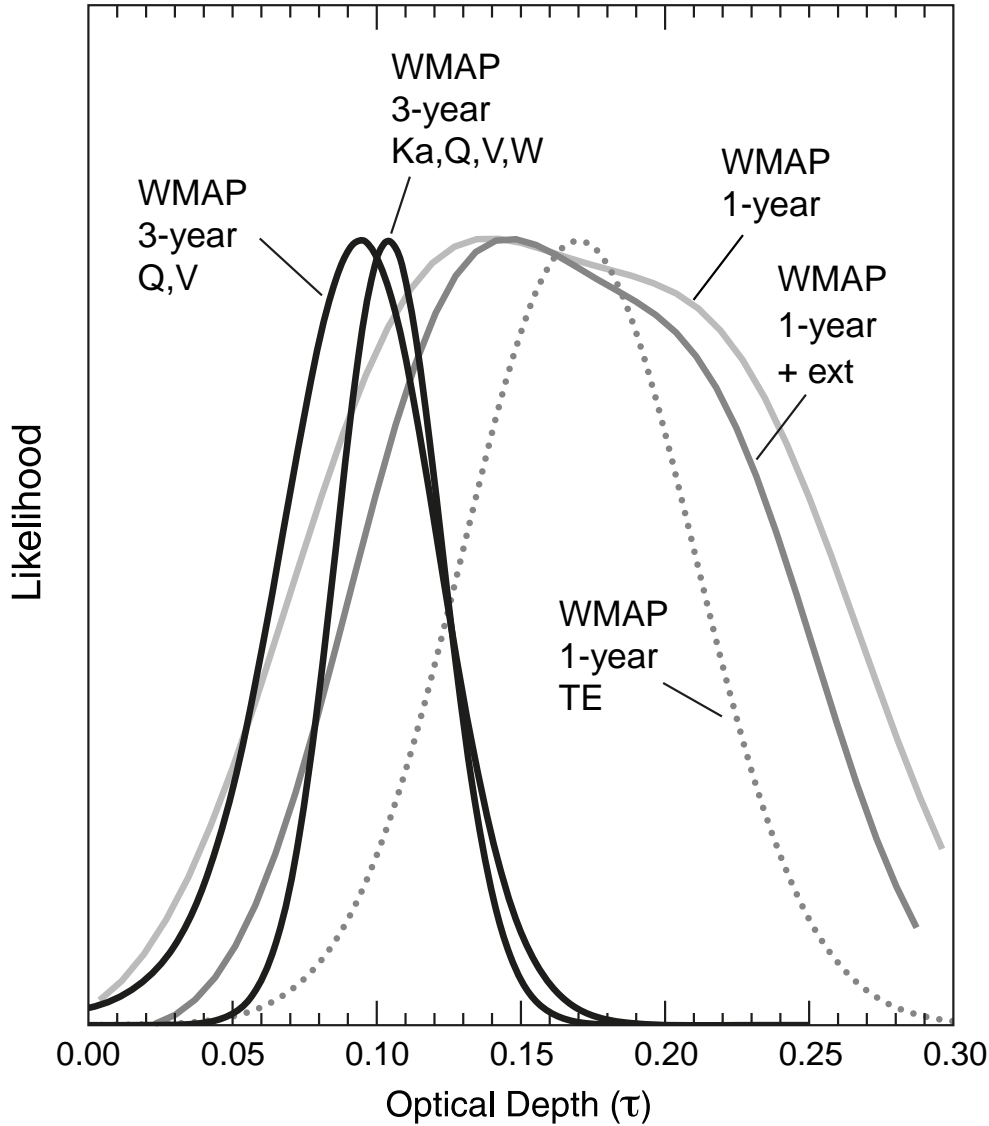}
\caption{The relative likelihoods of $\tau$
from the stand alone exact likelihood code and the first-year analysis.
For the three-year results, all parameters except $\tau$ and the 
scalar normalization, $A$, were held fixed as described in the text.
The solid curves (labeled ``\WMAP\ 3-years'') show 
the exact likelihood for the QV and KaQVW combinations for the combined
EE \& TE data. If the K-band directions had been used for the 
dust polarization template (\S 4.3), leading to inferior cleaning,
the likelihood curve would peak where the KaQVW does and have the width
of the QV curve. The similarity indicates that any foreground contamination 
is small. The two broadest curves are 
from \citet{spergel/etal:2003} and show the first-year 
likelihood for the \WMAP\ data alone and for \WMAP\ in combination 
with other data sets. The dotted line is $\tau$ likelihood for the first-year 
TE data as reported in \citet{kogut/etal:2003}. The curve has 
a mean of $\tau=0.17$ and width $\sigma=0.04$.
\label{fig:tau_likes} }
\end{figure*}
\clearpage 

\begin{figure*}[tb]
\epsscale{0.6}
\plotone{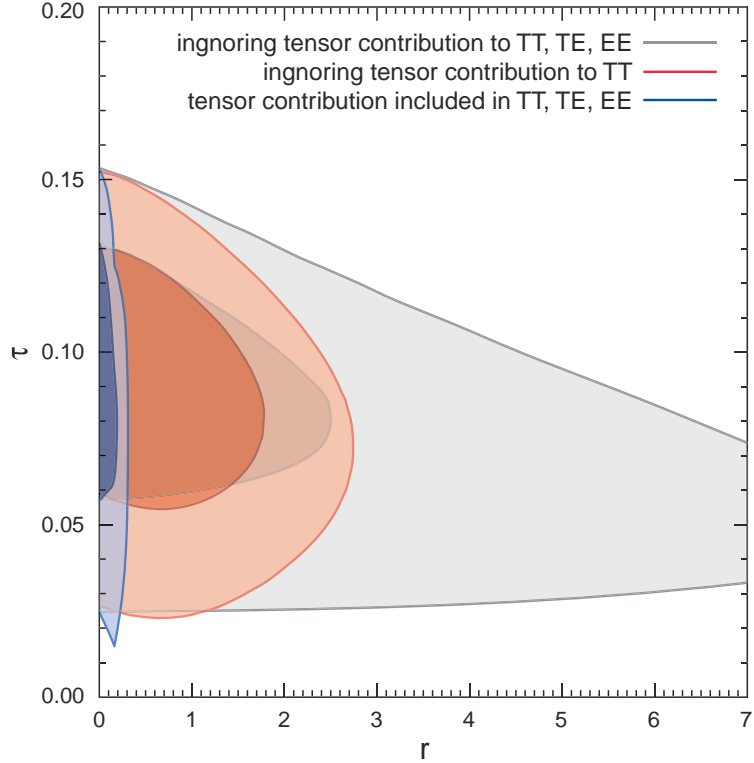}
\caption{The two dimensional likelihood as a function of
$\tilde\tau$ and $\tilde r$ for the BB spectrum. The contours 
indicate $1\sigma$ and $2\sigma$.
The $n_s$ parameter, which is degenerate with $\tau$ and $r$, has 
been set at $n_s=0.96$.
For the lightest contours, the tensor contribution to TT, TE, and EE 
is ignored.
Because $\tau$ is fully degenerate with $r$ when the data are restricted
to just BB, the limit is poor. 
The orange contours show the result when the TE and EE contributions 
are included, breaking the $r-\tau$ degeneracy.
The bluish contours show the result of including all data. The limit on $r$
is more restrictive than in \citet{spergel/etal:prep} because 
$n_s$ is fixed. When we marginalize over $\tilde\tau$, 
the 95\% upper limits on $\tilde r$ are 4.5, 2.2, and 0.27 for
the three cases respectively.
The plot shows that \WMAP's ability to constrain $r$ does not
yet come from the BB data. The plot also shows that \WMAP's ability
to limit $r$ depends critically on $\tau$. }
\label{fig:simple_rtau}
\end{figure*}
\clearpage 

\begin{figure}    
\epsscale{0.7}    
\plotone{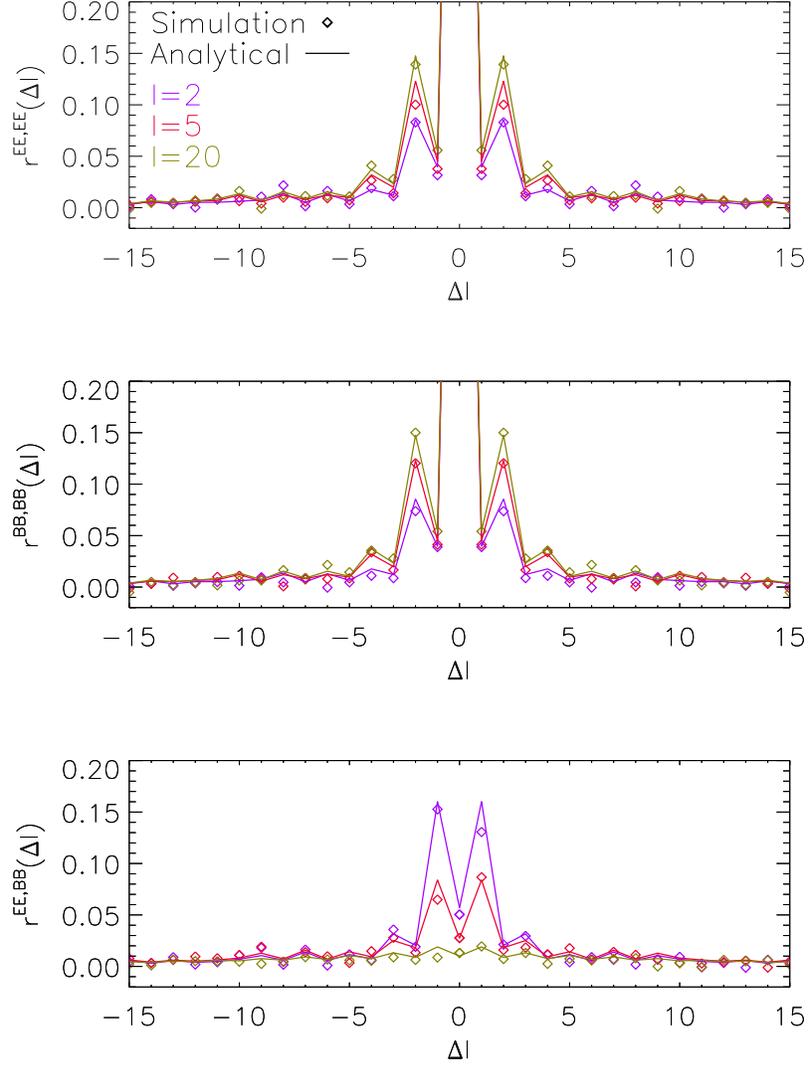}    
\caption{ The correlation coefficients of the Fisher matrices.     
The diamonds are derived from 100,000 Monte Carlo simulations, while    
the solid lines are the analytical formulae in the noise-dominated    
regime. In the simulations, the $B$-mode is noise only while the     
$E$-mode has some signals at low $l$ from reionization.}    
\label{fig:fisher}    
\end{figure}  
\clearpage 


\begin{thebibliography}{180}
\expandafter\ifx\csname natexlab\endcsname\relax\def\natexlab#1{#1}\fi

\bibitem[{Abbott \& Harari(1986)}]{abbott/harari:1986}
Abbott, L.~F. \& Harari, D.~D. 1986, Nucl. Phys., B264, 487

\bibitem[{{Baccigalupi} et~al.(2001){Baccigalupi}, {Burigana}, {Perrotta}, {De
  Zotti}, {La Porta}, {Maino}, {Maris}, \& {Paladini}}]{baccigalupi/etal:2001}
{Baccigalupi}, C., {Burigana}, C., {Perrotta}, F., {De Zotti}, G., {La Porta},
  L., {Maino}, D., {Maris}, M., \& {Paladini}, R. 2001, \aap, 372, 8

\bibitem[{{Banday} \& {Wolfendale}(1991)}]{banday/wolfendale:1991}
{Banday}, A.~J. \& {Wolfendale}, S. 1991, \mnras, 248, 705

\bibitem[{Barkats et~al.(2005)}]{barkats/etal:2005}
Barkats, D. et~al. 2005, \apjl, 619, L127

\bibitem[{{Barnes} et~al.(2002)}]{barnes/etal:2002}
{Barnes}, C., et~al. 2002, \apjs, 143, 567

\bibitem[{{Barnes} et~al.(2003)}]{barnes/etal:2003}
---. 2003, \apjs, 148, 51

\bibitem[{{Basko} \& {Polnarev}(1980)}]{basko/polnarev:1980}
{Basko}, M.~M. \& {Polnarev}, A.~G. 1980, \mnras, 191, 207

\bibitem[{Beck(2001)}]{beck:2001}
Beck, R. 2001, Space Science Reviews, 99, 243, kluwer Academic Publishers

\bibitem[{Beck(2006)}]{beck:2006}
Beck, R. 2006, in Polarisation 2005, ed. F.~B. .~M. Miville-Deschenes (EAS
  Publication Series)

\bibitem[{Beck et~al.(1996)Beck, Brandenburg, Moss, Shukurov, \&
  Sokoloff}]{beck/etal:1996}
Beck, R., Brandenburg, A., Moss, D., Shukurov, A., \& Sokoloff, D. 1996, Annu.
  Rev. Asron. Astrophys., 34, 155

\bibitem[{{Bennett} et~al.(2003{\natexlab{a}})}]{bennett/etal:2003c}
{Bennett}, C.~L., et~al. 2003{\natexlab{a}}, \apjs, 148, 97

\bibitem[{{Bennett} et~al.(2003{\natexlab{b}})}]{bennett/etal:2003}
---. 2003{\natexlab{b}}, \apj, 583, 1

\bibitem[{{Beno{\^i}t} et~al.(2004)}]{benoit/etal:2004}
{Beno{\^i}t}, A., et~al. 2004, \aap, 424, 571

\bibitem[{{Berdyugin} et~al.(2004){Berdyugin}, {Piirola}, \&
  {Teerikorpi}}]{berdyugin/piirola/teerikorpi:2004}
{Berdyugin}, A., {Piirola}, V., \& {Teerikorpi}, P. 2004, \aap, 424, 873

\bibitem[{{Berdyugin} \& {Teerikorpi}(2001)}]{berdyugin/teerikorpi:2001}
{Berdyugin}, A. \& {Teerikorpi}, P. 2001, \aa, 368, 635

\bibitem[{{Berdyugin} \& {Teerikorpi}(2002)}]{berdyugin/teerikorpi:2002}
---. 2002, \aap, 384, 1050

\bibitem[{{Berdyugin} et~al.(2001){Berdyugin}, {Teerikorpi}, {Haikala},
  {Hanski}, {Knude}, \& {Markkanen}}]{berdyugin/etal:2001}
{Berdyugin}, A., {Teerikorpi}, P., {Haikala}, L., {Hanski}, M., {Knude}, J., \&
  {Markkanen}, T. 2001, \aap, 372, 276

\bibitem[{{Berkhuijsen} et~al.(1971){Berkhuijsen}, {Haslam}, \&
  {Salter}}]{berkhuijsen/haslam/salter:1971}
{Berkhuijsen}, E.~M., {Haslam}, C.~G.~T., \& {Salter}, C.~J. 1971, \aap, 14,
  252

\bibitem[{{Bernardi} et~al.(2003){Bernardi}, {Carretti}, {Fabbri}, {Sbarra},
  {Poppi}, \& {Cortiglioni}}]{bernardi/etal:2003}
{Bernardi}, G., {Carretti}, E., {Fabbri}, R., {Sbarra}, C., {Poppi}, S., \&
  {Cortiglioni}, S. 2003, \mnras, 344, 347

\bibitem[{Blum(1959)}]{blum:1959}
Blum, E.~J. 1959, Annales D' Astrophysique, 22, 140

\bibitem[{Bond \& Efstathiou(1984)}]{bond/efstathiou:1984}
Bond, J.~R. \& Efstathiou, G. 1984, \apjl, 285, L45

\bibitem[{{Bond} \& {Efstathiou}(1987)}]{bond/efstathiou:1987}
{Bond}, J.~R. \& {Efstathiou}, G. 1987, \mnras, 226, 655

\bibitem[{Born \& Wolf(1980)}]{born/wolf:POO:6e}
Born, M. \& Wolf, E. 1980, Principles of Optics, sixth edn. (Pergamon Press)

\bibitem[{Boyle et~al.(2005)Boyle, Steinhardt, \&
  Turok}]{boyle/steinhardt/turok:2005}
Boyle, L., Steinhardt, P., \& Turok, N. 2005, \prl, \prl, submitted,
  (astro-ph/0507455)

\bibitem[{{Brouw} \& {Spoelstra}(1976)}]{brouw/spoelstra:1976}
{Brouw}, W.~N. \& {Spoelstra}, T.~A.~T. 1976, Astron. Astrophys. Suppl. Ser.,
  26, 129

\bibitem[{{Bunn} et~al.(2003){Bunn}, {Zaldarriaga}, {Tegmark}, \& {de
  Oliveira-Costa}}]{bunn/etal:2003}
{Bunn}, E.~F., {Zaldarriaga}, M., {Tegmark}, M., \& {de Oliveira-Costa}, A.
  2003, \prd, 67, 23501

\bibitem[{{Burn}(1966)}]{burn:1966}
{Burn}, B.~J. 1966, \mnras, 133, 67

\bibitem[{{Caderni} et~al.(1978){Caderni}, {Fabbri}, {Melchiorri},
  {Melchiorri}, \& {Natale}}]{caderni/etal:1978}
{Caderni}, N., {Fabbri}, R., {Melchiorri}, B., {Melchiorri}, F., \& {Natale},
  V. 1978, \prd, 17, 1908

\bibitem[{{Cioffi} \& {Jones}(1980)}]{cioffi/jones:1980}
{Cioffi}, D.~F. \& {Jones}, T.~W. 1980, \aj, 85, 368

\bibitem[{{Cortiglioni} \& {Spoelstra}(1995)}]{cortiglioni/spoelstra:1995}
{Cortiglioni}, S. \& {Spoelstra}, T.~A.~T. 1995, \aap, 302, 1

\bibitem[{{Coulson} et~al.(1994){Coulson}, {Crittenden}, \&
  {Turok}}]{coulson/crittenden/turok:1994}
{Coulson}, D., {Crittenden}, R.~G., \& {Turok}, N.~G. 1994, \prl, 73, 2390

\bibitem[{{Crittenden} et~al.(1993){Crittenden}, {Davis}, \&
  {Steinhardt}}]{crittenden/davis/steinhardt:1993}
{Crittenden}, R., {Davis}, R.~L., \& {Steinhardt}, P.~J. 1993, \apjl, 417, L13

\bibitem[{{Crittenden} et~al.(1995){Crittenden}, {Coulson}, \&
  {Turok}}]{crittenden/coulson/turok:1995}
{Crittenden}, R.~G., {Coulson}, D., \& {Turok}, N.~G. 1995, \prd, 52, 5402

\bibitem[{{Davis} \& {Greenstein}(1951)}]{davis/greenstein:1951}
{Davis}, L.~J. \& {Greenstein}, J.~L. 1951, \apj, 114, 206

\bibitem[{{de Oliveira-Costa} et~al.(1997){de Oliveira-Costa}, {Kogut},
  {Devlin}, {Netterfield}, {Page}, \& {Wollack}}]{deoliveira-costa/etal:1997}
{de Oliveira-Costa}, A., {Kogut}, A., {Devlin}, M.~J., {Netterfield}, C.~B.,
  {Page}, L.~A., \& {Wollack}, E.~J. 1997, \apjl, 482, L17+

\bibitem[{de~Oliveira-Costa et~al.(1998)de~Oliveira-Costa, Tegmark, Page, \&
  Boughn}]{deoliveira-costa/etal:1998a}
de~Oliveira-Costa, A., Tegmark, M., Page, L., \& Boughn, S. 1998, \apj, 509, L9

\bibitem[{de~Oliveira-Costa et~al.(1999)}]{deoliveira-costa/etal:1999}
de~Oliveira-Costa, A. et~al. 1999, \apj, 527, L9

\bibitem[{de~Oliveira-Costa et~al.(2003)}]{deoliveira-costa/etal:2003a}
---. 2003, \prd, 68

\bibitem[{de~Oliveira-Costa et~al.(2004)}]{deoliveira-costa/etal:2004}
---. 2004, Astrophys. J., 606, L89

\bibitem[{Dodelson(2003)}]{dodelson:MC}
Dodelson, S. 2003, Modern Cosmology (Academic Press)

\bibitem[{{Draine} \& {Lazarian}(1998)}]{draine/lazarian:1998a}
{Draine}, B.~T. \& {Lazarian}, A. 1998, \apjl, 494, L19

\bibitem[{{Draine} \& {Lazarian}(1999)}]{draine/lazarian:1999}
---. 1999, \apj, 512, 740

\bibitem[{{Drimmel} \& {Spergel}(2001)}]{drimmel/spergel:2001}
{Drimmel}, R. \& {Spergel}, D.~N. 2001, \apj, 556, 181

\bibitem[{Dumke et~al.(1995)Dumke, Krause, Wielebinski, \&
  Klein}]{dumke/etal:1995}
Dumke, M., Krause, M., Wielebinski, R., \& Klein, U. 1995, Astro. \& Astro.,
  302, 691

\bibitem[{Duncan et~al.(1999)Duncan, Reich, Reich, \& Furst}]{duncan/etal:1999}
Duncan, A., Reich, P., Reich, W., \& Furst, E. 1999, \aap, 350, 447

\bibitem[{Duncan et~al.(1995)Duncan, Haynes, Jones, \&
  Stewart}]{duncan/etal:1995}
Duncan, A.~R., Haynes, R.~F., Jones, K.~L., \& Stewart, R.~T. 1995, \mnras,
  277, 36

\bibitem[{{Efstathiou}(2004)}]{efstathiou:2004a}
{Efstathiou}, G. 2004, \mnras, 348, 885

\bibitem[{{Eisenhauer} et~al.(2003){Eisenhauer}, {Schoedel}, {Genzel}, {Ott},
  {Tecza}, {Abuter}, {Eckart}, \& {Alexander}}]{eisenhauer/etal:2003}
{Eisenhauer}, F., {Schoedel}, R., {Genzel}, R., {Ott}, T., {Tecza}, M.,
  {Abuter}, R., {Eckart}, A., \& {Alexander}, T. 2003, \apjl, 597, 121

\bibitem[{{Enomoto} et~al.(2002)}]{enomoto/etal:2002}
{Enomoto}, R., et~al. 2002, \nat, 416, 823

\bibitem[{En{\ss}lin et~al.(2006)En{\ss}lin, Waelkens, Vogt, \&
  Schekochihin}]{ensslin/etal:2006}
En{\ss}lin, T.~A., Waelkens, A., Vogt, C., \& Schekochihin, A.~A. 2006,
  Astronomische Nachrichten, 327, 626

\bibitem[{{Erickson}(1957)}]{erickson:1957}
{Erickson}, W.~C. 1957, \apj, 126, 480

\bibitem[{Faris(1967)}]{faris:1967}
Faris, J. 1967, J. Res. Nat. Bur. Stand., Engr. \& Instr., 71C, 153

\bibitem[{{Fernandez-Cerezo} et~al.(2006)}]{fernandez-cerezo/etal:2006}
{Fernandez-Cerezo}, S., et~al. 2006, ArXiv Astrophysics e-prints

\bibitem[{{Finkbeiner}(2003)}]{finkbeiner:2003}
{Finkbeiner}, D.~P. 2003, \apjs, 146, 407, accepted (astro-ph/0301558)

\bibitem[{{Finkbeiner}(2004)}]{finkbeiner:2004}
---. 2004, \apj, 614, 186

\bibitem[{Finkbeiner et~al.(1999)Finkbeiner, Davis, \&
  Schlegel}]{finkbeiner/davis/schlegel:1999}
Finkbeiner, D.~P., Davis, M., \& Schlegel, D.~J. 1999, \apj, 524, 867

\bibitem[{{Finkbeiner} et~al.(2004){Finkbeiner}, {Langston}, \&
  {Minter}}]{finkbeiner/langston/minter:2004}
{Finkbeiner}, D.~P., {Langston}, G.~I., \& {Minter}, A.~H. 2004, \apj, 617, 350

\bibitem[{{Finkbeiner} et~al.(2002){Finkbeiner}, {Schlegel}, {Frank}, \&
  {Heiles}}]{finkbeiner/etal:2002}
{Finkbeiner}, D.~P., {Schlegel}, D.~J., {Frank}, C., \& {Heiles}, C. 2002,
  \apj, 566, 898

\bibitem[{{Flett} \& {Henderson}(1979)}]{flett/henderson:1979}
{Flett}, A.~M. \& {Henderson}, C. 1979, \mnras, 189, 867

\bibitem[{{Flett} \& {Murray}(1991)}]{flett/murray:1991}
{Flett}, A.~M. \& {Murray}, A.~G. 1991, \mnras, 249, 4P

\bibitem[{{Fosalba} et~al.(2002){Fosalba}, {Lazarian}, {Prunet}, \&
  {Tauber}}]{fosalba/etal:2002}
{Fosalba}, P., {Lazarian}, A., {Prunet}, S., \& {Tauber}, J.~A. 2002, \apj,
  564, 762

\bibitem[{{Frewin} et~al.(1994){Frewin}, {Polnarev}, \&
  {Coles}}]{frewin/polnarev/coles:1994}
{Frewin}, R.~A., {Polnarev}, A.~G., \& {Coles}, P. 1994, \mnras, 266, L21+

\bibitem[{{Giardino} et~al.(2002){Giardino}, {Banday}, {G{\' o}rski},
  {Bennett}, {Jonas}, \& {Tauber}}]{giardino/etal:2002}
{Giardino}, G., {Banday}, A.~J., {G{\' o}rski}, K.~M., {Bennett}, K., {Jonas},
  J.~L., \& {Tauber}, J. 2002, \aap, 387, 82

\bibitem[{{G{\'o}rski} et~al.(1998){G{\'o}rski}, {Hivon}, \&
  {Wandelt}}]{gorski/hivon/wandelt:1998}
{G{\'o}rski}, K.~M., {Hivon}, E., \& {Wandelt}, B.~D. 1998, in Evolution of
  Large-Scale Structure: From Recombination to Garching

\bibitem[{{Greaves} et~al.(2003)}]{greaves/etal:2003}
{Greaves}, J.~S., et~al. 2003, \mnras, 340, 353

\bibitem[{Grishchuk(2001)}]{grishchuk:2001}
Grishchuk, L.~P. 2001, Lect. Notes Phys., 562, 167

\bibitem[{{Han}(2006)}]{han:2006b}
{Han}, J.-L. 2006, Chinese Journal of Astronony and Astrophysics, submitted
  (astro-ph/0603512)

\bibitem[{{Han} et~al.(2006){Han}, Manchester, Lyne, Qiao, \& van
  Straten}]{han/etal:2006}
{Han}, J.-L., Manchester, R.~N., Lyne, A.~G., Qiao, G.~J., \& van Straten, W.
  2006, \apj, accepted (astro-ph/0601357)

\bibitem[{{Harari} \& {Zaldarriaga}(1993)}]{harari/zaldarriaga:1993}
{Harari}, D.~D. \& {Zaldarriaga}, M. 1993, Physics Letters B, 319, 96

\bibitem[{Haslam et~al.(1982)Haslam, Stoffel, Salter, \&
  Wilson}]{haslam/etal:1982}
Haslam, C. G.~T., Stoffel, H., Salter, C.~J., \& Wilson, W.~E. 1982, \aaps, 47,
  1

\bibitem[{{Hedman} et~al.(2002){Hedman}, {Barkats}, {Gundersen}, {McMahon},
  {Staggs}, \& {Winstein}}]{hedman/etal:2002}
{Hedman}, M.~M., {Barkats}, D., {Gundersen}, J.~O., {McMahon}, J.~J., {Staggs},
  S.~T., \& {Winstein}, B. 2002, \apjl, 573, L73

\bibitem[{{Heiles}(2000)}]{heiles:2000}
{Heiles}, C. 2000, \aj, 119, 923

\bibitem[{{Hinshaw} et~al.(2003{\natexlab{a}})}]{hinshaw/etal:2003b}
{Hinshaw}, G., et~al. 2003{\natexlab{a}}, \apjs, 148, 63

\bibitem[{{Hinshaw} et~al.(2003{\natexlab{b}})}]{hinshaw/etal:2003}
---. 2003{\natexlab{b}}, \apjs, 148, 135

\bibitem[{Hinshaw et~al.(2006)}]{hinshaw/etal:prep}
Hinshaw, G. et~al. 2006, \apj, submitted

\bibitem[{{Hivon} et~al.(2002){Hivon}, {G{\' o}rski}, {Netterfield}, {Crill},
  {Prunet}, \& {Hansen}}]{hivon/etal:2002}
{Hivon}, E., {G{\' o}rski}, K.~M., {Netterfield}, C.~B., {Crill}, B.~P.,
  {Prunet}, S., \& {Hansen}, F. 2002, \apj, 567, 2

\bibitem[{{Hobbs} et~al.(1978){Hobbs}, {Maran}, \&
  {Brown}}]{hobbs/maran/brown:1978}
{Hobbs}, R.~W., {Maran}, S.~P., \& {Brown}, L.~W. 1978, \apj, 223, 373

\bibitem[{{Holder} et~al.(2003){Holder}, {Haiman}, {Kaplinghat}, \&
  {Knox}}]{holder/etal:2003}
{Holder}, G.~P., {Haiman}, Z., {Kaplinghat}, M., \& {Knox}, L. 2003, \apj, 595,
  13

\bibitem[{{Hu} et~al.(2003){Hu}, {Hedman}, \&
  {Zaldarriaga}}]{hu/hedman/zaldarriaga:2003}
{Hu}, W., {Hedman}, M.~M., \& {Zaldarriaga}, M. 2003, \prd, 67, 43004

\bibitem[{{Hu} \& {White}(1997)}]{hu/white:1997}
{Hu}, W. \& {White}, M. 1997, \prd, 56, 596

\bibitem[{{Hummel} et~al.(1991){Hummel}, {Dahlem}, {van der Hulst}, \&
  {Sukumar}}]{hummel/dahlem/vanderhulst:1991}
{Hummel}, E., {Dahlem}, M., {van der Hulst}, J.~M., \& {Sukumar}, S. 1991,
  \aap, 246, 10

\bibitem[{{Jarosik} et~al.(2003{\natexlab{a}})}]{jarosik/etal:2003}
{Jarosik}, N., et~al. 2003{\natexlab{a}}, \apjs, 145, 413

\bibitem[{{Jarosik} et~al.(2003{\natexlab{b}})}]{jarosik/etal:2003b}
---. 2003{\natexlab{b}}, \apjs, 148, 29

\bibitem[{Jarosik et~al.(2006)}]{jarosik/etal:prep}
Jarosik, N. et~al. 2006, \apj, submitted

\bibitem[{{Johnston} \& {Hobbs}(1969)}]{johnston/hobbs:1969}
{Johnston}, K.~J. \& {Hobbs}, R.~W. 1969, \apj, 158, 145

\bibitem[{Jones(1941)}]{jones:1941}
Jones, R.~C. 1941, J. Opt. Soc. Am., 31, 488

\bibitem[{{Jones} et~al.(1992){Jones}, {Klebe}, \&
  {Dickey}}]{jones/klebe/dickey:1992}
{Jones}, T.~J., {Klebe}, D., \& {Dickey}, J.~M. 1992, \apj, 389, 602

\bibitem[{{Junkes} et~al.(1993){Junkes}, {Haynes}, {Harnett}, \&
  {Jauncey}}]{junkes/etal:1993}
{Junkes}, N., {Haynes}, R.~F., {Harnett}, J.~I., \& {Jauncey}, D.~L. 1993,
  \aap, 269, 29

\bibitem[{{Kaiser}(1983)}]{kaiser:1983}
{Kaiser}, N. 1983, \mnras, 202, 1169

\bibitem[{{Kamionkowski} et~al.(1997){Kamionkowski}, {Kosowsky}, \&
  {Stebbins}}]{kamionkowski/kosowsky/stebbins:1997}
{Kamionkowski}, M., {Kosowsky}, A., \& {Stebbins}, A. 1997, \prd, 55, 7368

\bibitem[{{Keating} et~al.(2001){Keating}, {O'Dell}, {de Oliveira-Costa},
  {Klawikowski}, {Stebor}, {Piccirillo}, {Tegmark}, \&
  {Timbie}}]{keating/etal:2001}
{Keating}, B.~G., {O'Dell}, C.~W., {de Oliveira-Costa}, A., {Klawikowski}, S.,
  {Stebor}, N., {Piccirillo}, L., {Tegmark}, M., \& {Timbie}, P.~T. 2001,
  \apjl, 560, L1

\bibitem[{{Keating} et~al.(1998){Keating}, {Timbie}, {Polnarev}, \&
  {Steinberger}}]{keating/etal:1998}
{Keating}, B.~G., {Timbie}, P.~T., {Polnarev}, A., \& {Steinberger}, J. 1998,
  \apj, 495, 580

\bibitem[{Khoury et~al.(2002)Khoury, Ovrut, Seiberg, Steinhardt, \&
  Turok}]{khoury/etal:2002}
Khoury, J., Ovrut, B.~A., Seiberg, N., Steinhardt, P.~J., \& Turok, N. 2002,
  Phys. Rev., D65, 086007

\bibitem[{{Kogut} et~al.(1996){Kogut}, {Banday}, {Bennett}, {G{\'o}rski},
  {Hinshaw}, \& {Reach}}]{kogut/etal:1996a}
{Kogut}, A., {Banday}, A.~J., {Bennett}, C.~L., {G{\'o}rski}, K.~M., {Hinshaw},
  G., \& {Reach}, W.~T. 1996, \apj, 460, 1

\bibitem[{{Kogut} et~al.(2003)}]{kogut/etal:2003}
{Kogut}, A., et~al. 2003, \apjs, 148, 161

\bibitem[{Kosowsky(1996)}]{kosowsky:1996}
Kosowsky, A. 1996, Ann. Phys., 246, 49

\bibitem[{{Kovac} et~al.(2002){Kovac}, {Leitch}, {Pryke}, {Carlstrom},
  {Halverson}, \& {Holzapfel}}]{kovac/etal:2002}
{Kovac}, J.~M., {Leitch}, E.~M., {Pryke}, C., {Carlstrom}, J.~E., {Halverson},
  N.~W., \& {Holzapfel}, W.~L. 2002, \nat, 420, 772

\bibitem[{{Krauss} \& {White}(1992)}]{krauss/white:1992}
{Krauss}, L.~M. \& {White}, M. 1992, Physical Review Letters, 69, 869

\bibitem[{{Lagache}(2003)}]{lagache:2003}
{Lagache}, G. 2003, \aap, 405, 813

\bibitem[{{Lawson} et~al.(1987){Lawson}, {Mayer}, {Osborne}, \&
  {Parkinson}}]{lawson/etal:1987}
{Lawson}, K.~D., {Mayer}, C.~J., {Osborne}, J.~L., \& {Parkinson}, M.~L. 1987,
  \mnras, 225, 307

\bibitem[{{Lazarian} \& {Draine}(2000)}]{lazarian/draine:2000}
{Lazarian}, A. \& {Draine}, B.~T. 2000, \apjl, 536, L15

\bibitem[{{Leitch} et~al.(2005){Leitch}, {Kovac}, {Halverson}, {Carlstrom},
  {Pryke}, \& {Smith}}]{leitch/etal:2005}
{Leitch}, E.~M., {Kovac}, J.~M., {Halverson}, N.~W., {Carlstrom}, J.~E.,
  {Pryke}, C., \& {Smith}, M.~W.~E. 2005, \apj, 624, 10

\bibitem[{{Leitch} et~al.(1997){Leitch}, {Readhead}, {Pearson}, \&
  {Myers}}]{leitch/etal:1997}
{Leitch}, E.~M., {Readhead}, A.~C.~S., {Pearson}, T.~J., \& {Myers}, S.~T.
  1997, \apjl, 486, L23

\bibitem[{{Leitch} et~al.(2002)}]{leitch/etal:2002}
{Leitch}, E.~M., et~al. 2002, \nat, 420, 763

\bibitem[{{Lewis} et~al.(2000){Lewis}, {Challinor}, \&
  {Lasenby}}]{lewis/challinor/lasenby:2000}
{Lewis}, A., {Challinor}, A., \& {Lasenby}, A. 2000, \apj, 538, 473

\bibitem[{{Lewis} et~al.(2002){Lewis}, {Challinor}, \&
  {Turok}}]{lewis/challinor/turok:2002}
{Lewis}, A., {Challinor}, A., \& {Turok}, N. 2002, \prd, 65, 023505

\bibitem[{Liddle \& Lyth(2000)}]{liddle/lyth:CIALSS}
Liddle, A.~R. \& Lyth, D.~H. 2000, Cosmological inflation and large-scale
  structure (Cambridge University Press)

\bibitem[{{Limon} et~al.(2003)}]{limon/etal:2003}
{Limon}, M., et~al. 2003, {Wilkinson Microwave Anisotropy Probe ({\sl WMAP}):
  Explanatory Supplement},
  \texttt{http://lambda.gsfc.nasa.gov/data/map/doc/MAP\_supplement.pdf}

\bibitem[{Linde(1983)}]{linde:1983}
Linde, A.~D. 1983, Phys. Lett., B129, 177

\bibitem[{{Lubin} et~al.(1983){Lubin}, {Melese}, \&
  {Smoot}}]{lubin/melese/smoot:1983}
{Lubin}, P., {Melese}, P., \& {Smoot}, G. 1983, \apjl, 273, L51

\bibitem[{{Lubin} \& {Smoot}(1979)}]{lubin/smoot:1979}
{Lubin}, P.~M. \& {Smoot}, G.~F. 1979, Physical Review Letters, 42, 129

\bibitem[{{Lubin} \& {Smoot}(1981)}]{lubin/smoot:1981}
---. 1981, \apj, 245, 1

\bibitem[{{Matveenko} \& {Conklin}(1973)}]{matveenko/conklin:1973}
{Matveenko}, L.~I. \& {Conklin}, E.~K. 1973, Soviet Astronomy, 16, 726

\bibitem[{{Mayer} \& {Hollinger}(1968)}]{mayer/hollinger:1968}
{Mayer}, C.~H. \& {Hollinger}, J.~P. 1968, \apj, 151, 53

\bibitem[{Mitner \& Spangler(1996)}]{mitner/spangler:1996}
Mitner, A.~H. \& Spangler, S.~R. 1996, \apj, 458, 194

\bibitem[{Montgomery et~al.(1948)Montgomery, Dicke, \&
  Purcell}]{montgomery/dicke/purcell:POMC}
Montgomery, C.~G., Dicke, R.~H., \& Purcell, E.~M. 1948, MIT Radiation
  Laboratory Series, Vol.~8, Principles of Microwave Circuits (New York:
  McGraw-Hill)

\bibitem[{{Montgomery} et~al.(1971){Montgomery}, {Epstein}, {Oliver},
  {Dworetsky}, \& {Fogarty}}]{montgomery/etal:1971}
{Montgomery}, J.~W., {Epstein}, E.~E., {Oliver}, J.~P., {Dworetsky}, M.~M., \&
  {Fogarty}, W.~G. 1971, \apj, 167, 77

\bibitem[{Montroy et~al.(2005)}]{montroy/etal:2005}
Montroy, T.~E., et~al. 2005, \apj

\bibitem[{Mukhanov(2005)}]{mukhanov:PFC}
Mukhanov, V. 2005, Physical Foundations of Cosmology (Cambridge University
  Press)

\bibitem[{{Mukhanov} \& {Vikman}(2005)}]{mukhanov/vikman:2005}
{Mukhanov}, V. \& {Vikman}, A. 2005, LMU-ASC 78/05

\bibitem[{{Mukherjee} et~al.(2003){Mukherjee}, {Coble}, {Dragovan}, {Ganga},
  {Kovac}, {Ratra}, \& {Souradeep}}]{mukherjee/etal:2003}
{Mukherjee}, P., {Coble}, K., {Dragovan}, M., {Ganga}, K., {Kovac}, J.,
  {Ratra}, B., \& {Souradeep}, T. 2003, \apj, 592, 692

\bibitem[{{Nanos}(1979)}]{nanos:1979}
{Nanos}, G.~P. 1979, \apj, 232, 341

\bibitem[{Netterfield et~al.(1997)Netterfield, Devlin, Jarosik, Page, \&
  Wollack}]{netterfield/etal:1997}
Netterfield, C.~B., Devlin, M.~J., Jarosik, N., Page, L., \& Wollack, E.~J.
  1997, \apj, 474, 47

\bibitem[{Ng \& Ng(1995)}]{ng/ng:1995}
Ng, K.~L. \& Ng, K.-W. 1995, Phys. Rev., D51, 364

\bibitem[{Ng \& Speliotopoulos(1995)}]{ng/speliotopoulos:1995}
Ng, K.-W. \& Speliotopoulos, A.~D. 1995, Phys. Rev., D52, 2112

\bibitem[{{Page} et~al.(2003{\natexlab{a}})}]{page/etal:2003b}
{Page}, L., et~al. 2003{\natexlab{a}}, \apjs, 148, 39

\bibitem[{{Page} et~al.(2003{\natexlab{b}})}]{page/etal:2003}
---. 2003{\natexlab{b}}, \apj, 585, 566

\bibitem[{Peebles(1968)}]{peebles:1968}
Peebles, P. J.~E. 1968, \apj, 153, 1

\bibitem[{Peebles(1993)}]{peebles:POPC}
---. 1993, Principles of Physical Cosmology (Princeton, NJ: Princeton
  University Press)

\bibitem[{Peiris(2003)}]{peiris:phd}
Peiris, H. 2003, Ph.D. thesis, Princeton University

\bibitem[{{Peiris} et~al.(2003)}]{peiris/etal:2003}
{Peiris}, H.~V., et~al. 2003, \apjs, 148, 213

\bibitem[{{Penzias} \& {Wilson}(1965)}]{penzias/wilson:1965}
{Penzias}, A.~A. \& {Wilson}, R.~W. 1965, \apj, 142, 419

\bibitem[{{Piacentini} et~al.(2005)}]{piacentini/etal:2005}
{Piacentini}, F., et~al. 2005, ArXiv Astrophysics e-prints

\bibitem[{{Polnarev}(1985)}]{polnarev:1985}
{Polnarev}, A.~G. 1985, \azh, 62, 1041

\bibitem[{Ponthieu et~al.(2005)}]{ponthieu/etal:2005}
Ponthieu, N. et~al. 2005, Astron. \& Astro,, 607, 655

\bibitem[{Pritchard \& Kamionkowski(2005)}]{pritchard/kamionkowski:2005}
Pritchard, J.~R. \& Kamionkowski, M. 2005, Annals Phys., 318, 2

\bibitem[{{Prunet} \& {Lazarian}(1999)}]{prunet/lazarian:1999}
{Prunet}, S. \& {Lazarian}, A. 1999, in Microwave Foregrounds. Sloan Summit,
  Institute for Advanced Study, Princeton, New Jersey 14-15 November 1998. Eds:
  A. Rde Oliveira-Costa and M. Tegmark., Vol. 181 (Astronomical Society of the
  Pacific),  113

\bibitem[{{Readhead} et~al.(2004)}]{readhead/etal:2004}
{Readhead}, A.~C.~S., et~al. 2004, \apj, 609, 498

\bibitem[{Readhead et~al.(2004)}]{readhead/etal:2004b}
Readhead, A. C.~S., et~al. 2004, Science, 306, 761

\bibitem[{Rees(1968)}]{rees:1968}
Rees, M.~J. 1968, \apjl, 153, L1

\bibitem[{{Reich} \& {Reich}(1988)}]{reich/reich:1988}
{Reich}, P. \& {Reich}, W. 1988, \aaps, 74, 7

\bibitem[{Reich(2006)}]{reich:2006}
Reich, W. 2006, in Cosmic Polarization, Ed: Roberto Fabbri (Research Signpost)

\bibitem[{{Reid} \& {Brunthaler}(2005)}]{reid/brunthaler:2005}
{Reid}, M. \& {Brunthaler}, A. 2005, in Future Directions in High Resolution
  Astronomy: The 10th Anniversary of the VLBA, ASP Conference Proceedings, Eds:
  J. Romney and M. Reid., Vol. 340 (Astronomical Society of the Pacific),  253

\bibitem[{{Rybicki} \& {Lightman}(1979)}]{rybicki/lightman:1979}
{Rybicki}, G.~B. \& {Lightman}, A. 1979, {Radiative Processes in Astrophysics }
  (Wiley \& Sons: New York)

\bibitem[{{Sault} et~al.(1996){Sault}, {Hamaker}, \&
  {Bregman}}]{sault/hamaker/bregman:1996}
{Sault}, R.~J., {Hamaker}, J.~P., \& {Bregman}, J.~D. 1996, \aaps, 117, 149

\bibitem[{{Seljak}(1997)}]{seljak:1997}
{Seljak}, U. 1997, \apj, 482, 6

\bibitem[{{Seljak} \& {Zaldarriaga}(1996)}]{seljak/zaldarriaga:1996}
{Seljak}, U. \& {Zaldarriaga}, M. 1996, \apj, 469, 437

\bibitem[{{Sievers} et~al.(2005)}]{sievers/etal:2005}
{Sievers}, J.~L., et~al. 2005, ArXiv Astrophysics e-prints

\bibitem[{{Sironi} et~al.(1997){Sironi}, {Boella}, {Bonelli}, {Brunetti},
  {Cavaliere}, {Gervasi}, {Giardino}, \& {Passerini}}]{sironi/etal:1997}
{Sironi}, G., {Boella}, G., {Bonelli}, G., {Brunetti}, L., {Cavaliere}, F.,
  {Gervasi}, M., {Giardino}, G., \& {Passerini}, A. 1997, New Astronomy, 3, 1

\bibitem[{Slosar et~al.(2004)Slosar, Seljak, \&
  Makarov}]{slosar/seljak/makarov:2004}
Slosar, A., Seljak, U., \& Makarov, A. 2004, Phys. Rev., D69, 123003

\bibitem[{{Sofue} et~al.(1986){Sofue}, {Fujimoto}, \&
  {Wielebinski}}]{sofue/etal:1986}
{Sofue}, Y., {Fujimoto}, M., \& {Wielebinski}, R. 1986, \araa, 24, 459

\bibitem[{Spergel \& Zaldarriaga(1997)}]{spergel/zaldarriaga:1997}
Spergel, D.~N. \& Zaldarriaga, M. 1997, \prl, 79, 2180

\bibitem[{{Spergel} et~al.(2003)}]{spergel/etal:2003}
{Spergel}, D.~N., et~al. 2003, \apjs, 148, 175

\bibitem[{Spergel et~al.(2006)}]{spergel/etal:prep}
Spergel, D.~N. et~al. 2006, \apj, submitted

\bibitem[{{Starobinsky}(1979)}]{starobinsky:1979}
{Starobinsky}, A.~A. 1979, ZhETF Pis ma Redaktsiiu, 30, 719

\bibitem[{{Steinhardt} \& {Turok}(2002)}]{steinhardt/turok:2002}
{Steinhardt}, P.~J. \& {Turok}, N. 2002, Science, 296, 1436

\bibitem[{{Sukumar} \& {Allen}(1991)}]{sukumar/allen:1991}
{Sukumar}, S. \& {Allen}, R. 1991, \apj, 382, 100

\bibitem[{Taylor \& Cordes(1993)}]{taylor/cordes:1993}
Taylor, J.~H. \& Cordes, J.~M. 1993, \apj, 411, 674

\bibitem[{Tegmark et~al.(2000)Tegmark, Eisenstein, Hu, \&
  de~Oliveira-Costa}]{tegmark/etal:2000}
Tegmark, M., Eisenstein, D.~J., Hu, W., \& de~Oliveira-Costa, A. 2000, \apj,
  530, 133, astro-ps/9905257

\bibitem[{{Tinbergen}(1996)}]{tinbergen:AP}
{Tinbergen}, J. 1996, {Astronomical polarimetry} (Cambridge University Press)

\bibitem[{Torbet et~al.(1999)}]{torbet/etal:1999}
Torbet, E. et~al. 1999, \apjl, 521, L79

\bibitem[{{Tucci} et~al.(2002){Tucci}, {Carretti}, {Cecchini}, {Nicastro},
  {Fabbri}, {Gaensler}, {Dickey}, \& {McClure-Griffiths}}]{tucci/etal:2002}
{Tucci}, M., {Carretti}, E., {Cecchini}, S., {Nicastro}, L., {Fabbri}, R.,
  {Gaensler}, B.~M., {Dickey}, J.~M., \& {McClure-Griffiths}, N.~M. 2002, \apj,
  579, 607

\bibitem[{{Turner}(1997)}]{turner:1997}
{Turner}, M.~S. 1997, \prd, 55, 435

\bibitem[{Turner et~al.(1993)Turner, White, \&
  Lidsey}]{turner/white/lidsey:1993}
Turner, M.~S., White, M.~J., \& Lidsey, J.~E. 1993, Phys. Rev., D48, 4613

\bibitem[{{Turok}(1996)}]{turok:1996}
{Turok}, N. 1996, \apjl, 473, L5

\bibitem[{{Uyan{\i}ker} et~al.(1999){Uyan{\i}ker}, {Fuerst}, {Reich}, {Reich},
  \& {Wielebinski}}]{uyaniker/etal:1999}
{Uyan{\i}ker}, B., {Fuerst}, E., {Reich}, W., {Reich}, P., \& {Wielebinski}, R.
  1999, Astron. Astrophys. Suppl. Ser., 138, 31

\bibitem[{{Verde} et~al.(2006){Verde}, {Peiris}, \&
  {Jimenez}}]{verde/peiris/jimenez:2006}
{Verde}, L., {Peiris}, H., \& {Jimenez}, R. 2006, JCAP, 019

\bibitem[{{Verde} et~al.(2003)}]{verde/etal:2003}
{Verde}, L., et~al. 2003, \apjs, 148, 195

\bibitem[{Wang(1996)}]{wang:1996}
Wang, Y. 1996, Phys. Rev., D53, 639

\bibitem[{{Watson} et~al.(2005){Watson}, {Rebolo}, {Rubi{\~n}o-Mart{\'{\i}}n},
  {Hildebrandt}, {Guti{\'e}rrez}, {Fern{\'a}ndez-Cerezo}, {Hoyland}, \&
  {Battistelli}}]{watson/etal:2005}
{Watson}, R.~A., {Rebolo}, R., {Rubi{\~n}o-Mart{\'{\i}}n}, J.~A.,
  {Hildebrandt}, S., {Guti{\'e}rrez}, C.~M., {Fern{\'a}ndez-Cerezo}, S.,
  {Hoyland}, R.~J., \& {Battistelli}, E.~S. 2005, \apjl, 624, L89

\bibitem[{{Weiss}(1984)}]{weiss:1984}
{Weiss}, R. 1984, \araa, 12, 90

\bibitem[{Wielebinski(2005)}]{wielebinski:2005}
Wielebinski, R. 2005, in Cosmic Magnetic Fields, ed. R.~W. .~R. Reck, Vol. 664
  (LSP, Berlin: Springer), ~89

\bibitem[{Wollack et~al.(1993)Wollack, Jarosik, Netterfield, Page, \&
  Wilkinson}]{wollack/etal:1993}
Wollack, E.~B., Jarosik, N., Netterfield, C.~B., Page, L., \& Wilkinson, D.
  1993, \apj, 419, L49

\bibitem[{{Wolleben} et~al.(2005){Wolleben}, {Landecker}, {Reich}, \&
  {Wielebinski}}]{wolleben/etal:2005}
{Wolleben}, M., {Landecker}, T.~L., {Reich}, W., \& {Wielebinski}, R. 2005,
  ArXiv Astrophysics e-prints

\bibitem[{{Wright} et~al.(1996){Wright}, {Hinshaw}, \&
  {Bennett}}]{wright/hinshaw/bennett:1996}
{Wright}, E.~L., {Hinshaw}, G., \& {Bennett}, C.~L. 1996, \apjl, 458, L53

\bibitem[{{Wright} \& {Forster}(1980)}]{wright/forster:1980}
{Wright}, M. C.~H. \& {Forster}, J.~R. 1980, \apj, 239, 873

\bibitem[{Zaldarriaga \& Harari(1995)}]{zaldarriaga/harari:1995}
Zaldarriaga, M. \& Harari, D.~D. 1995, Phys. Rev., D52, 3276

\bibitem[{{Zaldarriaga} \& {Seljak}(1997)}]{zaldarriaga/seljak:1997}
{Zaldarriaga}, M. \& {Seljak}, U. 1997, \prd, 55, 1830

\bibitem[{{Zeldovich} et~al.(1969){Zeldovich}, {Kurt}, \&
  {Syunyaev}}]{zeldovich/kurt/sunyaev:1969}
{Zeldovich}, Y.~B., {Kurt}, V.~G., \& {Syunyaev}, R.~A. 1969, Journal of
  Experimental and Theoretical Physics, 28, 146

\bibitem[{Zhang et~al.(2005)Zhang, Yuan, Zhao, \& Chen}]{zhang/etal:2005}
Zhang, Y., Yuan, Y., Zhao, W., \& Chen, Y.-T. 2005, Class. Quant. Grav., 22,
  1383

\end{thebibliography}
\end{document}